\documentclass[preprint,12pt]{elsarticle}

\usepackage{amssymb}
\usepackage{amsmath}
\usepackage{amsthm}

\usepackage{booktabs}%
\usepackage{subcaption}
\usepackage{float}
\usepackage[most]{tcolorbox}
\usepackage{makecell}
\usepackage{caption}
\usepackage{listings}%
\usepackage{graphicx}
\usepackage{subcaption}
\usepackage{setspace}
\usepackage{multirow}%
\usepackage{url}

\journal{Computers and Fluids}

\begin{document}

\begin{frontmatter}



\title{Data-Driven RANS Closures Using a Relative Importance Term Analysis Based Classifier for 2D and 3D Separated Flows} %% Article title


\author[label_1,label_3]{Tyler Buchanan} %% Author name
\ead{t.s.b.buchanan@tudelft.nl}
\author[label_2]{Monica Lăcătuş}
\ead{m.i.l.lacatus@tudelft.nl}
\author[label_3]{Alastair West}
\ead{alastair.west@williamsf1.com}
\author[label_1]{Richard P. Dwight}
\ead{r.p.dwight@tudelft.nl}

\affiliation[label_1]{organization={Faculty of Aerospace Engineering, Delft University of Technology},%Department and Organization
            addressline={Kluyverweg 1}, 
            city={Delft},
            postcode={2600GB}, 
            state={Zuid-Holland},
            country={The Netherlands}}
            
\affiliation[label_2]{organization={Faculty of Electrical Engineering, Mathematics and Computer Science, Delft University of Technology},%Department and Organization
            addressline={Kluyverweg 1}, 
            city={Delft},
            postcode={2600GB}, 
            state={Zuid-Holland},
            country={The Netherlands}}
\affiliation[label_3]{organization={Williams Racing Formula One Team},%Department and Organization 
            city={Grove},
            postcode={OX12 0DQ}, 
            state={Oxfordshire},
            country={United Kingdom}}
%% Abstract
\begin{abstract}
This study presents a novel approach for enhancing Reynolds-averaged Navier-Stokes (RANS) turbulence modeling through the application of a Relative Importance Term Analysis (RITA) methodology to develop a new zonally-augmented $k-\omega$ SST model. Traditional Linear Eddy Viscosity Models often struggle with separated flows. Our approach introduces a physics-based binary classifier that systematically identifies separated shear layers requiring correction by analyzing the relative magnitudes of terms in the turbulence kinetic energy equation. Using symbolic regression, we develop compact correction terms for Reynolds stress anisotropy and turbulent kinetic energy production. Trained on two-dimensional configurations, our model demonstrates significant improvements in predicting separation dynamics while maintaining baseline performance and fully attached flows. Generalization tests on Ahmed body and Faith Hill three-dimensional configurations confirm robust transferability, establishing an effective methodology for targeted enhancement of RANS predictions in separated flows.
\end{abstract}

%\includegraphics[width=\textwidth]{Figures/Ahmed-Body/ahmed_wake_comparison_streamwise.png}

%\item Introduced RITA (Relative Importance Term Analysis), a physics-based methodology that identifies and corrects RANS model deficiencies in separated flows using dimensionless ratios from the turbulence kinetic energy equation.
%\item Developed a frame-invariant classifier that selectively activates correction terms in shear layers while maintaining baseline performance in well-predicted regions, demonstrating improved prediction accuracy across 2D benchmarks and complex 3D geometries with the zonally-augmented $k-\omega$ SST model.

%% Keywords

\end{frontmatter}



\section{Introduction}\label{sec1}
Turbulence modelling stands at a crossroads in Computational Fluid Dynamics (CFD). Industry-standard Reynolds-averaged Navier-Stokes (RANS) simulations are widely adopted for their computational efficiency but often lack accuracy in predicting complex flow phenomena. Conversely, high-fidelity methods such as Large Eddy Simulation (LES) and Direct Numerical Simulation (DNS) offer superior accuracy in resolving turbulent scales but remain computationally prohibitive for routine engineering applications. This dichotomy between computational efficiency and predictive accuracy motivates the development of innovative turbulence modelling techniques that bridge the gap between RANS simulations and high-fidelity methods.

Despite growing computational power, RANS is expected to remain the workhorse of engineering CFD for decades \cite{duraisamy_turbulence_2019}, primarily due to its ability to cheaply perform design optimization workflows. In contrast, LES remains computationally prohibitive for such iterative design processes. However, RANS approaches face critical limitations when predicting fundamental flow phenomena such as separation, reattachment, and strong adverse pressure gradients. Linear Eddy Viscosity Models (LEVMs), which assume a linear relationship between the Reynolds stress tensor and mean strain rate tensor, do not account for turbulence anisotropy \cite{wilcox_turbulence_2006,schmitt_about_2007}, resulting in insufficient turbulent mixing in shear layers \cite{rumsey_exploring_2009,li_aerodynamic_2020}.

These limitations have spurred the development of data-driven turbulence modelling approaches in recent years. Early data-driven RANS investigations focused primarily on addressing model coefficient optimization and uncertainty quantification \cite{platteeuw_uncertainty_2008,cheung_bayesian_2011,edeling_bayesian_2014,ray_bayesian_2016}. Subsequently, researchers tackled more fundamental structural uncertainties by developing corrective terms or new model structures from high-fidelity data \cite{tracey_machine_2015,gamahara_searching_2017,maulik_neural_2017,ling_reynolds_2016}.

While these data-driven methods improved accuracy, they often lacked physical interpretability. \citet{weatheritt_development_2017} addressed this limitation by developing a sparse symbolic regression framework using genetic programming (GEP). Their approach leveraged Pope's tensor basis theory to derive interpretable algebraic equations for Reynolds stress suitable for direct implementation in RANS solvers. \citet{schmelzer_discovery_2019} further improved this approach by developing SpaRTA (Sparse Regression of Turbulence Anisotropy), which introduced the k-corrective-frozen RANS approach that maintained interpretability while enabling efficient derivation of correction terms. Recent developments include Symbolic Bayesian Learning for SpaRTA (SBL-SpaRTA), which combines symbolic regression with Bayesian inference for comprehensive uncertainty quantification \cite{cherroud_sparse_2022}.

Despite these methodological advances, which improve upon generalizability, interoperability, and compute time, recent studies \citep{srivastava_generalizably_2024,wu_development_2024,nishi_generalization_2024,mandler_generalization_2024} have highlighted the critical importance of selectively applying corrections in separated flows, particularly in shear layers. \citet{rumsey_exploring_2009} and \citet{li_aerodynamic_2020} demonstrated that primary errors in RANS models for separated flows predominantly occur outside the boundary layer, especially in shear layers where turbulent flow behaviour exhibits strong anisotropy, non-equilibrium conditions, and coherent structure dynamics. Specifically, Rumsey's investigation of an ad hoc modification for $k - \omega $ models utilized a multiplier that adjusted the $\omega$-destruction term based on the ratio of turbulent production to dissipation to help correct underestimated turbulent shear stress in separation bubbles. While this approach showed promising improvements in specific cases, it lacked universal effectiveness across different separated flow scenarios.

Building on these insights, \citet{srivastava_generalizably_2024} developed refined FIML-based augmentation strategies for separated flows, while \citet{wu_development_2024} introduced the Conditioned Field Inversion method to maintain boundary layer accuracy while enhancing separated flow predictions. \citet{nishi_generalization_2024} proposed a closed-form model correction using Gaussian radial basis functions for incorporating local flow features into RANS models. The implementation of such corrections presents inherent challenges, particularly with the $ k-\omega$ SST model. As emphasized by \citet{wu_development_2024} and \citet{nishi_generalization_2024}, boundary and shear layer corrections must be implemented cautiously to preserve the model's accuracy in the log-law region, while still improving predictions in separated flow regions.

These implementation challenges highlight a broader issue in the field: the significant difficulty in developing corrections using full-field data that generalize effectively across different flow configurations \citep{mandler_generalization_2024}. While various methods have demonstrated success in specific classes of flows, achieving broader applicability remains elusive due to the complexity of training with diverse flow conditions and the challenge of identifying consistent error patterns across different geometries.

With the aim of addressing these fundamental limitations, the present study introduces the Relative Importance Term Analysis (RITA) methodology inspired by \citet{brunton_machine_2020}. Rather than attempting to develop a universally generalizable correction model, we provide a systematic method for identifying and classifying regions where RANS models require improvement in separated flows. RITA uniquely focuses on quantitatively analyzing the ratios of terms in the turbulence kinetic energy equation, which allowed the development of a physics-based classifier that restricts corrections only to regions where model deficiencies are identified. This approach preserves baseline RANS accuracy elsewhere while using correction terms derived through the SpaRTA approach introduced by \citet{schmelzer_discovery_2019} on regions that need corrections.

The methodology leverages a physics-based binary classifier that uses RITA-derived ratios to identify shear layers through three key criteria: (1) destruction-to-production ratio, (2) Turbulent-to-Total Kinetic Energy Ratio (TTKER), and (3) vorticity-based Reynolds number. By focusing on the relative magnitudes of terms in the turbulence kinetic energy equation, RITA facilitates more targeted and efficient development of correction strategies. Crucially, the approach preserves the accuracy of existing models in well-predicted flow regions, particularly in fundamental cases such as boundary layers and channel flows where the $k-\omega$ SST model already performs well, while providing a systematic method for identifying where improvements are most needed in separated regions. We validated RITA on canonical separated flows, including the periodic hill, NASA wall-mounted hump, and curved backwards-facing step, before extending to generalization tests of more complex three-dimensional geometries such as the Faith Hill and Ahmed body, where we demonstrated good performance on 3D geometries.

The remainder of this manuscript is organized as follows: Section 2 presents the formulation of the RITA methodology and its theoretical development. Section 3 provides a comprehensive evaluation of RITA's performance through training and generalization tests in  2D benchmark cases, along with an extension to 3D geometries. Finally, Section 4 summarizes the key findings and identifies promising directions for future investigation.

\section{Methodology}\label{sec2}
This section presents our framework for enhancing RANS turbulence modeling via a new zonally-augmented $k-\omega$ SST model. We begin by quantifying model-form errors between baseline RANS solutions and high-fidelity data using the k-corrective-frozen-RANS method (Section \ref{sec2.1}). Observing that correction fields predominantly localize in shear layers and wake regions, we develop the Relative Importance Term Analysis (RITA) methodology (Section \ref{sec2.2}) to systematically identify these regions through physics-based classification criteria. Following classifier validation, we employ symbolic regression through SpaRTA (Section \ref{sec2.3}) to derive compact correction terms applied exclusively in classified regions. Finally, we outline our comprehensive dataset selection and numerical setup (Section \ref{sec2.4} and \ref{sec2.5}), featuring both 2D training cases and 3D generalization test cases to evaluate the robustness of our approach. Throughout this section, we emphasize how each component contributes to our goal of improving predictions in separated flows while preserving baseline model performance in well-resolved regions.

\subsection{Extracting Model-form Errors of RANS}\label{sec2.1}

This study starts from the SpaRTA framework developed by \citet{schmelzer_discovery_2019}. The SpaRTA framework operates by $(i)$  injecting high-fidelity data into RANS closure equations to obtain residual correction fields using the k-corrective-frozen-RANS method and $(ii)$ approximating these fields as a function of local flow quantities.  The high-fidelity data used for this study consists of  DNS and LES results, providing accurate representations of turbulent flows and capturing complex phenomena that RANS models often struggle with, particularly in separated flow regions. 

Following \citet{schmelzer_discovery_2019}, we apply SpaRTA to our baseline $ k - \omega $ SST model. This enables the identification and correction of model-form errors in both the Reynolds stress tensor and the RANS closure equations. Within the RANS closure equations, the Reynolds stress $\tau^{B}_{ij}$ based on Boussinesq approximation is approximated as:
\begin{equation}
    \tau_{ij} \simeq \tau^{B}_{ij} := 2k ( \frac{1}{3}\delta_{ij} + b^B_{ij} ),\label{eq1}
\end{equation}
where  $k$ is the turbulent kinetic energy, $\delta_{ij}$ is the Kronecker delta, $b_{ij} = -\frac{\nu_t}{k}S_{ij}$, $\nu_t$ is the eddy viscosity, and $S_{ij} := \frac{1}{2}(\partial_jU_i + \partial_iU_j)$ is the mean-strain tensor. This formulation divides the Reynolds stress tensor into an isotropic part, $\frac{2}{3}k\delta_{ij}$, and an anisotropic part, $a_{ij} = 2kb_{ij}$. The anisotropic component $a_{ij}$  captures momentum transport in different flow directions. In contrast, the isotropic component is absorbed into a modified mean pressure, contributing to the pressure-like behaviour of turbulence. The Boussinesq hypothesis approximates the anisotropic component in LEVMs as a linear function of the mean velocity gradient:

\begin{equation}
     2kb^B_{ij} = -2\nu_t S_{ij},\label{eq2}
\end{equation} 

where $b_{ij}^B$ denotes the anisotropy tensor in the baseline $k - \omega$ SST model. To capture the model-form error between the high-fidelity data's Reynolds stress $\tau^*_{ij}$ and the baseline $k - \omega $ STT model, a residual term $b^\Delta_{ij}$ is introduced into \eqref{eq1} as 
\begin{equation} \tau^*_{ij} = 2k(\frac{1}{3}\delta_{ij} + b^B_{ij} + b^\Delta_{ij}),\label{eq3} 
\end{equation} 
where $b^\Delta_{ij}$ represents the additional anisotropy correction needed to account for the nonlinear viscosity relationship \citep{wu_reynolds-averaged_2019}. This correction requires specification of both $\nu_t$ and $\omega$. We employ the k-corrective-frozen-RANS method \citep{schmelzer_discovery_2019}, which iteratively solves the $\omega$ transport equation while "freezing" the high-fidelity variables $U^*$, $k^*$, and $b^*_{ij}$. The augmented \( k \) and \( \omega \) transport equations are:
\begin{equation}
    \underbrace{\frac{\partial k^*}{\partial t} + U^*_i \frac{\partial k^*}{\partial x_i}}_{ C_k - \text{Convection} } = P_k + R -\underbrace{\beta^*wk^*}_{D_k - \text{Destruction}} + \underbrace{\frac{\partial}{\partial x_i}[(\nu + \sigma^*_k\nu_t) \frac{\partial k^*}{\partial x_i}]}_{d_{k} - \text{Diffusion}}, \label{keq}
\end{equation} 
\begin{equation}
    \frac{\partial \omega}{\partial t} + U^*_i \frac{\partial \omega}{\partial x_i} = \frac{\gamma}{\nu_t}(P_k + R) -\beta w^2 +\frac{\partial}{\partial x_i}[(\nu + \sigma_{\omega} \nu_t) \frac{\partial \omega}{\partial x_i}] + CD_{kw}, \label{eq5}
\end{equation} 
here, $\omega$ represents the specific dissipation rate determining the conversion of turbulent energy to thermal energy, $\beta$ is a model coefficient, and $\sigma_k$ and $\sigma_{\omega}$ are turbulent Prandtl numbers controlling the diffusive transport of $k$ and $\omega$ respectively. Since the model-form error in the $k$ equation cannot be fully addressed by modifying the baseline model's production term with $b^\Delta_{ij}$ alone, a residual correction term $R$ is introduced to address errors in both the production and dissipation terms of the $k$ and $\omega$ equations. This correction can be represented either as a field $R(\mathbf{x})$ or as a function of local flow variables $R(\nabla U, k, \omega, ...)$. The production term $P_k$ in the augmented $k-\omega$ SST model incorporates $b^\Delta_{ij}$ to account for the anisotropy correction. The augmented production term $P_k$ is defined as:

\begin{equation}
    P_k = \min \left( -2k\left(b_{ij}^{B} + b^\Delta_{ij}\right) \frac{\partial U_i}{\partial x_j}, 10 \beta^{*} \omega k^* \right). \label{eq6}
\end{equation}

\begin{figure}[H]
    \centering
    \begin{subfigure}[b]{0.99\textwidth} 
        \centering
        \includegraphics[width=\textwidth]{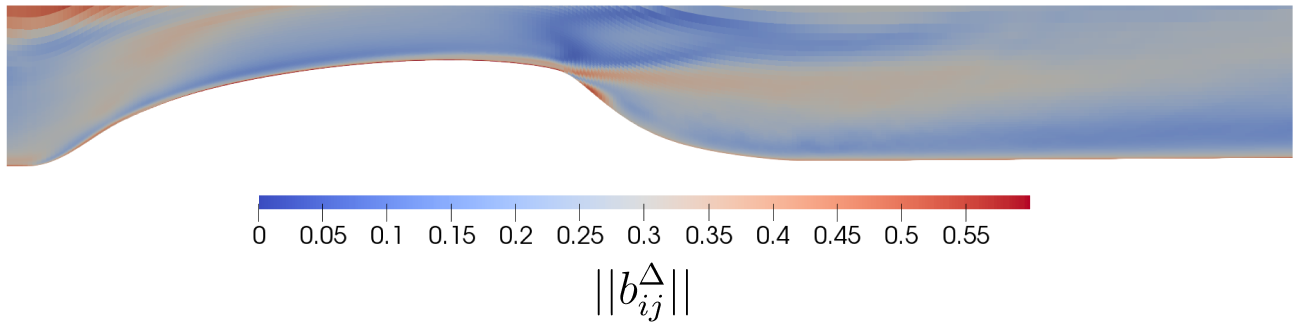}
        \caption{$b_{ij}^\Delta$ correction field.}
        \label{fig:bij-frozen-hump-v}
    \end{subfigure}
    \vfill
    \begin{subfigure}[b]{0.99\textwidth}
        \centering
        \includegraphics[width=\textwidth]{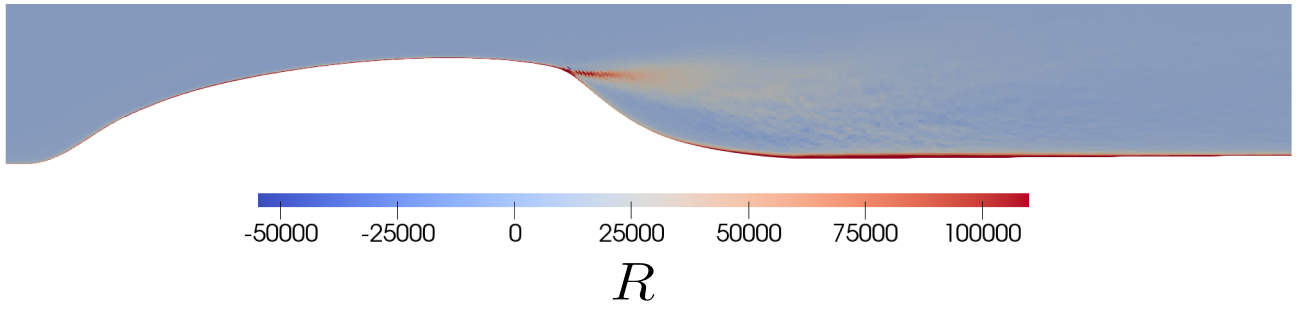} 
        \caption{$R$ correction field.}
        \label{fig:R-frozen-hump-v}
    \end{subfigure}
    
    \caption{Correction fields obtained using the $k$-corrective frozen approach on the NASA-Hump training case. The $b_{ij}^{\Delta}$ field (top) shows the Frobenius norm of anisotropic stress correction, while the $R$ field (bottom) indicates additional turbulence defect in the k-budget. }
    \label{fig:frozen-hump-v}
\end{figure}

The $k$-corrective-frozen-RANS method identifies and addresses model-form errors in the $k$ and $\omega$ equations by generating correction fields. This methodology is validated using the 2D NASA wall-mounted hump case \cite{seifert_active_2002}, a benchmark configuration for evaluating turbulence models under flow separation and reattachment conditions with adverse pressure gradients. Figure \ref{fig:frozen-hump-v} presents the correction fields generated by the $k$-corrective-frozen-RANS method, highlighting regions where the baseline $k-\omega$ SST model deviates from high-fidelity DNS/LES data.

The magnitude and spatial distribution of the $b_{ij}^\Delta$ correction field (Figure \ref{fig:bij-frozen-hump-v}) reveals two critical regions requiring significant adjustment to the baseline $k-\omega$ SST model's linear eddy-viscosity assumptions: the shear layer emanating from the hump's apex and the near-wall recovery region downstream. The most substantial corrections occur in the separated shear layer, indicating significant anisotropy that the baseline model fails to capture. This observation is reinforced by the corresponding $R$ correction field (\ref{fig:R-frozen-hump-v}), which shows strong positive corrections concentrated in similar regions. The spatial correlation between these two correction fields suggests that the baseline model mis-predicts the Reynolds stress anisotropy and, even taking that into account, still significantly underestimates TKE generation in the separated region. Most notably, the corrections persist well into the recovery region downstream of the hump, highlighting the baseline model's limitations in predicting both the immediate separation and subsequent flow reattachment physics. These results demonstrate the necessity for non-linear turbulence modeling approaches that can better capture both the turbulent anisotropy and corrections to the \textit{k-}budget in separated flows.

Despite requiring corrections in separated flows, The $k - \omega$ SST model, has robust performance in predicting wall-bounded turbulence. However, implementing corrections within the boundary layer requires caution to maintain the model's accuracy in predicting the log-law of the wall \citep{pope_turbulent_2000}. \citet{wu_development_2024} emphasize in their work on conditioned field inversion (FI-CND) that corrections, within the boundary should be minimized or avoided to maintain the log-law's accuracy. These recent studies, including the k-corrective-frozen-RANS method, collectively highlight the importance of selectively applying corrections in separated flows, particularly in shear layers. However, accurately identifying these critical regions while preserving the model's performance in well-predicted areas remains challenging. To address this, we introduce RITA, aligning with the strategies proposed by \citet{srivastava_generalizably_2024} and \citet{wu_development_2024}, as highlighted next.

\subsection{Relative Importance Term Analysis (RITA)}\label{sec2.2}

To accurately target corrections within shear layers identified by the k-corrective-frozen-RANS method, we developed the RITA technique. RITA serves as a physics-based classification method designed to isolate flow phenomena, in this case, shear layers, based on the relative importance of terms in the $k$-equation of the $k - \omega$ SST model.

RITA draws inspiration from the concept of learning dominant physical processes introduced by \citet{brunton_machine_2020}, who analyzed momentum equation terms to characterize boundary layer flows. In our approach, we examined the $k$-equation terms across training cases and found that term ratios provide more robust flow physics indicators than absolute magnitudes. The balance between destruction and production terms shows distinctive patterns: in shear layers, $\phi_{D_k/P_k}$ consistently falls below 0.55, compared to boundary layer regions where destruction dominates (exceeding 0.55) and free-stream regions where this ratio approaches 1.0 due to minimal production \citep{lacatus_improving_2024}.

Based on these observations, the $k$-equation, defined in \eqref{keq}, was selected for the RITA method because it reveals the dominant production of turbulent kinetic energy in shear layers and its corresponding destruction in recirculation regions. The terms in this equation—convection ($C_k$), production ($P_k$), destruction ($D_k$), and diffusion ($d_k$)—form the basis of our analysis. The terms we found most effective for characterizing different flow regions are:
\begin{equation} \phi_{D_k/P_k} = \frac{|D_k|}{|P_k| + |D_k|}, \label{eq11} \end{equation}
\begin{equation} \phi_{C_k/D_k} = \frac{|D_k|}{|C_k| + |D_k|}, \label{eq12} \end{equation}
\begin{equation} \phi_{d_{k}/D_k} = \frac{|D_k|}{|d_{k}| + |D_k|}. \label{eq13} \end{equation}

When analyzing separated flows, $\phi_{D_k/P_k}$ emerges as particularly significant as expected. However, this ratio alone is insufficient for robust shear layer identification. In our analysis of canonical separated flows (the periodic hill, NASA wall-mounted hump, and curved backward-facing step), we found that additional criteria related to turbulence energy content and rotational effects were necessary to accurately isolate shear layers. Based on these observations, we developed a binary classifier $\sigma_{SL}$ using three physically-motivated criteria:
\begin{equation} \phi_{D_k/P_k} < 0.55, \phi_{k} = \frac{k}{k + 0.5|U|^2} \geq 0.12 \text{ and } Re_{\Omega} = d_{w}^{2}\Omega/\nu \geq 0.02.\label{eq19} \end{equation}

Here, $\phi_{k}$ represents the Turbulent-to-Total Kinetic Energy Ratio (TTKER), and $Re_{\Omega}$ is the vorticity Reynolds number, where $d_w$ is the wall distance and $\Omega$ is the vorticity magnitude. Through analysis of these benchmark cases, we established three physically-motivated criteria for shear layer identification. The ratio $\phi_{D_k/P_k} < 0.55$ identifies regions where dissipation does not overwhelmingly dominate production, characteristic of shear layers with significant turbulent kinetic energy production. This threshold emerged from observing that shear layers consistently maintain a production-to-dissipation balance where dissipation accounts for less than $55\%$ of the combined magnitude. The TTKER criterion $\phi_{k} \geq 0.12$ ensures the identification of regions with substantial turbulent fluctuations relative to the mean flow, distinguishing energetic shear layers characterized by enhanced mixing and momentum transfer between different velocity streams. The vorticity Reynolds number criterion $Re_{\Omega} \geq 0.02$ complements these parameters by identifying regions of strong rotational flow, with a quadratic wall distance dependence ($d_w^2$) making it particularly effective at detecting separated shear layers as it naturally scales with the growth of vortical structures away from the wall.

\begin{figure}[H]
\centering
\begin{subfigure}[b]{\textwidth}
\centering
\includegraphics[width=\textwidth]{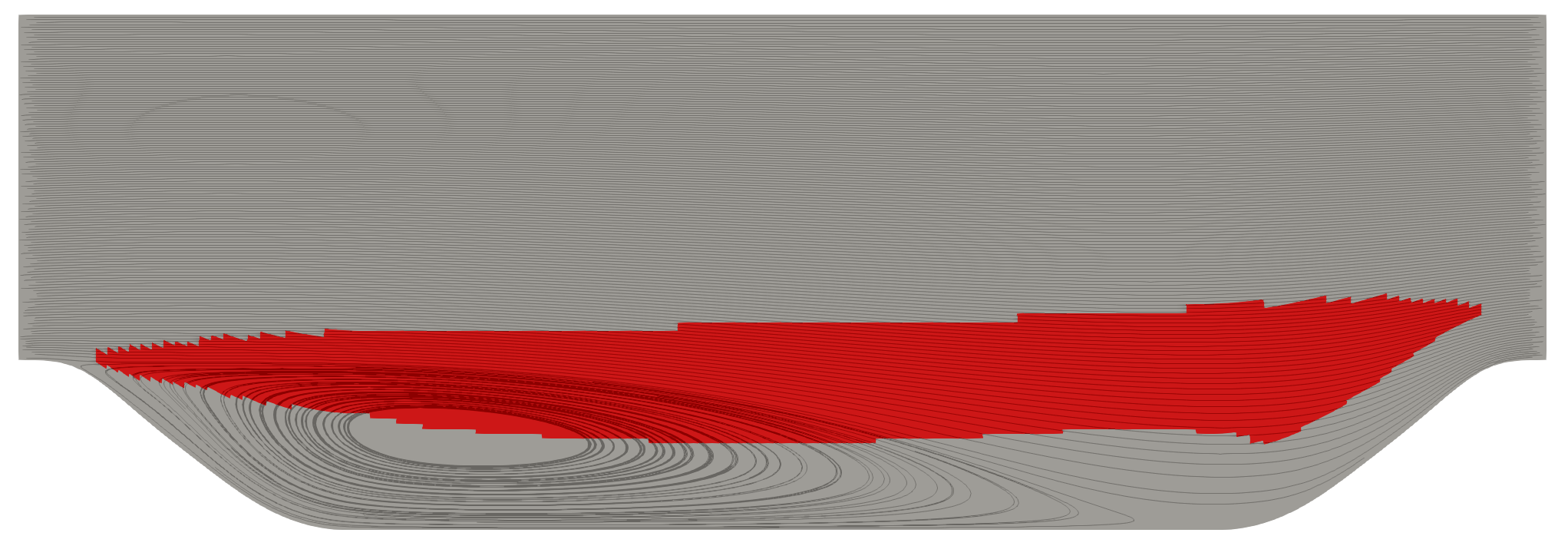}
\caption{Periodic-Hill case.}
\label{fig:sigma_PH}
\end{subfigure}

\begin{subfigure}[b]{0.46\textwidth}
    \centering
    \includegraphics[width=\textwidth]{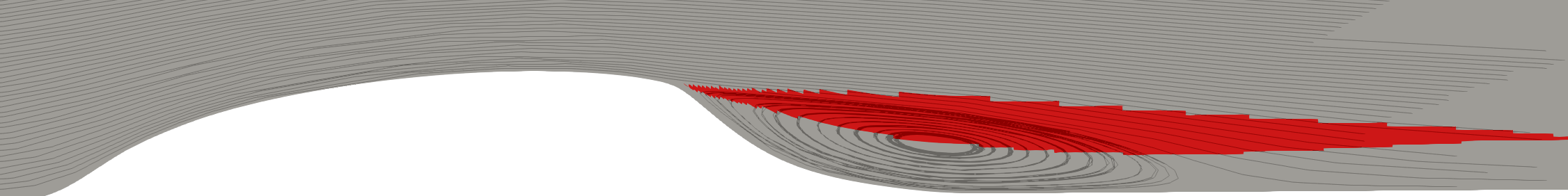}
    \caption{NASA-Hump case.}
    \label{fig:sigma_hump}
\end{subfigure}
\hfill
\begin{subfigure}[b]{0.44\textwidth}
    \centering
    \includegraphics[width=\textwidth]{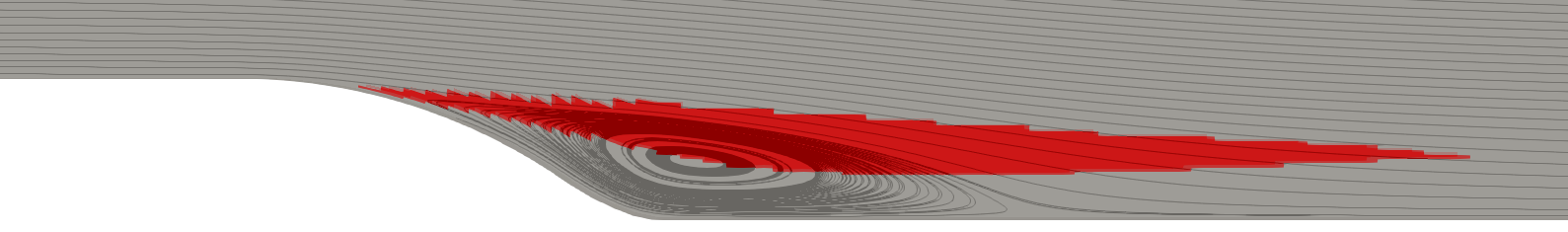} 
    \caption{CBFS case.}
    \label{fig:sigma_CBFS}
\end{subfigure}

\caption{Classification outcome of the $\sigma_{SL}$ classifier on the baseline training cases: red region: $\sigma_{SL}$ = 1, gray region: $\sigma_{SL}$ = 0.}
\label{fig:RITA-classifier-1}

\end{figure}

Together, these criteria form a physics-based classifier that identifies regions where production mechanisms are significant ($\phi_{D_k/P_k}$), turbulent fluctuations are energetic ($\phi_{k}$), and rotational effects are prominent ($Re_{\Omega}$). Cells meeting these criteria are classified as part of the shear layer and assigned $\sigma_{SL} = 1$, receiving full $b^\Delta_{ij}$ and $R$ corrections, while cells outside the shear layer ($\sigma_{SL} = 0$) receive no corrections. Notably, this classification approach ensures that corrections are not applied to fundamental flow cases such as attached boundary layers and channel flows, where the baseline $k-\omega$ SST model already provides accurate predictions. This selective correction strategy maintains the established performance of the baseline model in these canonical cases. Figure \ref{fig:RITA-classifier-1} demonstrates the reasonableness of this classification approach across our training cases, clearly identifying the shear layer regions in the periodic hill (Figure \ref{fig:sigma_PH}), NASA hump (Figure \ref{fig:sigma_hump}), and curved backwards-facing step (Figure \ref{fig:sigma_CBFS}) configurations.

To assess the practical impact of the classifier, we define Shear Layer Propagation (SL-Propagation) as the approach where correction terms from the $k-$corrective-frozen-RANS method are applied exclusively to cells where $\sigma_{SL} = 1$, while the baseline model remains unchanged in regions where $\sigma_{SL} = 0$. This results in the modified $k-\omega$ SST model equations:

\begin{equation}
    \frac{\partial k}{\partial t} + U_j \frac{\partial k}{\partial x_j} = P_k + \sigma_{SL}R -\beta^*wk + \frac{\partial}{\partial x_j}[(\nu + \sigma_k\nu_t) \frac{\partial k}{\partial x_j}], \label{eq:SL_keq_model}
\end{equation} 

\begin{equation}
\frac{\partial \omega}{\partial t} + U_j \frac{\partial \omega}{\partial x_i} = \frac{\gamma}{\nu_t}(P_k + \sigma_{SL}R) -\beta \omega^2 + \frac{\partial}{\partial x_i}[(\nu + \sigma_{\omega}\nu_t) \frac{\partial \omega}{\partial x_i}] + CD_{k\omega},
\label{eq:SL_omega_augmented}
\end{equation}

\begin{equation}
P_k = \min\left(-2k(b_{ij}^B + \sigma_{SL}b_{ij}^{\Delta})\frac{\partial U_i}{\partial x_j}, 10P_{\omega}k\right).
\label{eq:SL_Pk_augmented}
\end{equation}

\begin{figure}[H]
    \centering
    \begin{subfigure}[H]{0.8\textwidth}
        \centering
        \includegraphics[width=\textwidth]{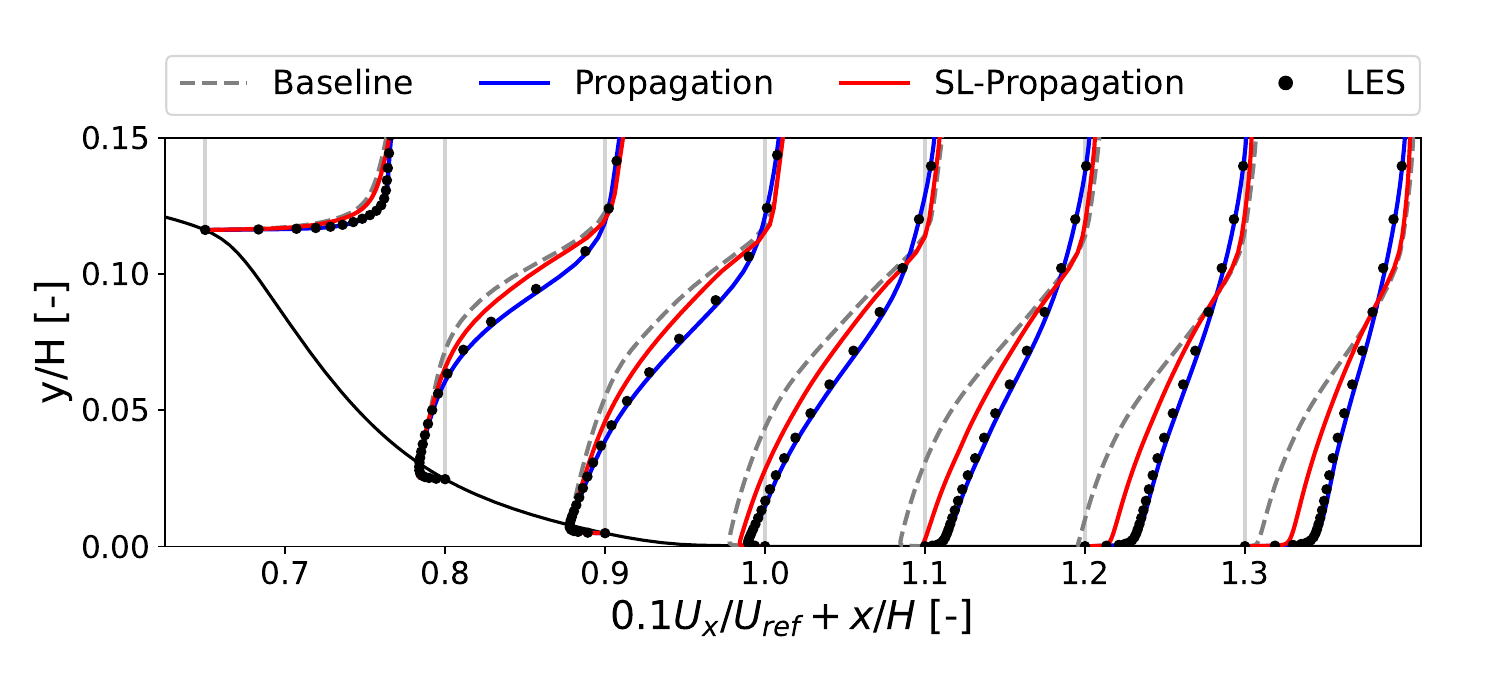} 
        \caption{Axial velocity profiles.}
        \label{fig:ux-prop-NASA-final}
    \end{subfigure}
    
    \begin{subfigure}[H]{0.49\textwidth}
        \centering
        \includegraphics[width=\textwidth]{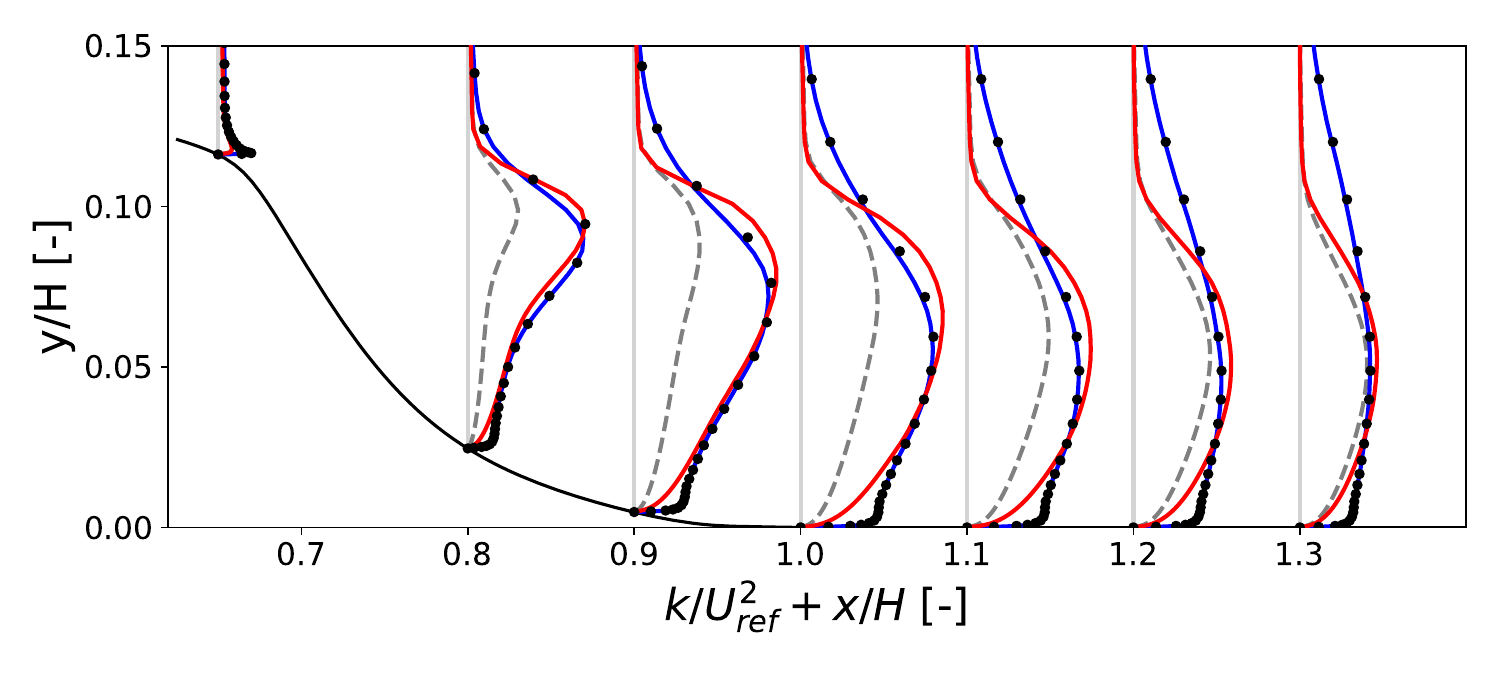}
        \caption{Turbulent kinetic energy profiles.}
        \label{fig:k-prop-NASA-final}
    \end{subfigure}
    \hfill
    \begin{subfigure}[H]{0.49\textwidth}
        \centering
        \includegraphics[width=\textwidth]{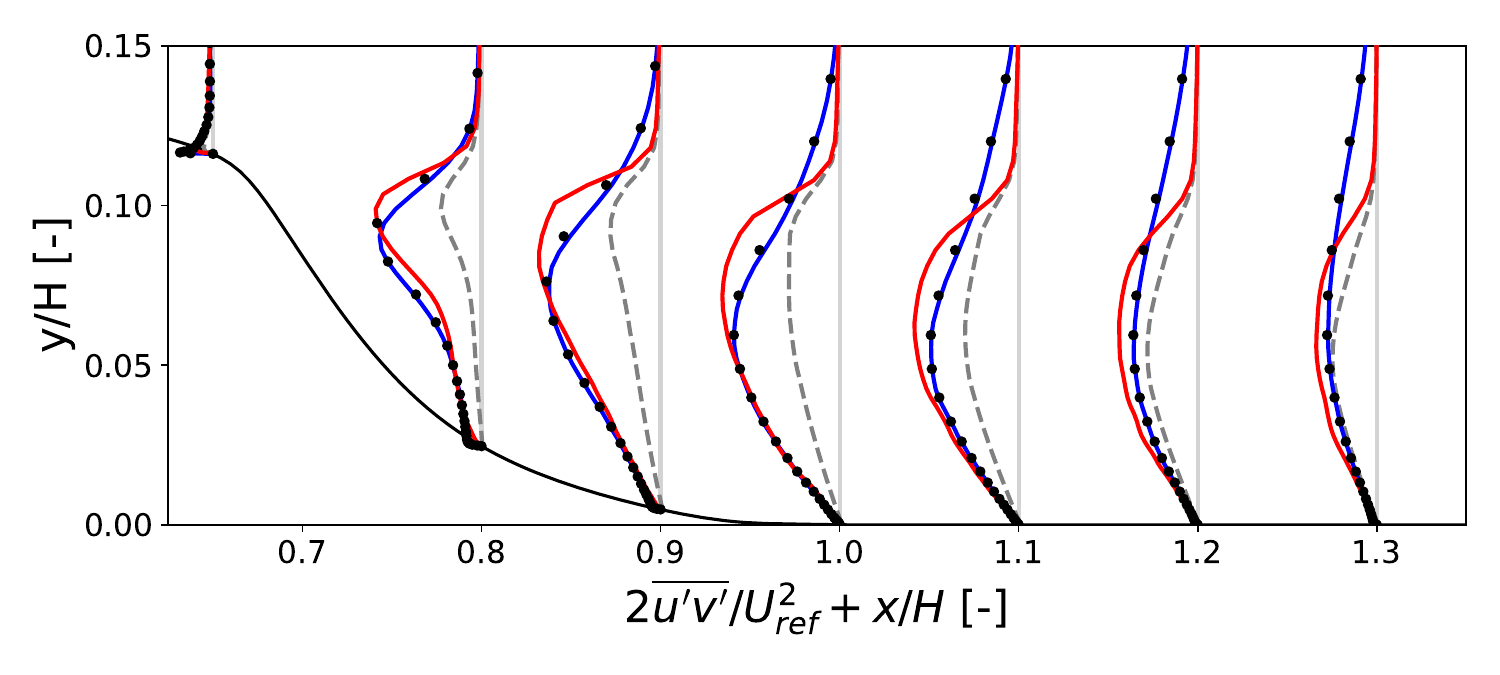} 
        \caption{Reynolds shear stress profiles.}
        \label{fig:upwp-prop-NASA-final}
    \end{subfigure}
    
    \caption{Performance comparison between full field propagation and shear layer propagation on the NASA-Hump training case. }
    \label{fig:prop-NASA-final}
\end{figure}

As a note, the binary classifier $\sigma_{SL}$ is never differentiated in our formulation, neither in the transport equations nor in the momentum equation. Consequently, its discrete nature (0 or 1) does not introduce immediate numerical instabilities at the interfaces between corrected and uncorrected regions. We compared SL-Propagation with full-field correction propagation for the NASA wall-mounted hump case, where the former applies corrections exclusively to regions identified by $\sigma_{SL}=1$. Figure \ref{fig:prop-NASA-final} presents axial velocity, turbulent kinetic energy, and Reynolds shear stress profiles at multiple streamwise locations. While SL-Propagation shows reduced accuracy compared to full-field propagation, primarily due to the absence of corrections in the region above the hill, it still substantially improves upon the baseline model in capturing the shear-layer development and reattachment characteristics.

\begin{figure}[H]
    \centering
    \includegraphics[width=0.9\linewidth]{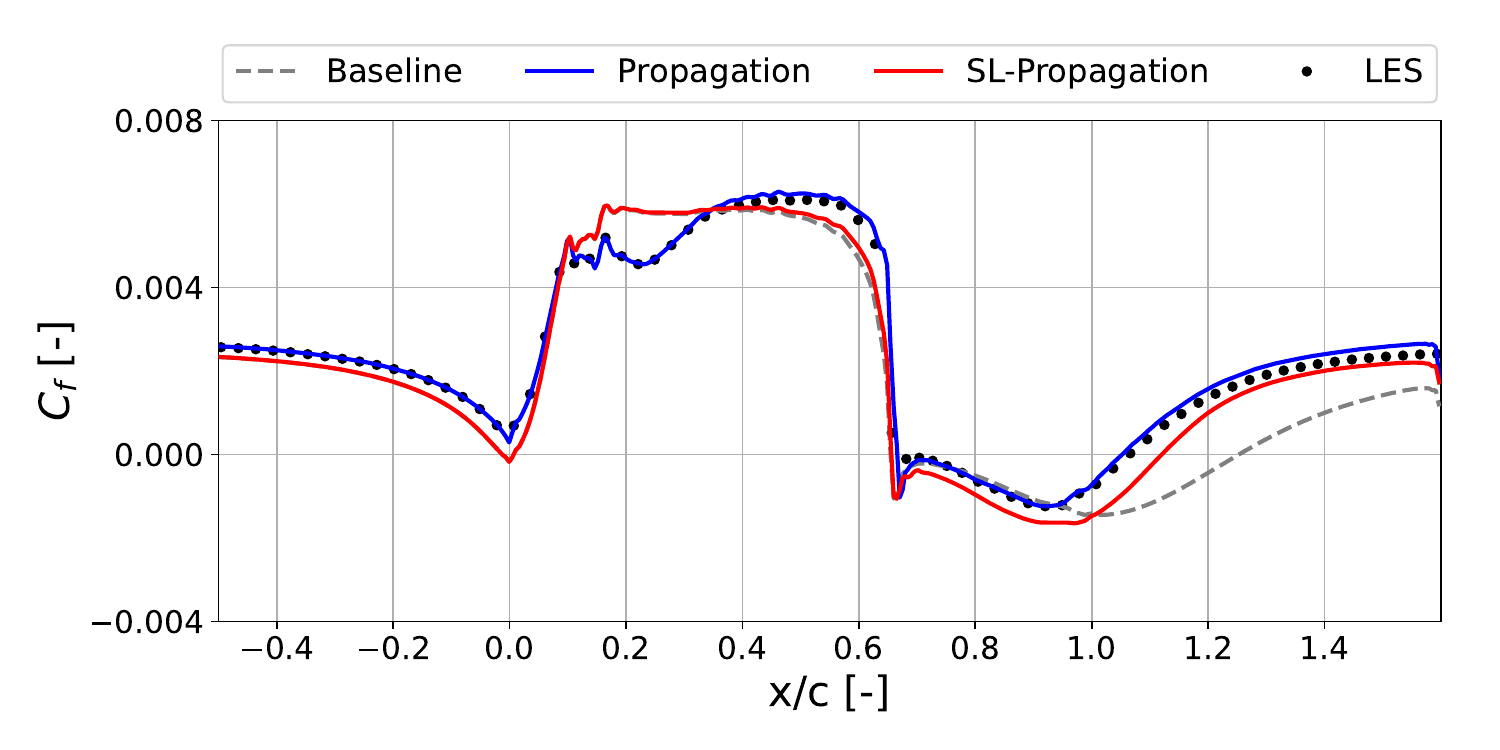}
    \caption{Skin friction comparison plot between full field propagation and shear layer
propagation on the Periodic-Hill training case.}
    \label{fig:Cf-Prop-NASA-final}
\end{figure}

Further validation of the method's performance is provided by the skin friction coefficient comparison shown in Figure \ref{fig:Cf-Prop-NASA-final}. Although SL-Propagation shows some deviation in the recovery region $(0.7 \leq x/c \leq 1.0)$, it captures the key separation and reattachment features well, and notably improves prediction in the wake region ($x/c > 1.0$) compared to the baseline model. This trade-off in accuracy is offset by RITA's ability to preserve the baseline model's behaviour in well-predicted regions, such as the upstream boundary layer and freestream flow. The targeted strategy improves computational efficiency while maintaining the model's established capabilities where corrections are unnecessary.

Having established a zonal classifier for identifying regions requiring correction, the next challenge is developing generalizable models for these corrections. This motivates our application of symbolic regression to discover compact, physics-based expressions for the $b_{ij}^{\Delta}$ and $R$ correction terms within $\sigma_{SL}$ identified regions.

\subsection{Model Discovery}\label{sec2.3}
The symbolic regression method SpaRTA \cite{schmelzer_discovery_2019} is employed to formulate algebraic models for the $b_{ij}^{\Delta}$ and $R$ correction fields within the identified $\sigma_{SL}$ regions. Following \citet{pope_more_1975}, the anisotropic Reynolds stress correction adopts a nonlinear generalization of the eddy viscosity concept, where $b_{ij}$ depends only on the strain rate tensor $S_{ij} = \tau\frac{1}{2}(\partial_jU_i + \partial_iU_j)$ and rotational rate tensor $\Omega_{ij} = \tau\frac{1}{2}(\partial_jU_i - \partial_iU_j)$ with timescale $\tau = 1/\omega$. Using the Cayley-Hamilton theorem, the anisotropic part of the Reynolds stress takes the general form:

\begin{align}
b_{ij}(S_{ij},\Omega_{ij}) = \sum_{n=1}^{10}  \alpha_n (I_1,I_2, ...,I_{m})T_{ij}^{(n)},
\label{eq:pope-formulation}
\end{align}
where $\alpha_n$ is a set of coefficients for a given value of $n$ invariants. The terms $T_{ij}^{(n)}$ and $I_m$ correspond to the ten non-linear basis tensors and five invariants, as defined in detail in \ref{app:invariants}. This formulation, with its basis tensors (\ref{eq:appendix-pope-bases}) and invariants (\ref{eq:appendix-pope-bases-invars}), provides the foundation for expressing the correction terms $b_{ij}^{\Delta}$ and $R$. The library extends beyond Pope's original basis tensors and invariants by incorporating additional scalar invariants ($q_m$) introduced by \citet{wang_physics-informed_2017}, RITA terms
as defined in the following table.

\begin{table}[H]
\centering
\caption{ Additional scalar features ($q_m$) added to $b_{ij}^{\Delta}$ and $R$ feature libraries.}
\label{tab:newfeatures}
\begin{tabular}{@{}lll@{}}
\toprule \toprule
Description & Symbol & Equation \\ \midrule \midrule
\begin{tabular}[c]{@{}l@{}}Time Scale Ratio\end{tabular}   & $\phi_{TS}$          & $\frac{k\|S\|}{\epsilon + k\|S\|}$  \\ \midrule
\begin{tabular}[c]{@{}l@{}}Q-criterion \end{tabular}              & $Q_{\text{criterion}}$ & $\frac{\|\Omega\|^2-\|S\|^2}{\|\Omega\|^2+ \|S\|^2}$      \\ \midrule
\begin{tabular}[c]{@{}l@{}} Ratio of Pressure Normal \\ Stresses to Shear Stresses \end{tabular}              & $\phi_{PS}$      & $\frac{\|\nabla P \|}{\|\nabla P \| + \|U\nabla U\|}$        \\ \midrule
\begin{tabular}[c]{@{}l@{}}Ratio of Total to Normal \\ Reynolds Stresses\end{tabular} & $\tau_{ij,ratio}$        & $\frac{\|\tau_{ij}\|}{10k + \|\tau_{ij}\|}$ \\ \midrule
\begin{tabular}[c]{@{}l@{}} Shear Parameter\end{tabular}                                                            & $S_k$   & $\frac{k\|\nabla U\|}{\epsilon}$  \\ \midrule
\begin{tabular}[c]{@{}l@{}} RITA: $D_k / P_k$ Ratio\end{tabular}                                                            & $\phi_{D_{k}/P_{k}}$   & $\frac{|D_k|}{(|D_k| + |P_k|)}$    \\ \midrule
\begin{tabular}[c]{@{}l@{}} RITA: $D_k / C_k$ Ratio\end{tabular}                                                            & $\phi_{D_{k}/C_{k}}$   & $\frac{|D_k|}{(|D_k| + |C_k|)}$    \\ \midrule
\begin{tabular}[c]{@{}l@{}} RITA: $D_k / d_k$ Ratio\end{tabular}                                                            & $\phi_{D_{k}/d_{k}}$   & $\frac{|D_k|}{(|D_k| + |d_k|)}$    \\ \midrule
\begin{tabular}[c]{@{}l@{}} RITA: $C_k / d_k$ Ratio\end{tabular}                                                            & $\phi_{C_{k}/d_{k}}$   & $\frac{|C_k|}{(|C_k| + |d_k|)}$    \\ \midrule
\begin{tabular}[c]{@{}l@{}} RITA: $P_k / d_k$ Ratio\end{tabular}                                                            & $\phi_{P_{k}/d_{k}}$   & $\frac{|P_k|}{(|P_k| + |d_k|)}$    \\ \midrule
\begin{tabular}[c]{@{}l@{}} RITA: $P_k / C_k$ Ratio\end{tabular}                                                            & $\phi_{P_{k}/C_{k}}$   & $\frac{|P_k|}{(|P_k| + |C_k|)}$    \\ \bottomrule \bottomrule
\end{tabular}
\end{table}

For the scalar correction $R$, the tensor basis functions are converted to scalar features through double dot products with the mean velocity gradient tensor. In addition, the turbulence dissipation rate ($\epsilon$) is added as a basis function in the $R$ correction. The final form for $b_{ij}^\Delta$ and $R$ is expressed

\begin{align}
b_{ij}^{\Delta}(S_{ij},\Omega_{ij},q_m) = \sum_{n=1}^{10}  \beta_n (I_1, ..., I_5, q_m) T_{ij}^{(n)},
\label{eq:pope-bijD}
\end{align}
\begin{align}
R(S_{ij},\Omega_{ij},q_m) = (2k\sum_{n=1}^{10}  \alpha_{n} (I_1, ..., I_5, q_m)T_{ij}^{(n)}\partial_{j}U_i) + a_{n}(I_n,q_m)\epsilon,
\label{eq:pope-R}
\end{align}
where $\beta_n$ and $\alpha_n$ are the coefficient functions for the respective corrections, $\beta_n : \mathbb{R}^5 \to \mathbb{R}$ and $\alpha_n : \mathbb{R}^5 \to \mathbb{R}$. Given the distinct nature of our two correction terms, separate candidate \textit{libraries} are constructed. Following \citet{schmelzer_discovery_2019}, the libraries forming the columns of $C_{b_{ij}^\Delta}$ and $C_R$ are generated using the FFX algorithm \citep{mcconaghy_ffx_2011}. In this approach, the invariants and basis functions are combined using unary functions to generate the candidate functions that are regressed to the target data. The library degree for both corrections is set to 1, resulting in monovariate candidate functions (single invariant transformed by a unary function and multiplied by a basis function). This design choice achieves an optimal balance between model complexity, computational efficiency, and physical interpretability. The resulting libraries contain 2510 candidate functions for $b_{ij}^{\Delta}$ and 399 candidate functions for $R$, utilizing the unary functions $(f_i)$ detailed in ~\ref{app:unary_functions}. All features are normalized by their standard-deviations to ensure numerical stability during regression. The regression problem is subsequently formulated in matrix form:

\begin{align}
b_{ij}^{\Delta} = C_{b_{ij}^\Delta} \Theta_{b_{ij}^\Delta} = \begin{bmatrix} 
f_1(I_1)T^{(1)}_{ij}, f_2(I_1)T^{(1)}_{ij}, \cdots, f_6(q_m) T^{(10)}_{ij} \\
\end{bmatrix}
\begin{bmatrix} 
\theta_1 \\
\theta_2 \\
\vdots \\
\theta_{2510}
\end{bmatrix},
\label{eq:bijD-matrix}
\end{align}

\begin{align}
R = C_R \Theta_R = \begin{bmatrix} 
f_1(I_1)T^{(1)}_{ij}\partial_{j}U_i, \cdots, f_6(q_m) T^{(10)}_{ij}\partial_{j}U_i, \cdots, f_6(q_m)\epsilon \\
\end{bmatrix}
\begin{bmatrix} 
\theta_1 \\
\vdots \\
\theta_{399}
\end{bmatrix},
\label{eq:R-matrix}
\end{align}
where $\Theta_{(b_{ij}^\Delta)}$ and $\Theta_R$ are the coefficient matrices for each correction term. These are determined through elastic net optimization:
\begin{equation}
\Theta = \underset{\hat{\Theta}}{argmin} \left|\left|  C_\Delta\hat{\Theta} - \Delta\right|\right|^2_2 + \lambda\rho \left|\left| \hat{\Theta} \right|\right|_1   + 0.5\lambda(1 -\rho) \left|\left| \hat{\Theta} \right|\right|_2^2
\end{equation}
where $\lambda >0$ controls the regularization weight and $1 > \rho >0$ determines the mixing parameter between Lasso and Ridge regression terms. This formulation promotes model sparsity while ensuring numerical stability \citep{zou_regularization_2005}.

To ensure physical interpretability and numerical stability, the regression process employs a two-step approach \citet{schmelzer_discovery_2019}. First, candidates undergo standardization before elastic-net regression to assess their relative significance independent of magnitude. Second, Ridge regression is applied using the original, unstandardized candidate functions to maintain appropriate units and small refit coefficients in the OpenFOAM solver \cite{schmelzer_discovery_2019}. Through systematic evaluation of various basis function and feature combinations, we identified optimal formulations for both R and $b_{ij}^{\Delta}$ corrections.

While \textit{a priori} regression metrics ($R^2$ values) are commonly used for initial model validation, they often fail to capture critical aspects of model performance in practice. This limitation stems from fundamental challenges mentioned in \citet{mandler_generalization_2024}: the disconnect between model development and application environments \cite{sotgiu_turbulent_2018, duraisamy_perspectives_2021, kurz_investigating_2021}, and the compounding effects of numerical uncertainties during simulations \cite{thompson_methodology_2016, wu_reynolds-averaged_2019, melchers_comparison_2023}. Therefore, we focus our analysis on comprehensive \textit{a posteriori} testing, beginning with the training cases themselves to establish baseline performance before examining generalization capabilities.

\subsection{Dataset Selection and Description} \label{sec2.4}

The development and validation of our correction terms requires careful selection of training cases that span diverse separation mechanisms. As summarized in Table \ref{tab:training_cases}, we selected three fundamental 2D separated flow configurations for model training: the periodic hill, NASA wall-mounted hump, and curved backward-facing step. Each case presents distinct flow physics - the periodic hill ($Re_h = 1.0 \times 10^4$) features sustained separation and reattachment cycles, the NASA hump ($Re_h = 9.3 \times 10^5$) introduces high Reynolds number effects such as thin turbulent boundary layers and smooth-surface separation under adverse pressure gradients, while the curved backward-facing step ($Re_h = 1.3 \times 10^4$) combines geometric and curvature-induced separation mechanisms.

\begin{table}[H]
\centering
\caption{Training and test cases with their numerical configurations: dimensionality $Dim$, Reynolds number Re$_h$, hill width-scaling factor $\alpha$, cell count $N$, and source of the reference data.}
\label{tab:training_cases}
\begin{tabular}{@{}llllll@{}}
\toprule
Case & $Dim$ & Re$_h$ & $\alpha$ & $N$ & Reference data \\ \midrule
\multicolumn{6}{c}{Training Cases} \\ \midrule
HUMP & 2D & $9.3 \times 10^5$ & --  & $5.1 \times 10^4$ & \citet{uzun_wall-resolved_2017} \\ 
PH & 2D & $1.0 \times 10^4$ & 1.0  & $1.5 \times 10^4$  & \citet{breuer_flow_2009} \\
CBFS & 2D & $1.3 \times 10^4$ & --  & $2.1 \times 10^4$  & \citet{bentaleb_large-eddy_2012} \\
\midrule
\multicolumn{6}{c}{Test Cases} \\ \midrule
PH & 2D & $5.6 \times 10^3$ & 0.5   & $1.5 \times 10^4$  & \citet{xiao_flows_2020} \\
PH & 2D & $5.6 \times 10^3$ & 0.8   & $1.5 \times 10^4$  & \citet{xiao_flows_2020} \\
PH & 2D & $5.6 \times 10^3$ & 1.0   & $1.5 \times 10^4$ & \citet{xiao_flows_2020} \\
PH & 2D & $5.6 \times 10^3$ & 1.2   & $1.5 \times 10^4$ & \citet{xiao_flows_2020} \\
PH & 2D & $5.6 \times 10^3$ & 1.5   & $1.5 \times 10^4$ & \citet{xiao_flows_2020} \\
Faith Hill & 3D & $5.0 \times 10^5$ & --  & $1.5 \times 10^6$ & \citet{bell_surface_2012} \\
Ahmed & 3D & $7.6 \times 10^5$ & -- & $1.1 \times 10^7$ & \citet{wagner_flow_2002} \\
\bottomrule
\end{tabular}
\end{table}

To assess model generalization, we constructed a test suite that systematically increases complexity beyond the training cases. The parametric periodic hill series varies the hill width-scaling factor ($\alpha = 0.5-1.5$), testing robustness to geometric modifications while maintaining similar flow physics. This generalization capability of SpaRTA models has already been demonstrated in a non-zonal model framework by \cite{zhang_customized_2021}. 

\begin{figure}[H]
    \centering
    \begin{subfigure}[b]{0.49\textwidth}
        \centering
        \includegraphics[width=\textwidth]{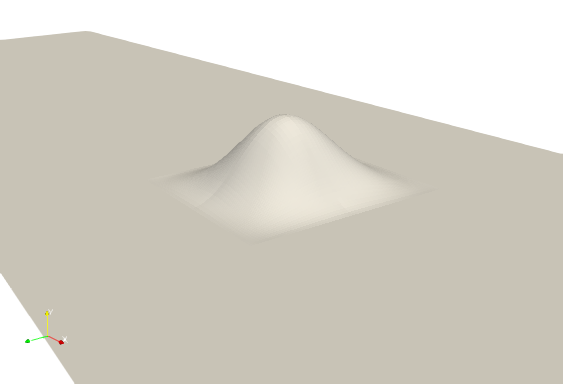}
        \caption{Faith-Hill Geometry.}
        \label{fig:faith-hill-geo}
    \end{subfigure}
    \hfill
    \begin{subfigure}[b]{0.49\textwidth}
        \centering
        \includegraphics[width=\textwidth]{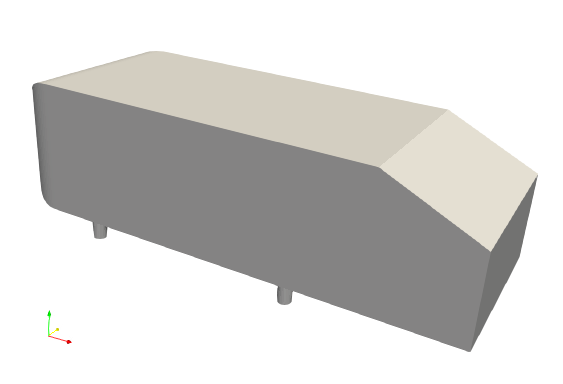} 
        \caption{Ahmed Body Geometry.}
        \label{fig:Ahmed-body-geo}
    \end{subfigure}

    \caption{ Geometries of 3D test cases: (a) Faith Hill, (b) Ahmed body.}
    \label{fig:3D-test-cases}
\end{figure}

We include two fully three-dimensional flow cases: the Faith Hill (Figure \ref{fig:faith-hill-geo}) and the Ahmed body (Figure \ref{fig:Ahmed-body-geo}). These configurations significantly differ from the 2D training cases in both geometric complexity and flow physics, providing a rigorous evaluation of the model’s ability to extend learned corrections beyond 2D.

The Faith Hill case ($Re_h = 5.0 \times 10^5$) presents a three-dimensional separation scenario driven by an adverse pressure gradient over a smooth surface, similar in principle to the NASA hump training case but with the added complexity of a necklace vortex system, spiral nodes in the wake region, and significant spanwise flow variations. The higher Reynolds number produces stronger shear layers with varying skin friction intensities, complex saddle points, and three-dimensional breakdown of coherent structures at reattachment.

The Ahmed body ($Re_h = 7.6 \times 10^5$) represents a bluff-body wake flow, characterized by strong counter-rotating C-pillar vortices from the slant edges, large-scale recirculation regions behind the vertical base, and unsteady shear layer development. Unlike the training cases, which predominantly feature wall-bounded separation, the Ahmed body introduces a distinctly three-dimensional wake structure where the interaction between longitudinal vortices and separated flow creates asymmetric pressure distributions and a complex dynamic wake topology.

\subsection{Numerical Setup} \label{sec2.5}

All simulations were performed using OpenFOAM with second-order accurate discretization schemes. Grid independence was ensured through systematic refinement studies, with final mesh resolutions detailed in Table \ref{tab:training_cases}. Detailed numerical setup parameters, grid descriptions and boundary conditions for each configuration are provided in \ref{secA1}.

\section{Results and Discussion}
\subsection{RITA-identified resulting model}
The application of SpaRTA to the identified shear layer regions for training cases listed in \ref{tab:training_cases} yielded compact, physics-based corrections for both the Reynolds stress anisotropy and turbulent kinetic energy production. These corrections are specifically activated within RITA-identified shear layer regions, where the baseline $k-\omega$ SST model typically underpredicts turbulent mixing and kinetic energy in separated flows. The scalar correction term $R$ takes the form:
\begin{equation}
R = 0.426\left(\frac{\phi_{D_k/C_k}/0.1248}{1 + (\phi_{D_k/C_k}/0.1248)^2}\right)\epsilon.
\label{eq:R_model}
\end{equation}
The structure of this scalar correction directly addresses the dissipation-production imbalance in separated shear flows. The correction scales with the turbulence dissipation rate $\epsilon$ and effectively reduces dissipation in the turbulent kinetic energy equation, similar to how \citet{rumsey_exploring_2009} adjusts the $\omega$-destruction term in separation regions. It is modulated by the ratio of destruction to convection $\phi_{D_k/C_k}$, capturing the local balance between turbulent transport and dissipation. The nonlinear limiting term prevents excessive corrections in regions of extreme destruction-convection imbalance. The anisotropic Reynolds stress correction is expressed as:

\begin{equation}
b_{ij}^{\Delta} = 3.69\phi_{D_k/P_k}T_{ij}^{(2)} - 5.092\left(\frac{I_{2}/0.01247}{1 + (I_{2}/0.01247)^2}\right)T_{ij}^{(2)}.
\label{eq:bij_model}
\end{equation}

This Reynolds stress formulation combines two physically distinct mechanisms: the first term captures shear-layer anisotropy through the destruction-production ratio $\phi_{D_k/P_k}$, while the second term, scaled by the rotation invariant $I_2$, accounts for the effects of strong rotation on turbulent stress alignment. Both terms utilize Pope's second basis tensor $T_{ij}^{(2)}$, which represents the interaction between mean strain and rotation rate.

The complete proposed, zonally-augmented $k-\omega$ SST model, incorporating these corrections through the shear layer classifier $\sigma_{SL}$, is presented in Box \ref{box:New_Model}. This framework shows how the correction terms are integrated with the baseline model equations while maintaining the original model structure through appropriate limiters and blending functions. The standard $k-\omega$ SST model coefficients, blending functions and auxiliary are provided in \ref{app:coef}.

\begin{tcolorbox}[colback=white,colframe=gray,title=\textbf{Zonally Augmented $k-\omega$ SST Model }]
\setlength{\abovedisplayskip}{4pt}
\setlength{\belowdisplayskip}{4pt}
\setlength{\abovedisplayshortskip}{2pt}
\setlength{\belowdisplayshortskip}{2pt}
\textbf{$\sigma_{SL}$ Classifier Criteria:}
\begin{equation}
\sigma_{SL} = \begin{cases}
1 & \text{if } \phi_{D_k/P_k} < 0.55 \text{ and } \phi_{k} \geq 0.12 \text{ and } Re_{\Omega} \geq 0.02 \\
0 & \text{otherwise}
\end{cases}
\label{eq:classifier}
\end{equation}
\textbf{Correction Term Equations:}
\begin{equation}
R = 0.426\left(\frac{\phi_{D_k/C_k}/0.1248}{1 + (\phi_{D_k/C_k}/0.1248)^2}\right)\epsilon.
\label{eq:R_model_box}
\end{equation}
\begin{equation}
b_{ij}^{\Delta} = 3.69\phi_{D_k/P_k}T_{ij}^{(2)} - 5.092\left(\frac{I_{2}/0.01247}{1 + (I_{2}/0.01247)^2}\right)T_{ij}^{(2)}.
\label{eq:bij_model_box}
\end{equation}
\textbf{Continuity and Momentum Equation:}
\begin{equation}
\frac{\partial U_i}{\partial x_i} = 0
\label{eq:continuity}
\end{equation}
\begin{equation}
\begin{split}
\frac{\partial U_i}{\partial t} + U_j\frac{\partial U_i}{\partial x_j} = -\frac{1}{\rho}\frac{\partial P}{\partial x_i} + \frac{\partial}{\partial x_j}\Big[\nu\Big(\frac{\partial U_i}{\partial x_j} + \frac{\partial U_j}{\partial x_i}\Big)-2k(b_{ij}^B )\Big] \\
+ \sigma_{SL}\frac{\partial}{\partial x_j}(2k b_{ij}^{\Delta})
\label{eq:momentum}
\end{split}
\end{equation}
\textbf{Turbulent Kinetic Energy Equation:}
\begin{equation}
    \frac{\partial k}{\partial t} + U_j \frac{\partial k}{\partial x_j} = P_k + \sigma_{SL}R -\beta^*wk + \frac{\partial}{\partial x_j}[(\nu + \sigma_k\nu_t) \frac{\partial k}{\partial x_j}] \label{keq_model}
\end{equation} 
\textbf{Specific Dissipation Rate Equation:}
\begin{equation}
\frac{\partial \omega}{\partial t} + U_j \frac{\partial \omega}{\partial x_i} = \frac{\gamma}{\nu_t}(P_k + \sigma_{SL}R) -\beta \omega^2 + \frac{\partial}{\partial x_i}[(\nu + \sigma_{\omega}\nu_t) \frac{\partial \omega}{\partial x_i}] + CD_{k\omega}
\label{eq:omega_augmented}
\end{equation}
\textbf{Eddy Viscosity Definition:}
\begin{equation}
\label{eq:vtkomegasst}
      \nu_t = \frac{a_1 k}{\max{(a_1 \omega,S F_2})}
\end{equation}
\textbf{Turbulent Production Term:}
\begin{equation}
P_k = \min\left(-2k(b_{ij}^B + \sigma_{SL}b_{ij}^{\Delta})\frac{\partial U_i}{\partial x_j}, 10P_{\omega}k\right)
\label{eq:Pk_augmented}
\end{equation} 

\end{tcolorbox}

Having developed these physically-motivated corrections that leverage RITA parameters for local flow characterization, we now turn to their validation through comprehensive \textit{a posteriori} testing. Our analysis focuses particularly on the model's ability to predict separation and reattachment dynamics in complex geometries, first examining performance on training configurations before assessing generalization capabilities on independent test cases.

\subsection{Performance on Training Cases}
Model performance was evaluated through systematic \textit{a posteriori} testing using three configurations. The baseline $k-\omega$ SST model serves as the reference point. The SL-Propagation case implements frozen corrections with dynamically updated $\sigma_{SL}$ classification, representing optimal performance achievable with zonal corrections. The SL-Model case applies the discovered $R$ and $b_{ij}^{\Delta}$ models with dynamic $\sigma_{SL}$ classification. All predictions are validated against high-fidelity DNS/LES data to assess both the RITA classifier's effectiveness and the correction terms' performance.

\subsubsection{NASA HUMP}

\begin{figure}[H]
    \centering
    \begin{subfigure}[b]{0.78\textwidth}
        \centering
        \includegraphics[width=\textwidth]{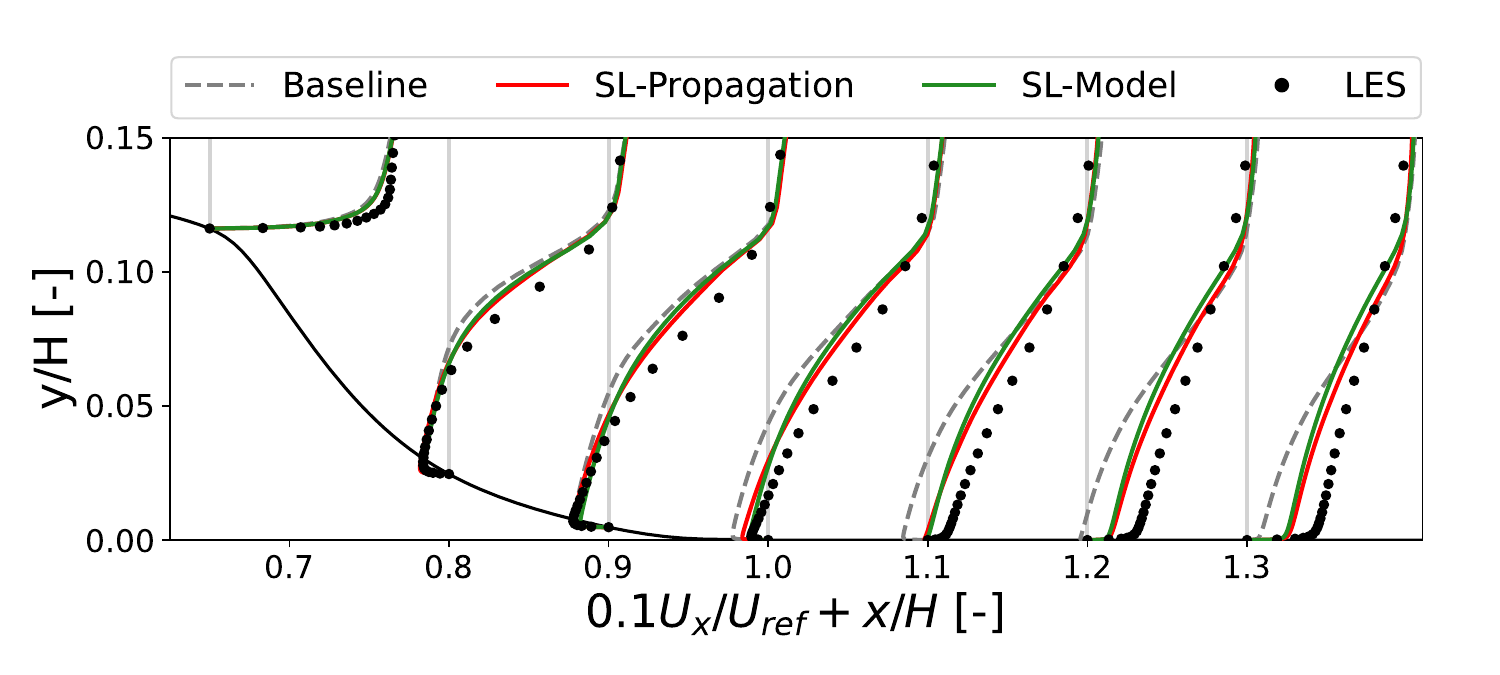} 
        \caption{Axial velocity profiles.}
        \label{fig:ux-prop-NASA-Model-final}
    \end{subfigure}
    
    \begin{subfigure}[b]{0.75\textwidth}
        \centering
        \includegraphics[width=\textwidth]{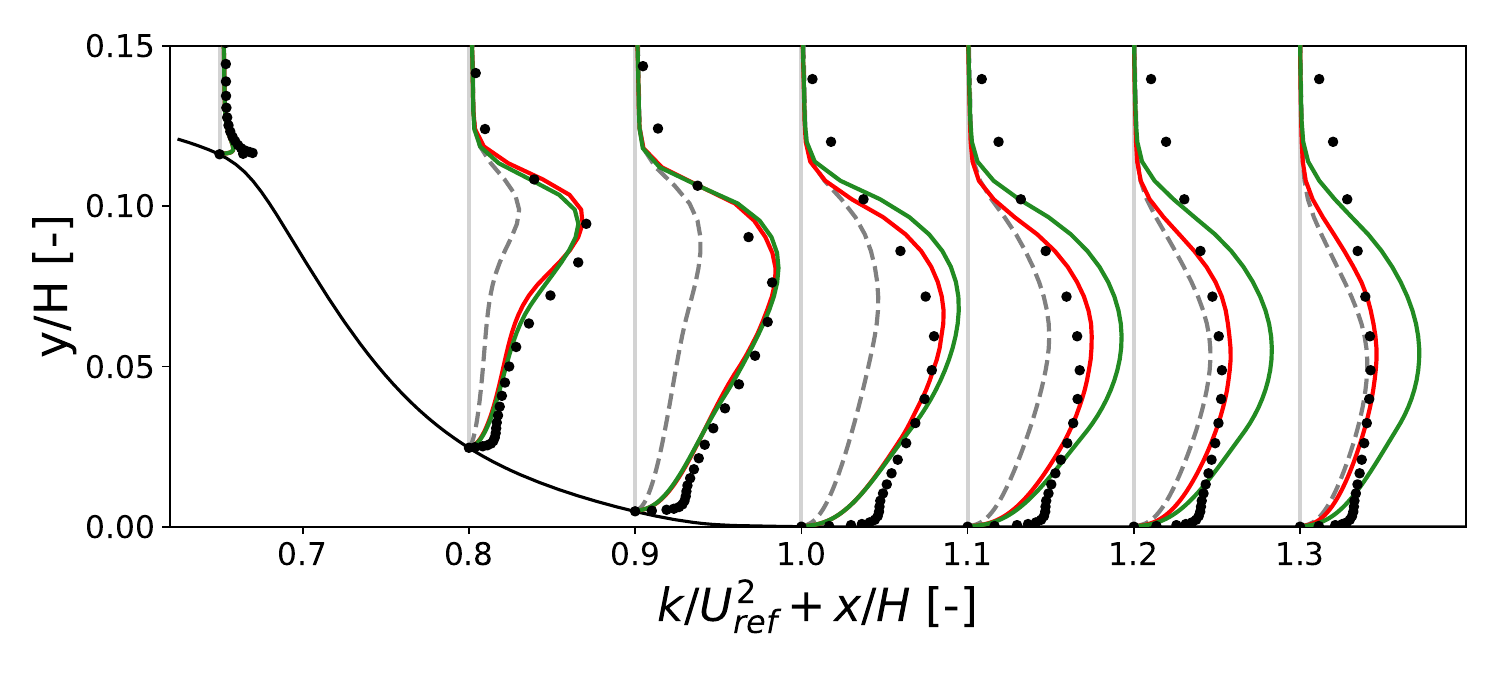}
        \caption{Turbulent kinetic energy profiles.}
        \label{fig:k-prop-NASA-Model-final}
    \end{subfigure}

    \begin{subfigure}[b]{0.75\textwidth}
        \centering
        \includegraphics[width=\textwidth]{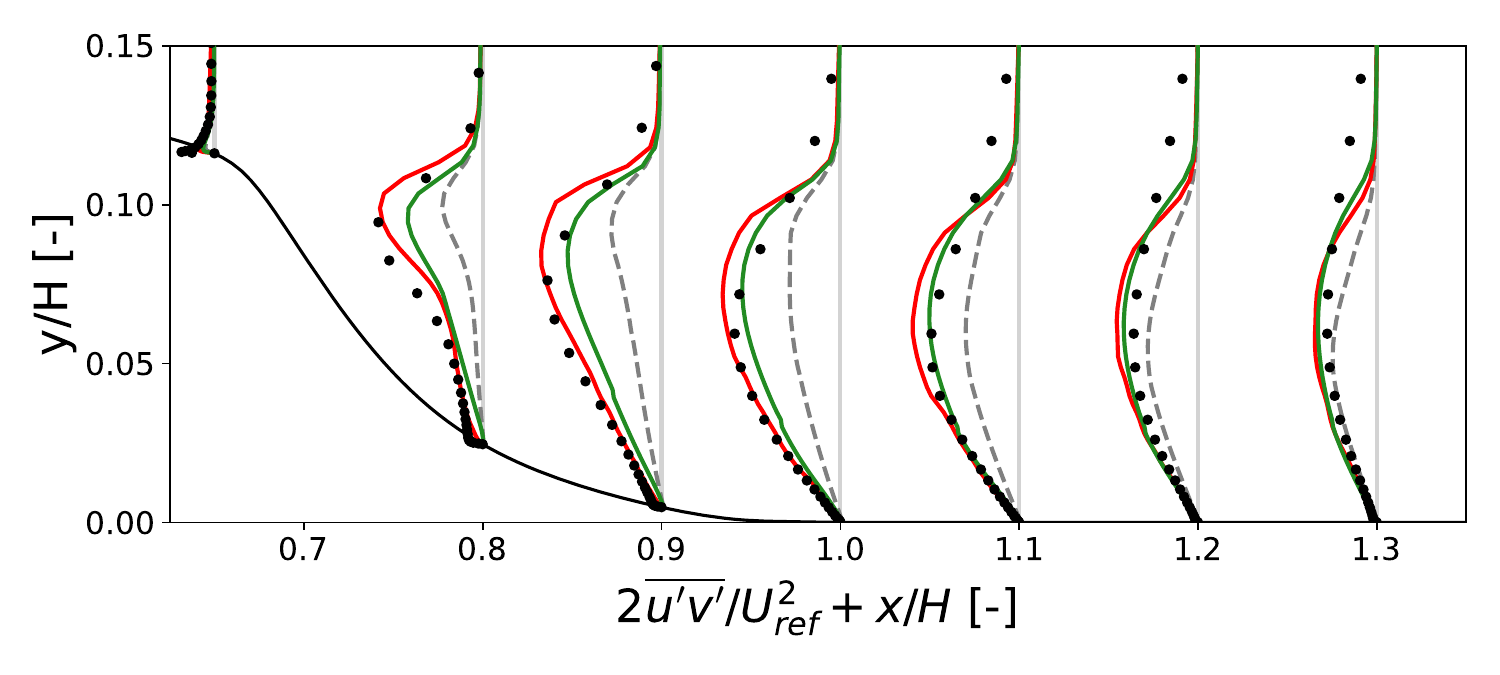} 
        \caption{Reynolds shear stress profiles.}
        \label{fig:upwp-prop-NASA-Model-final}
    \end{subfigure}
    
    \caption{Performance of the shear layer SpaRTA model on the NASA-Hump training case.}
    \label{fig:prop-NASA-Model-final}
\end{figure}

As demonstrated in Figures \ref{fig:prop-NASA-final} and \ref{fig:Cf-Prop-NASA-final}, the RITA classifier effectively identifies shear layer regions for correction, showing comparable performance to full-field propagation. Building on this foundation, Figure \ref{fig:prop-NASA-Model-final} presents the performance of the complete SL-Model implementation. The axial velocity profiles in Figure \ref{fig:ux-prop-NASA-Model-final} demonstrate that the SL-Model reduces the separation bubble length overprediction characteristic of the baseline $k-\omega$ SST model, showing improved agreement with LES data, particularly in the recovery region ($x/H \geq 1.2$), where the baseline model typically overpredicts the separation bubble extent. The consistency between SL-Model and SL-Propagation results indicates that the correction terms effectively capture the key separation flow physics.

The turbulent kinetic energy profiles (Figure \ref{fig:k-prop-NASA-Model-final}) reveal an overprediction compared to both the baseline model and LES data, particularly in the wake region ($x/H \geq 1.0$). While the SL-Model achieves improved mean flow predictions, it suggests the correction mechanisms may compensate for other model deficiencies through enhanced turbulent mixing. The Reynolds shear stress distributions (Figure \ref{fig:upwp-prop-NASA-Model-final}) show better alignment with LES data in terms of peak locations, though their magnitudes are influenced by these elevated turbulence levels. The simple algebraic correction model reproduces much of the performance of optimal propagated corrections, while also revealing a limitation in the current approach, where improved mean flow prediction comes at the cost of turbulence quantity accuracy.

\begin{figure}[H]
    \centering
    \includegraphics[width=\linewidth]{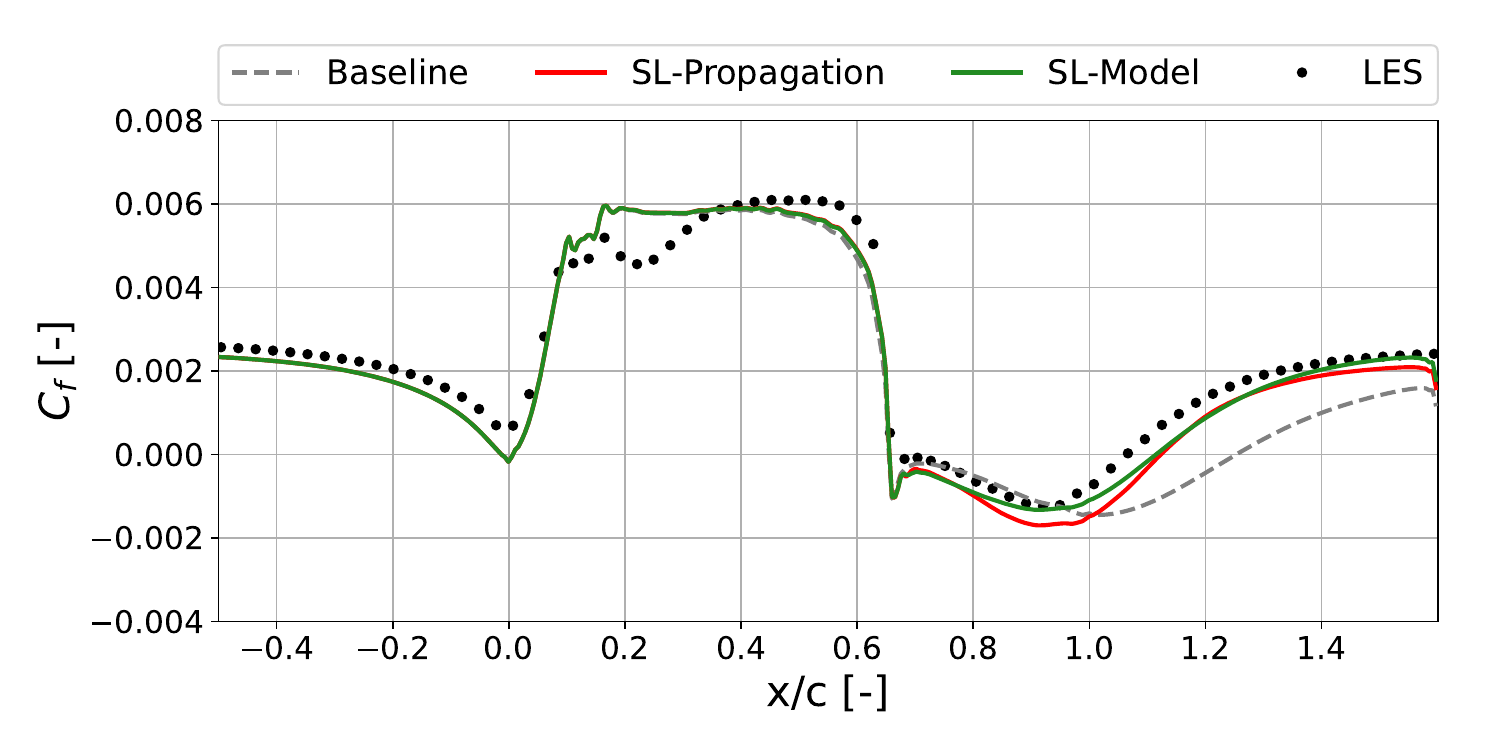}
    \caption{Skin friction comparison plot for the shear layer SpaRTA model on the NASA-Hump training case.}
    \label{fig:Cf-Model-NASA-final}
\end{figure}

The skin friction distribution (Figure \ref{fig:Cf-Model-NASA-final}) quantitatively validates separation and reattachment prediction. The SL-Model shows improved prediction of skin friction through separation ($x/C \approx 0.65$) and in the far wake region ($x/C > 1.0$), reducing the reattachment length overprediction characteristic of the baseline model. A slight degradation in prediction is observed in the immediate post-separation region ($0.7 < x/C < 0.9$), which can be attributed to the vorticity Reynolds number criterion limiting corrections near the wall. Despite this local trade-off, the RITA classifier successfully preserves the baseline model's accurate prediction of the attached boundary layer upstream of separation ($x/C < 0.5$) while achieving better overall wake recovery predictions. Similar validation was performed for the Periodic Hill and Curved Backwards Facing Step cases, with detailed results presented in \ref{app:Train_Cases}. These cases demonstrate consistent model performance across varied separation mechanisms.

\subsubsection{Other Training Cases}

The validation across three distinct separation mechanisms demonstrates both the capabilities and limitations of the zonally-augmented $k-\omega$ SST model. The RITA classifier successfully identifies shear layer regions requiring correction across varying geometric configurations, from smooth-surface separation (NASA Hump) to periodic separation-reattachment cycles (Periodic Hill) to combined geometric and curvature effects (CBFS), while appropriately avoiding corrections in attached boundary layers and channel flows where the baseline model already performs well. The discovered correction terms work in tandem to improve flow prediction through complementary mechanisms: the anisotropic stress correction ($b_{ij}^{\Delta}$) enhances shear layer prediction through the $T_{ij}^{(2)}$ tensor, which better captures the interaction between mean strain and rotation rates, while the scalar correction ($R$) augments turbulent transport through increased production. A significant achievement across all training cases is that this simple algebraic formulation nearly reproduces the performance of optimal propagated corrections, demonstrating the efficacy of the physics-based approach in capturing the essential correction mechanisms.

The model's behavior reflects the interplay between these corrections in the transport equations. The $b_{ij}^{\Delta}$ term, modulated by both the destruction-production ratio ($\phi_{D_k/P_k}$) and rotation invariant ($I_2$), improves Reynolds stress alignment in shear layers by modifying how momentum transfer occurs through regions of high strain rate. This effect is particularly evident in the Reynolds stress profiles across all cases. Meanwhile, in the $k-$equation, the $R$ term introduces additional production modulated by the destruction-to-convection ratio ($\phi_{D_k/C_k}$) and scaled by the turbulence dissipation rate ($\epsilon$). The nonlinear form of R creates a more complex feedback mechanism where the correction magnitude depends on the local balance between destruction and convection, not just a simple scaling of $\epsilon$. These modifications propagate through the $\omega$-equation via the enhanced production term $\gamma/\nu_t(P_k + \sigma_{SL}R)$, leading to the systematically elevated TKE levels observed in comparison to high-fidelity data.

Despite this trade-off in turbulence prediction, the model demonstrates robust performance in capturing separation locations and recovery characteristics across all training configurations. The combined effect of enhanced turbulent mixing from $R$ and improved directional distribution from $b_{ij}^{\Delta}$ successfully improves mean flow predictions, though through a simplified mechanism that may not fully capture the complex physics of turbulent transport in separated flows. The zonal framework's success in handling various separation mechanisms while maintaining appropriate behavior in attached regions suggests that the combination of physically-motivated corrections with selective application provides a viable approach for enhancing RANS predictions of separated flows.

\subsection{Parametrized Periodic Hills (Generalization Test)}

The RITA classifier's response to changing geometry provides a critical test of the model's generalization capability. By examining its behavior across weakly varying hill width-scaling factors, we can assess whether the classification criteria remain physically meaningful beyond the reference case.

\begin{figure}[H]
    \centering
    \includegraphics[width=0.9\textwidth]{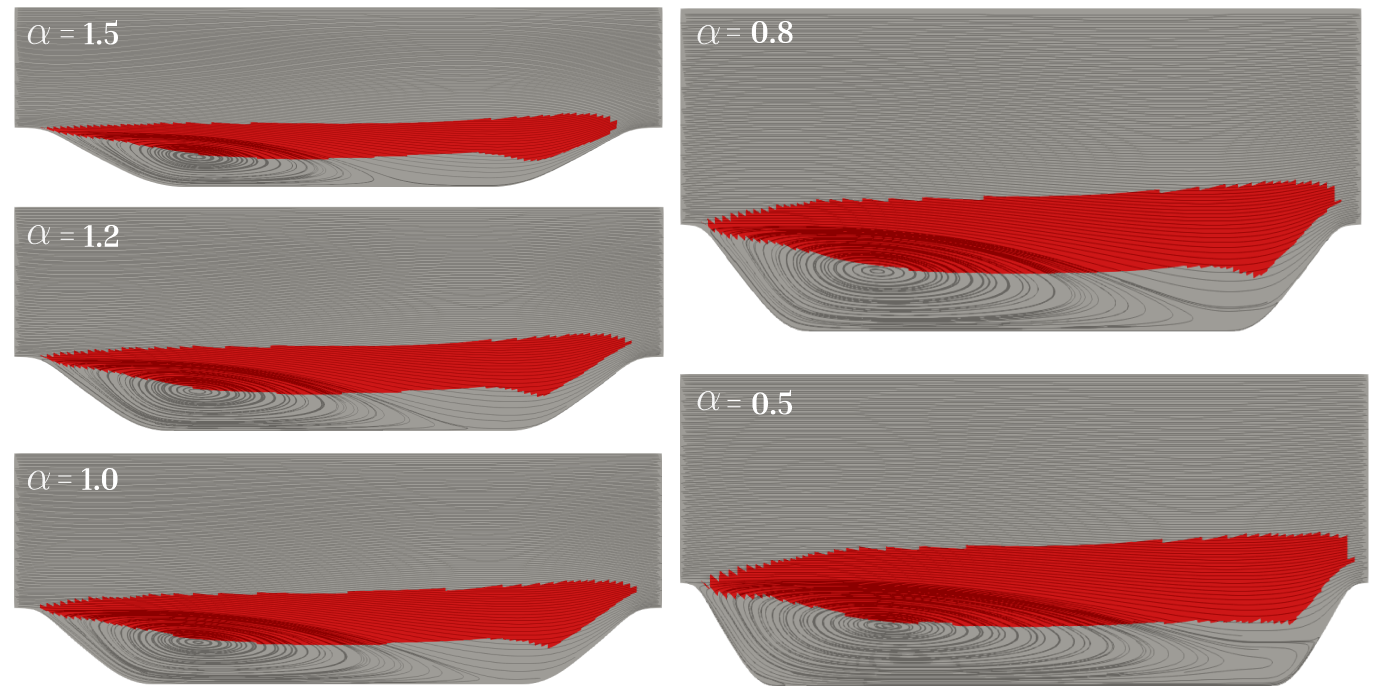}
    \caption{Classification outcome of the $\sigma_{SL}$ classifier on the baseline Parametrized Periodic-Hill cases: red region: $\sigma_{SL}$ = 1, gray region: $\sigma_{SL}$ = 0 }
    \label{fig:sigma_parametrized-final}
\end{figure}

Figure \ref{fig:sigma_parametrized-final} presents the classification results of the $\sigma_{SL}$ classifier on the periodic hill cases. The classifier effectively identifies shear layer regions requiring correction, as indicated by the red regions corresponding to $\sigma_{SL} = 1$. The spatial extent of these regions systematically evolves with $\alpha$: for larger hill spacing ($\alpha = 1.5$), the shear layer region extends further downstream due to delayed reattachment, while for smaller spacing ($\alpha = 0.5$), the classifier captures the more compact separation region and stronger interaction between successive hills. This systematic adaptation demonstrates the classifier's ability to identify physically relevant correction regions across weakly varying geometries.

\begin{figure}[htbp]

    \centering
    
    \begin{subfigure}[b]{\textwidth}
        \centering
        \includegraphics[width=0.9\textwidth]{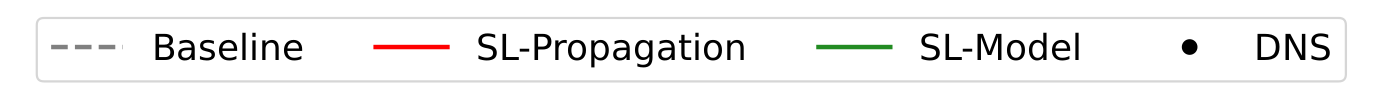} 
       % \caption{Axial velocity profiles.}
        %\label{fig:ux-PH15}
    \end{subfigure}
    
    \begin{subfigure}[b]{0.48\textwidth}
        \centering
        \includegraphics[width=\textwidth]{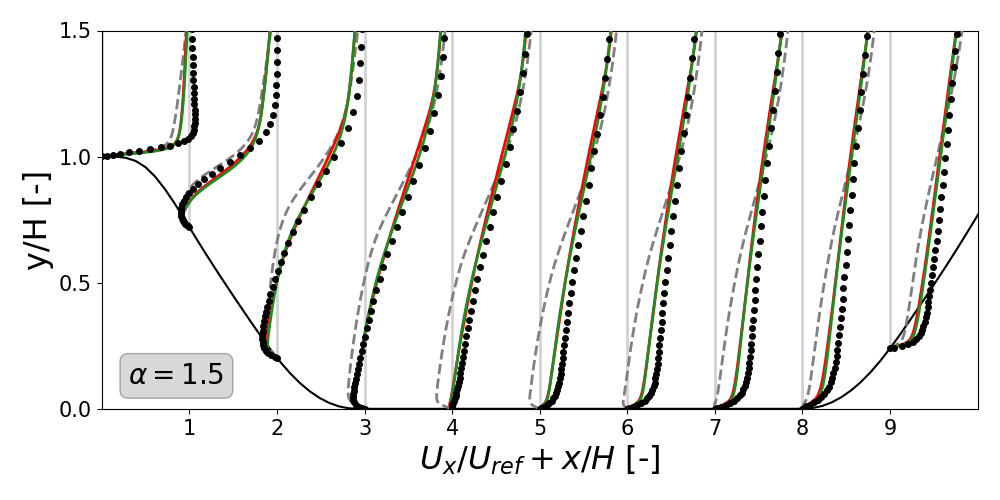} 
       % \caption{Axial velocity profiles.}
        %\label{fig:ux-PH15}
    \end{subfigure}
     \hfill
    \begin{subfigure}[b]{0.48\textwidth}
        \centering
        \includegraphics[width=\textwidth]{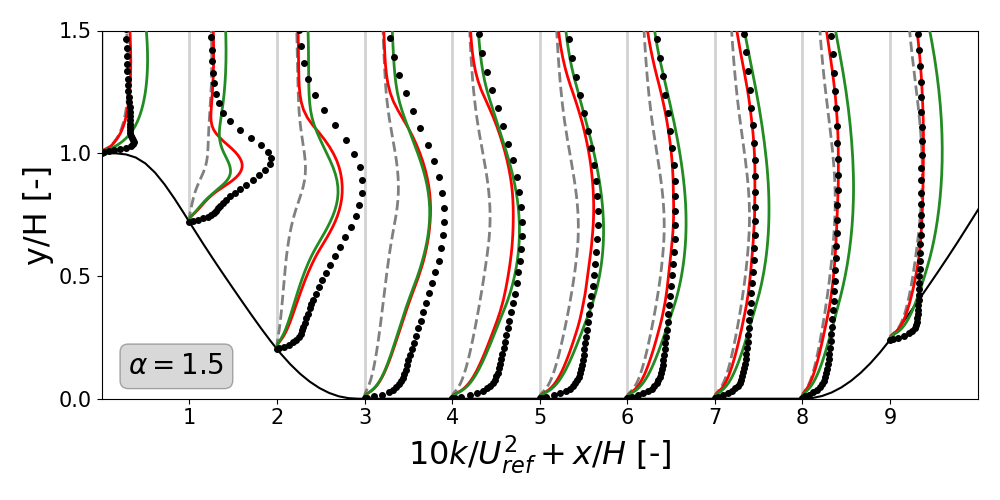}
       % \caption{Turbulent kinetic energy profiles.}
        %\label{fig:k-PH15}
    \end{subfigure}

    %\begin{comment}
    \begin{subfigure}[b]{0.48\textwidth}
        \centering
        \includegraphics[width=\textwidth]{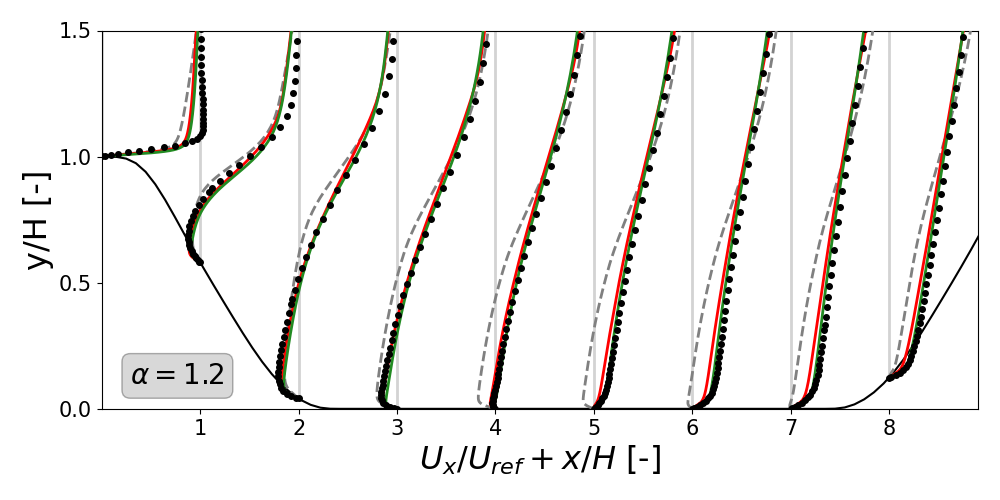} 
        %\caption{Axial velocity profiles.}
        %\label{fig:ux-PH12}
    \end{subfigure}
     \hfill
    \begin{subfigure}[b]{0.48\textwidth}
        \centering
        \includegraphics[width=\textwidth]{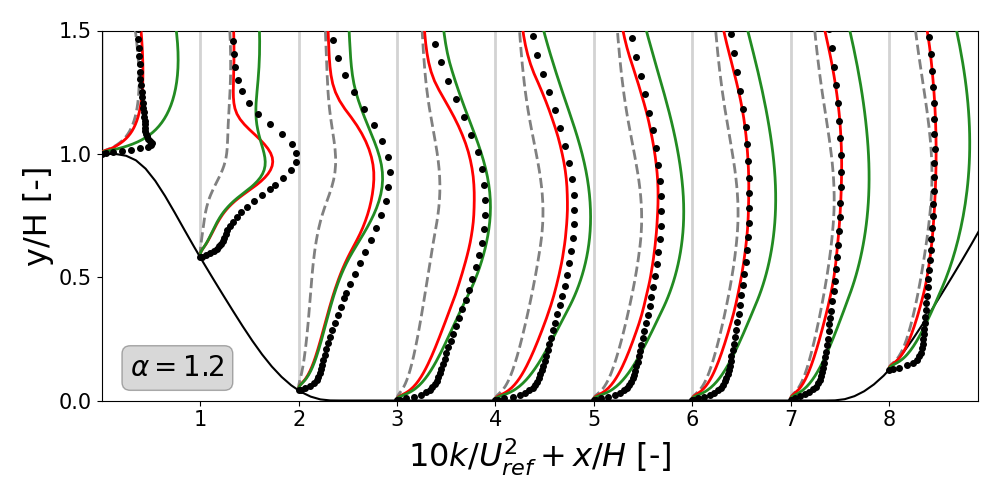}
       % \caption{Turbulent kinetic energy profiles.}
        %\label{fig:k-PH12}
    \end{subfigure}        
    %\end{comment}

     \begin{subfigure}[b]{0.48\textwidth}
        \centering
        \includegraphics[width=\textwidth]{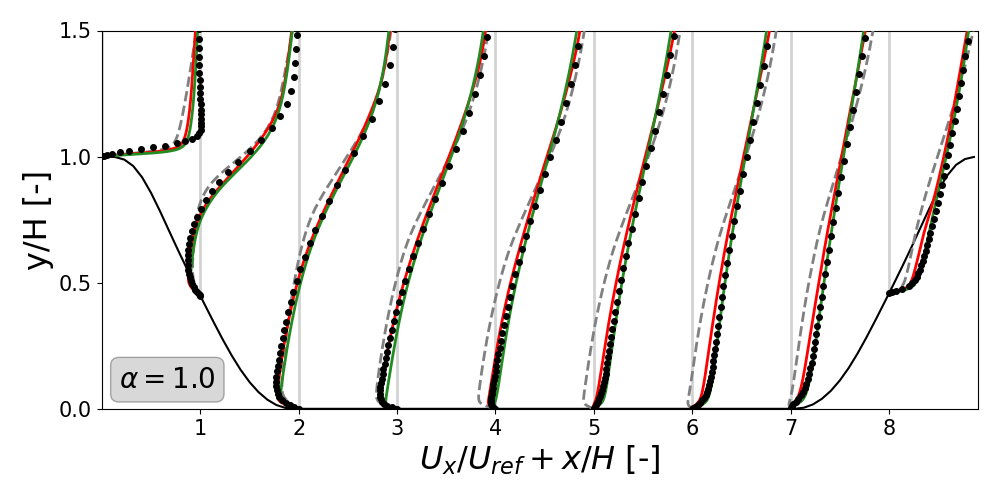} 
       % \caption{Axial velocity profiles.}
       % \label{fig:ux-PH10}
    \end{subfigure}
     \hfill
    \begin{subfigure}[b]{0.48\textwidth}
        \centering
        \includegraphics[width=\textwidth]{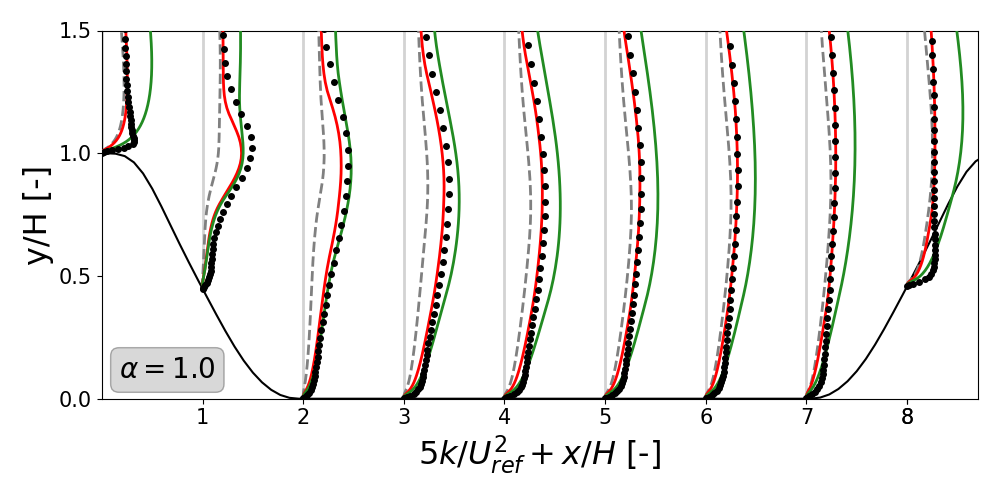}
        %\caption{Turbulent kinetic energy profiles.}
        %\label{fig:k-PH10}
    \end{subfigure}

    %\begin{comment}
    \begin{subfigure}[b]{0.48\textwidth}
        \centering
        \includegraphics[width=\textwidth]{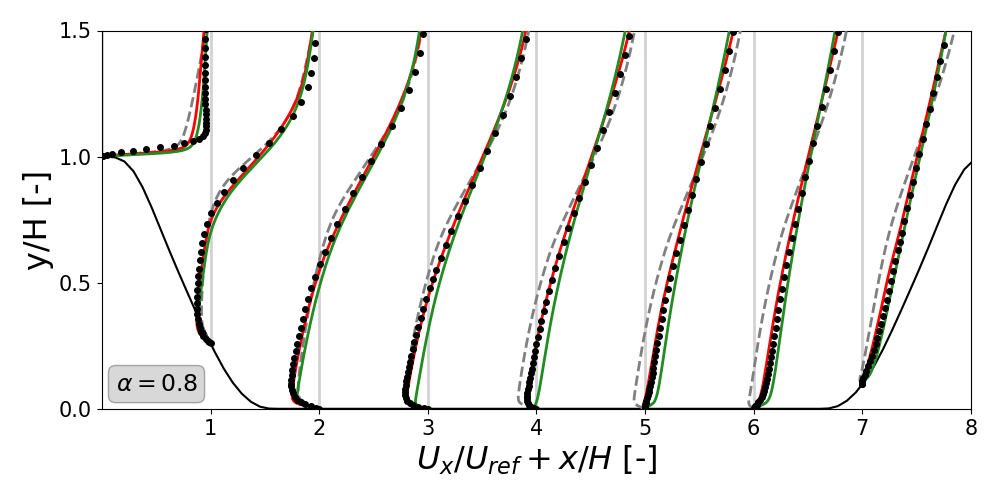} 
       % \caption{Axial velocity profiles.}
       % \label{fig:ux-PH08}
    \end{subfigure}
     \hfill
    \begin{subfigure}[b]{0.48\textwidth}
        \centering
        \includegraphics[width=\textwidth]{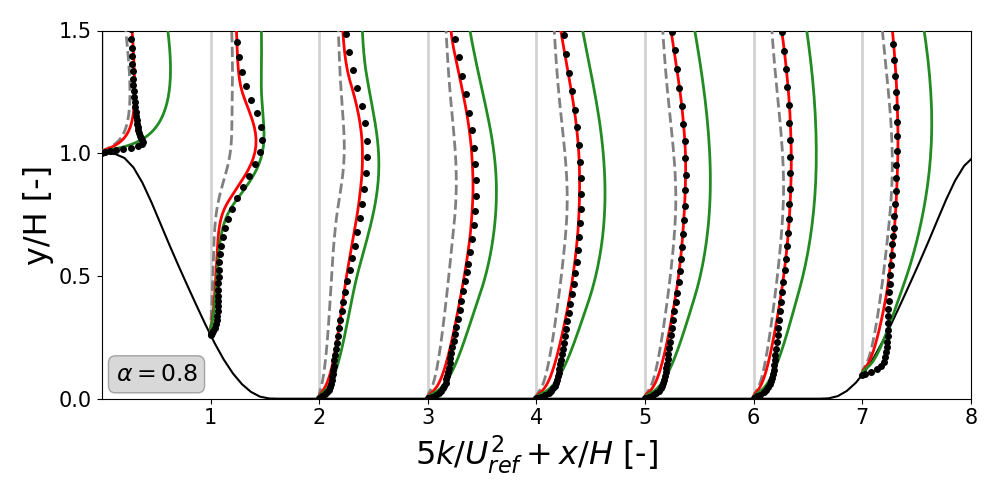}
       % \caption{Turbulent kinetic energy profiles.}
        %\label{fig:k-PH08}
    \end{subfigure}
    %\end{comment}

    \begin{subfigure}[b]{0.48\textwidth}
        \centering
        \includegraphics[width=\textwidth]{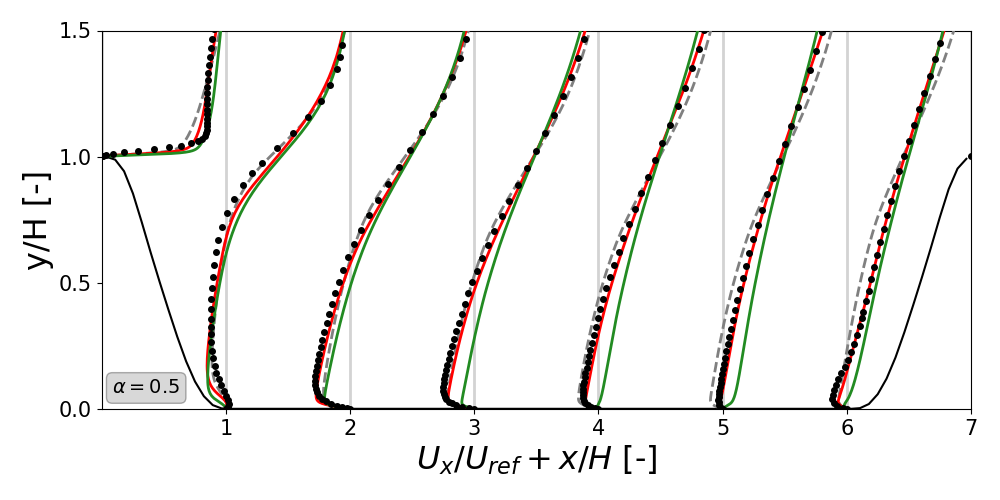} 
        \caption{Axial velocity profiles.}
        \label{fig:ux-PH05}
    \end{subfigure}
     \hfill
    \begin{subfigure}[b]{0.48\textwidth}
        \centering
        \includegraphics[width=\textwidth]{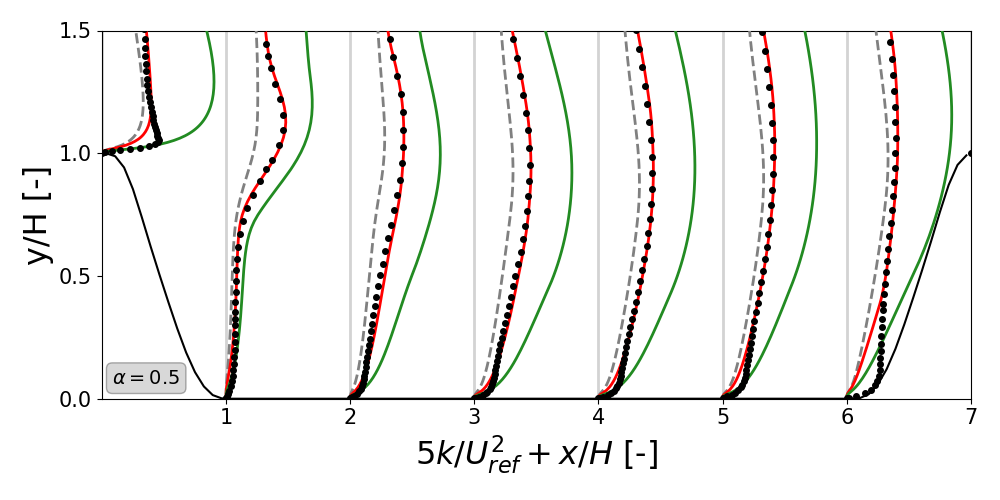}
        \caption{Turbulent kinetic energy profiles.}
        \label{fig:k-PH05}
    \end{subfigure}
    
    \caption{Performance of the shear layer SpaRTA model across parameterized Periodic Hill cases with varying width-scaling factors ($\alpha = 0.5-1.5$). DNS data (black dots), baseline $k$-$\omega$ SST (dashed), SL-Propagation (red), and SL-Model (green).}
    \label{fig:parametrized-results-final}
\end{figure}

The performance of the shear layer SpaRTA model on the periodic hill test cases is summarized in Figure \ref{fig:parametrized-results-final}, comparing axial velocity and turbulent kinetic energy (TKE) profiles for various $\alpha$ values. The SL-Model consistently improves upon the baseline $k-\omega$ SST predictions, with the magnitude of improvement varying systematically with geometry.

For larger hill spacing ($\alpha = 1.5$), where the flow has more room to develop, the velocity profiles show significant improvement in the extended separation region ($4 \leq x/H \leq 6$). The SL-Model reduces the baseline's underprediction of velocity within the recirculation zone, better capturing the delayed reattachment characteristic of widely-spaced hills. The TKE overprediction is most pronounced in the extended shear layer region, where the correction term R enhances mixing to achieve improved mean flow prediction.

For intermediate cases ($\alpha = 1.2$ and $\alpha = 1.0$), where the separation bubble size gradually decreases, the SL-Model shows particular improvement in capturing the separation point and initial shear layer development. At $x/H \approx 2.5$, the model better captures the velocity inflection point, reducing the overestimation of recirculation intensity observed in the baseline model. The TKE predictions show better spatial distribution compared to larger $\alpha$ cases, though still elevated compared to HF data.

For the smallest hill spacing ($\alpha = 0.8$ and $\alpha = 0.5$), where hill-to-hill interaction becomes stronger, the SL-Model shows more variation in performance. While it improves the prediction of the compact separation region, there is a tendency to overpredict velocity ($x/H > 3$) due to enhanced mixing. The TKE profiles reflect the intensified interaction between successive separations, with elevated levels particularly evident in the regions between hills. Despite these challenges, the model maintains better prediction of shear layer development compared to the baseline.

\begin{figure}[H]
\centering
\begin{subfigure}{0.48\textwidth}
    \centering
    \includegraphics[width=\textwidth]{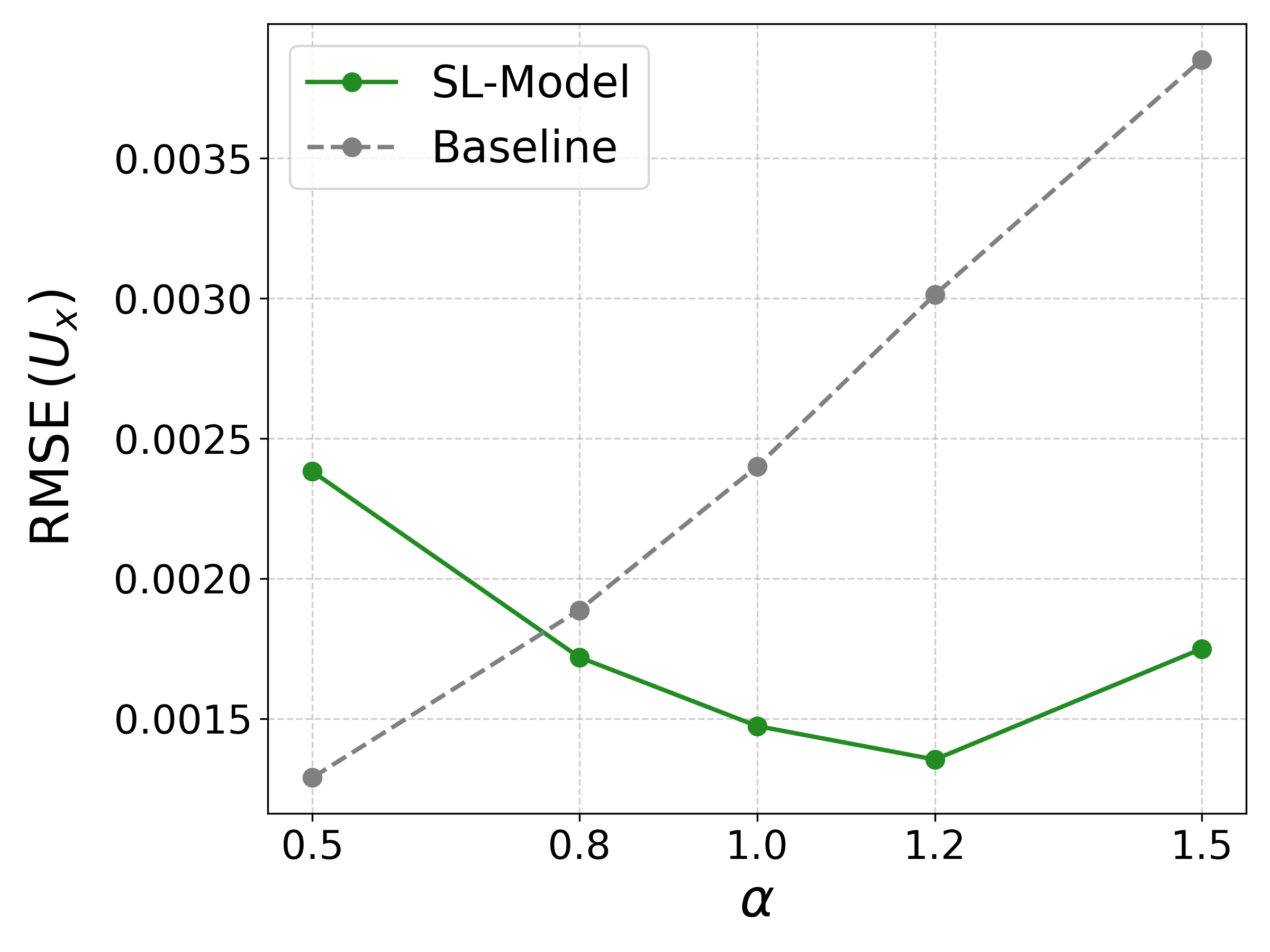}
    \caption{Root mean square error (RMSE) of axial velocity predictions for baseline $k-\omega$ SST and SL-Model.}
    \label{fig:RMSE_UX}
\end{subfigure}
\hfill
\begin{subfigure}{0.48\textwidth}
    \centering
    \includegraphics[width=\textwidth]{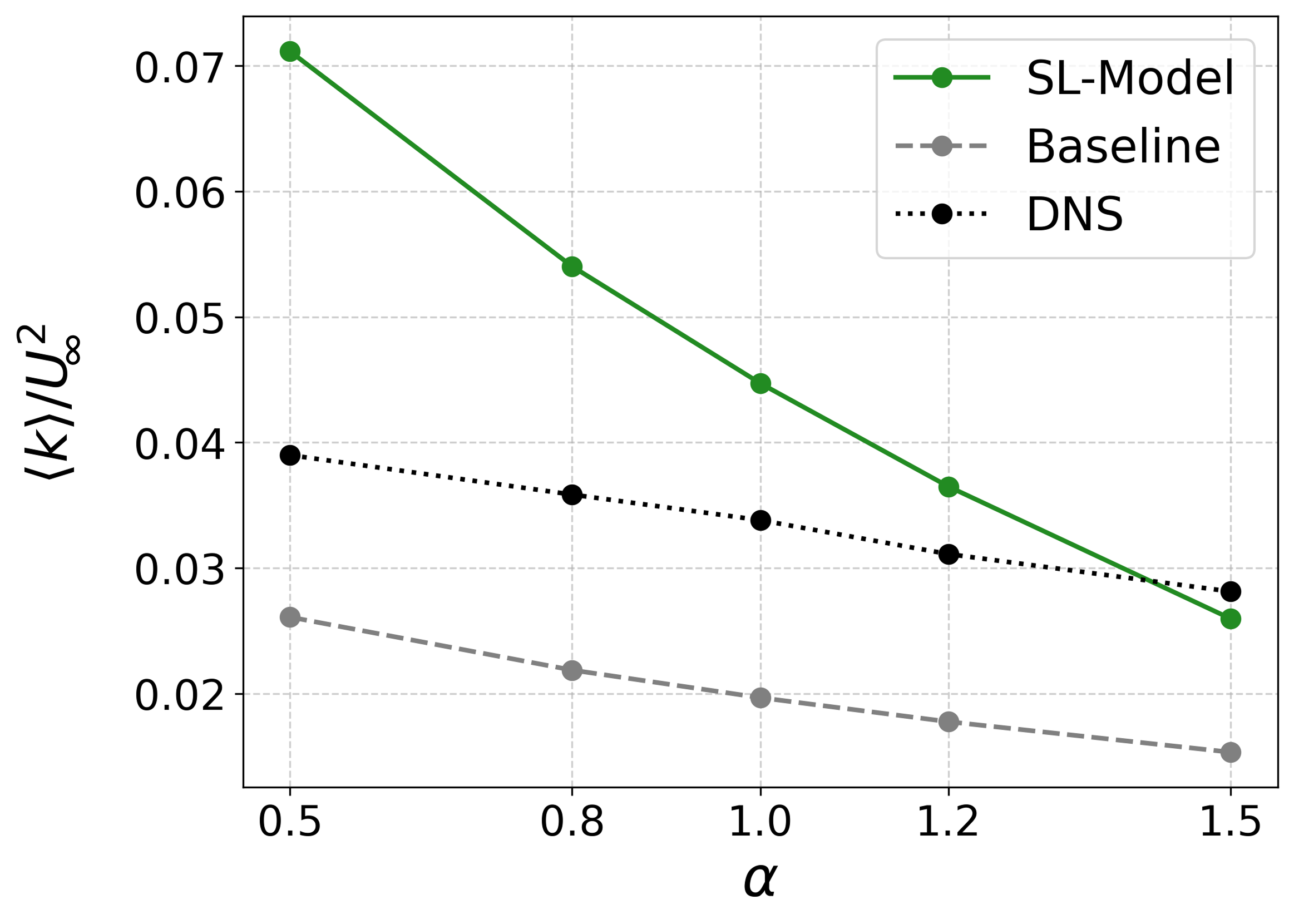}
    \caption{Average TKE in the shear layer region for baseline $k-\omega$ SST, SL-Model, and DNS data.}
    \label{fig:AVG_TKE}
\end{subfigure}
\caption{Comparison of RANS model performance metrics across different hill geometries of hill width-scaling factors ($\alpha$).}
\label{fig:combined_metrics}
\end{figure}

%\begin{figure}[H]
%\centering
%\includegraphics[width=0.8\textwidth]{Figures/Periodic-Hill-Parametrized/RMSE_U_PHcases.png}
%\caption{Root mean square error (RMSE) of axial velocity predictions for baseline k-$\omega$ SST and SL-Model across different hill width-scaling factors ($\alpha$). Lower values indicate better agreement with DNS data.}
%\label{fig:RMSE_UX}
%\end{figure}

To quantitatively evaluate prediction accuracy across geometric configurations, Figure \ref{fig:RMSE_UX} presents the root mean square error (RMSE) of axial velocity as a function of $\alpha$. The baseline $k-\omega$ SST model shows steadily increasing error with $\alpha$, indicating degrading performance as hill spacing increases and the separation region extends. The SL-Model exhibits a more nuanced performance pattern across the parametric range. Notably, at $\alpha = 0.5$, the SL-Model actually performs worse than the baseline, reflecting the challenges in capturing the complex hill-to-hill interactions at small spacings where enhanced mixing leads to velocity overprediction. However, as $\alpha$ increases, the SL-Model demonstrates progressively better performance with optimal accuracy at $\alpha = 1.2$, followed by a slight degradation at $\alpha = 1.5$ while still maintaining substantial improvement over the baseline. This pattern suggests that the correction mechanisms are most effective for intermediate to large hill spacings where flow structures have sufficient room to develop, while struggling with the highly interactive flows at the smallest spacing.

%\begin{figure}[H]
%\centering
%\includegraphics[width=0.8\textwidth]%{Figures/Periodic-Hill-%Parametrized/mean_k_PHCases.png}
%\caption{Average turbulent kinetic energy in the %shear layer region normalized by reference %velocity squared for baseline k-$\omega$ SST, SL-%Model, and DNS data across different hill width-%scaling factors ($\alpha$).}
%\label{fig:AVG_TKE}
%\end{figure}

Figure \ref{fig:AVG_TKE} examines average TKE levels within the shear layer region across the parametric space. The DNS data shows gradually decreasing TKE with increasing $\alpha$, reflecting reduced interaction between successive hills in widely-spaced configurations. The baseline model systematically underpredicts turbulence intensity across all cases, while the SL-Model shows elevated TKE that progressively converges with DNS data as $\alpha$ increases. This convergence trend suggests that the correction terms are most effective in well-defined, extended shear layers characteristic of larger $\alpha$ values, where flow structures develop more fully between hills.

These complementary metrics reveal an important trade-off in the model's behavior: improved mean flow prediction comes through enhanced turbulent mixing that sometimes overestimates TKE magnitude, particularly at smaller $\alpha$ values where complex hill-to-hill interactions present greater modeling challenges. As separation regions become more well-defined at larger $\alpha$ values, the SL-Model achieves both better mean flow prediction and more physically accurate turbulence levels, suggesting an optimal range for the current formulation.

This systematic variation in model performance with $\alpha$ demonstrates that the improvements are not coincidental but rather reflect the model's ability to adapt to different separation mechanisms. While this parametric study represents a constrained generalization test (maintaining the same fundamental flow physics with geometric variations), the consistent behavior of both the classifier and correction terms across these geometries provides valuable validation. This controlled test complements our more challenging 3D generalization cases by isolating the model's response to geometric changes while preserving similar underlying physics. The persistent TKE overprediction observed across all parametric cases indicates an area for potential model refinement that may be significant when extending to more diverse flow configurations.

\subsection{Faith Hill (3D Generalization Test)}

Moving beyond parametric variations of 2D geometries, we next examine the model's performance on the Faith Hill configuration - a fully three-dimensional separated flow. This case, which was not included in the training dataset, presents a more challenging test of generalization, combining smooth-surface separation with specific three-dimensional effects such as horseshoe vortex formation and spanwise flow variation that were not present in the training data.

\begin{figure}[H]
    \centering
    \includegraphics[width=\textwidth]{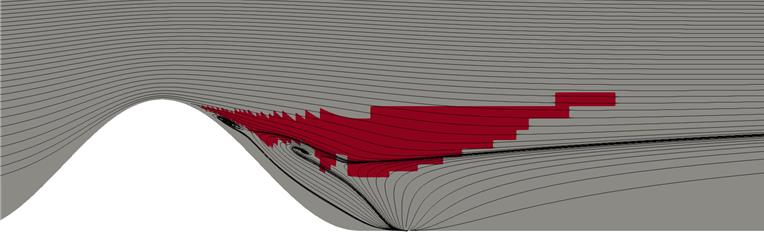}
    \caption{Classification outcome of the $\sigma_{SL}$ classifier on the baseline Faith Hill case: red region: $\sigma_{SL}$ = 1, gray region: $\sigma_{SL}$ = 0.}
    \label{fig:faith-hill-sigma}
\end{figure}

Figure \ref{fig:faith-hill-sigma} shows the classification outcome of the $\sigma_{SL}$ classifier at the symmetry plane. The classifier's parameters ($\phi_{D_k/P_k}$, $\phi_k$, $Re_\Omega$) successfully identify the separated shear layer region downstream of the hill's crest where the baseline $k-\omega$ SST model requires enhancement. The spatial extent of the identified region (shown in red) aligns with the expected separation zone, indicating that the classifier criteria remain physically relevant in 3D configurations.

\begin{figure}[H]
    \centering
    \begin{subfigure}[b]{0.49\textwidth}
        \centering
        \includegraphics[width=\textwidth]{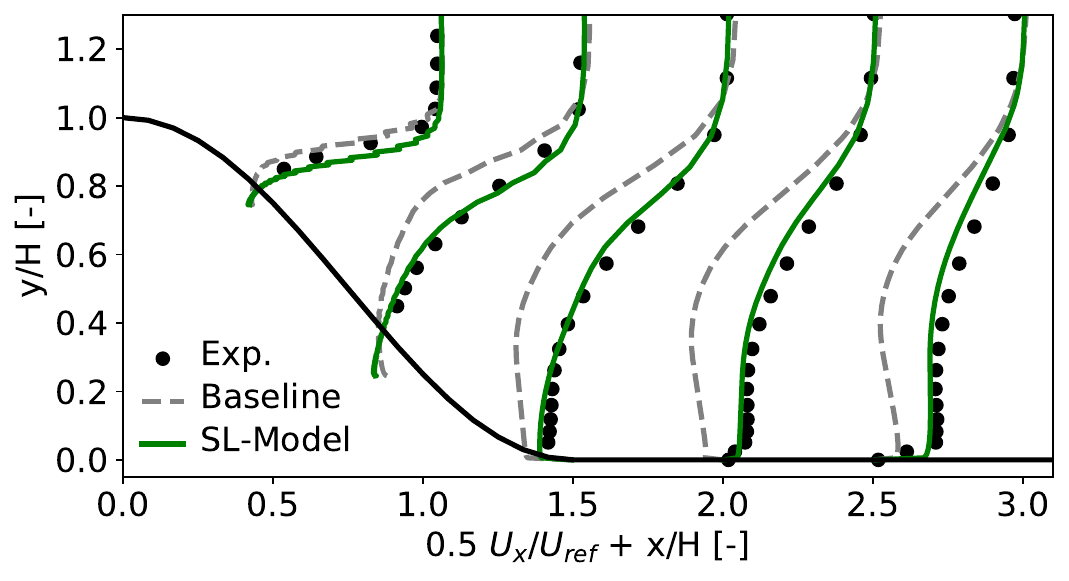}
        \caption{Axial velocity profiles.}
        \label{fig:faith-hill-Ux}
    \end{subfigure}
    \hfill
    \begin{subfigure}[b]{0.49\textwidth}
        \centering
        \includegraphics[width=\textwidth]{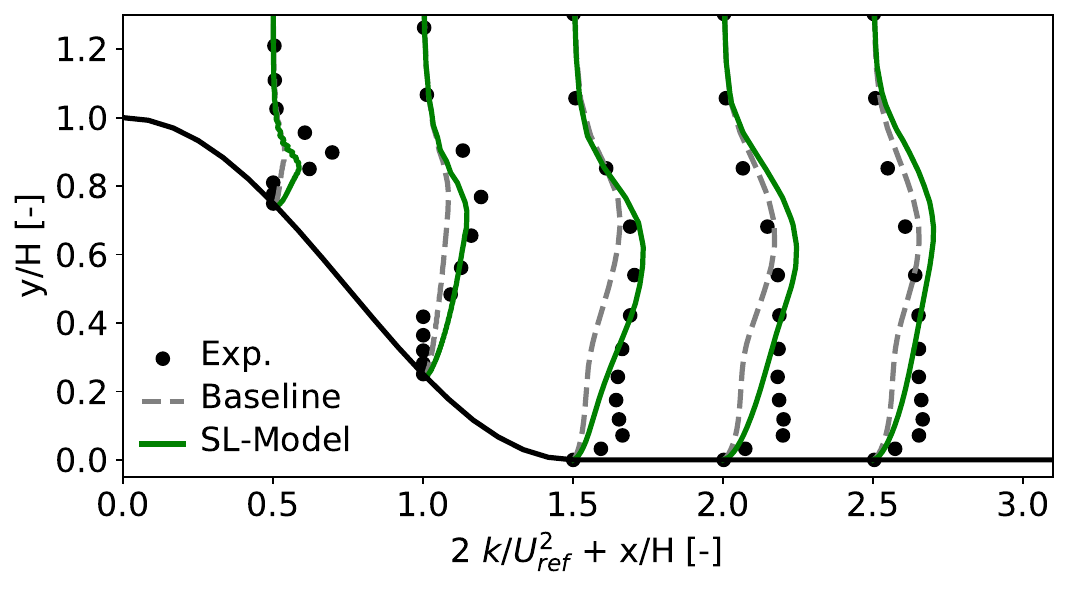} 
        \caption{Turbulent kinetic energy profiles.}
        \label{fig:faith-hill-k}
    \end{subfigure}
    
    \caption{Model performance comparison at the Faith Hill symmetry plane: experimental data (symbols), baseline $k$-$\omega$ SST (dashed), and SL-Model (solid) at five streamwise locations ($x/H$ = 0.5-3.0).}
    \label{fig:faith-hill-profiles}
\end{figure}

The performance of the SL-Model at the symmetry plane is presented in Figure \ref{fig:faith-hill-profiles}. The axial velocity profiles show significant improvement over the baseline $k-\omega$ SST predictions, particularly in the separated region ($1.5 < x/H < 2.5$). The SL-Model better captures both the extent of the separation bubble and the recovery of the velocity profiles downstream. This improvement demonstrates that the tensor-based corrections, formulated using Pope's invariant basis, maintain their effectiveness when applied to 3D flows.

The turbulent kinetic energy profiles in Figure \ref{fig:faith-hill-k} exhibit behavior consistent with previous test cases. While the SL-Model shows improved prediction of TKE distribution compared to the baseline, particularly in capturing peak locations, it maintains the characteristic overprediction in the shear layer region. This consistency between 2D and 3D predictions suggests the correction mechanisms remain physically relevant while trained on 2D data due to their invariant formulation.

The model's ability to improve predictions for this 3D configuration stems from two key aspects: the use of frame-invariant quantities in both the classifier and correction terms, ensuring their physical meaning translates across dimensionality; and the robust formulation of the correction terms using Pope's tensor basis, which maintains consistent physical behavior in both 2D and 3D flows. This successful application to the Faith Hill case demonstrates the model's capability to enhance predictions beyond its training scope.

\subsection{Ahmed Body (3D Generalization Test)}

We examine the Ahmed body at slant angle of 25 degrees \cite{wagner_flow_2002}, which features complex separation patterns including streamwise vortices shed from the sharp upper corners of the rear slant. At a 25 degree rear slant angle, these corner vortices have sufficient strength to reattach the flow half way down the slant. The Ahmed body, with its distinct slant angle and sharp edges, represents a significant departure from the training configurations, testing the model's ability to handle fully three-dimensional separation mechanisms.

\begin{figure}[H]
    \centering
    \begin{subfigure}[b]{\textwidth}
        \centering
        \includegraphics[width=0.9\textwidth]{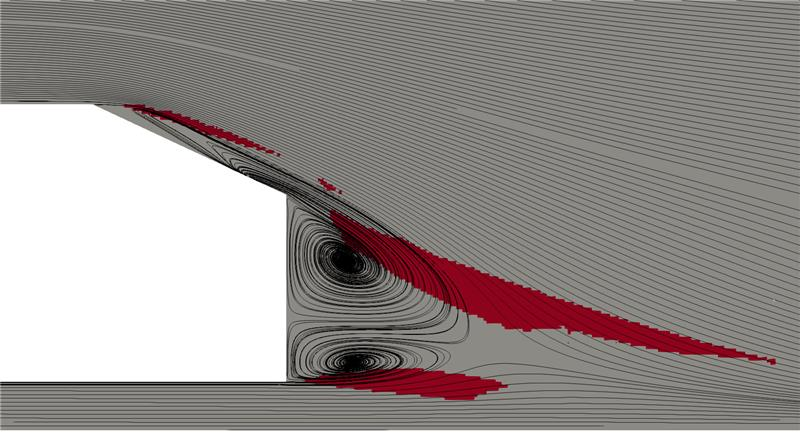}
        \caption{$\sigma_{SL}$ classifier at symmetry plane.}
        \label{fig:Ahmed-sigma-symm}
    \end{subfigure}
    \hfill
    \begin{subfigure}[b]{0.49\textwidth}
        \centering
        \includegraphics[width=\textwidth]{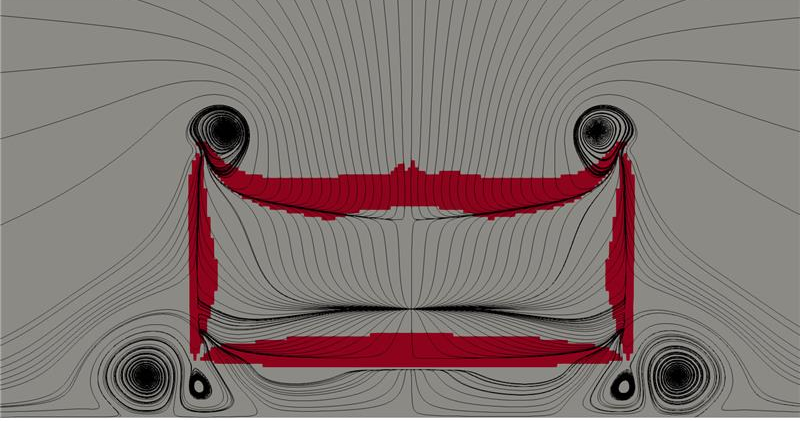} 
        \caption{$\sigma_{SL}$ classifier at x/H = 0.27}
        \label{fig:Ahmed-body-sigma-27}
    \end{subfigure}
    \hfill
    \begin{subfigure}[b]{0.49\textwidth}
        \centering
        \includegraphics[width=\textwidth]{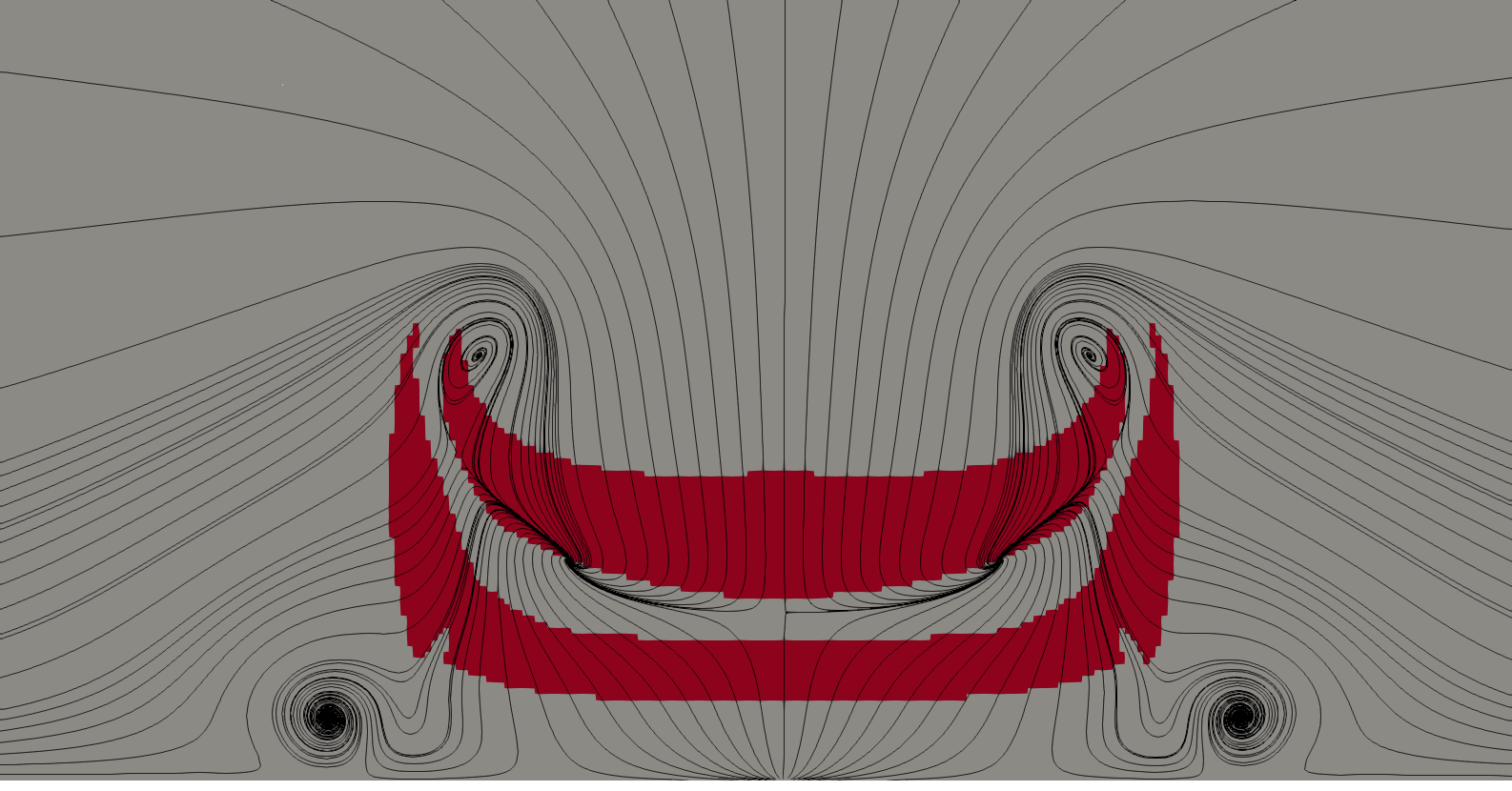} 
        \caption{$\sigma_{SL}$ classifier at x/H = 0.69}
        \label{fig:Ahmed-body-sigma-69}
    \end{subfigure}
    \caption{Classification outcome of the $\sigma_{SL}$ classifier on the Ahmed body: visualization of shear layer regions ($\sigma_{SL}$ = 1, red) at (a) symmetry plane showing separation from slant angle, (b,c) streamwise planes (x/H = 0.27, 0.69) showing how the classifier remains inactive in regions where three-dimensional vortical structures develop.}
    \label{fig:Ahmed-sigma}
\end{figure}

Figure \ref{fig:Ahmed-sigma} presents the $\sigma_{SL}$ classifier results across multiple planes. At the symmetry plane (Figure \ref{fig:Ahmed-sigma-symm}), it captures the separation from the slant angle and subsequent shear layer development. The streamwise planes (Figures \ref{fig:Ahmed-body-sigma-27} and \ref{fig:Ahmed-body-sigma-69}) reveal how the classifier appropriately remains inactive in regions where streamwise vortices develop from the slant edges. This selective activation demonstrates the classifier's ability to distinguish between shear layer separation and vortical regions.

\begin{figure}[H]
    \centering
    \includegraphics[width=\textwidth]{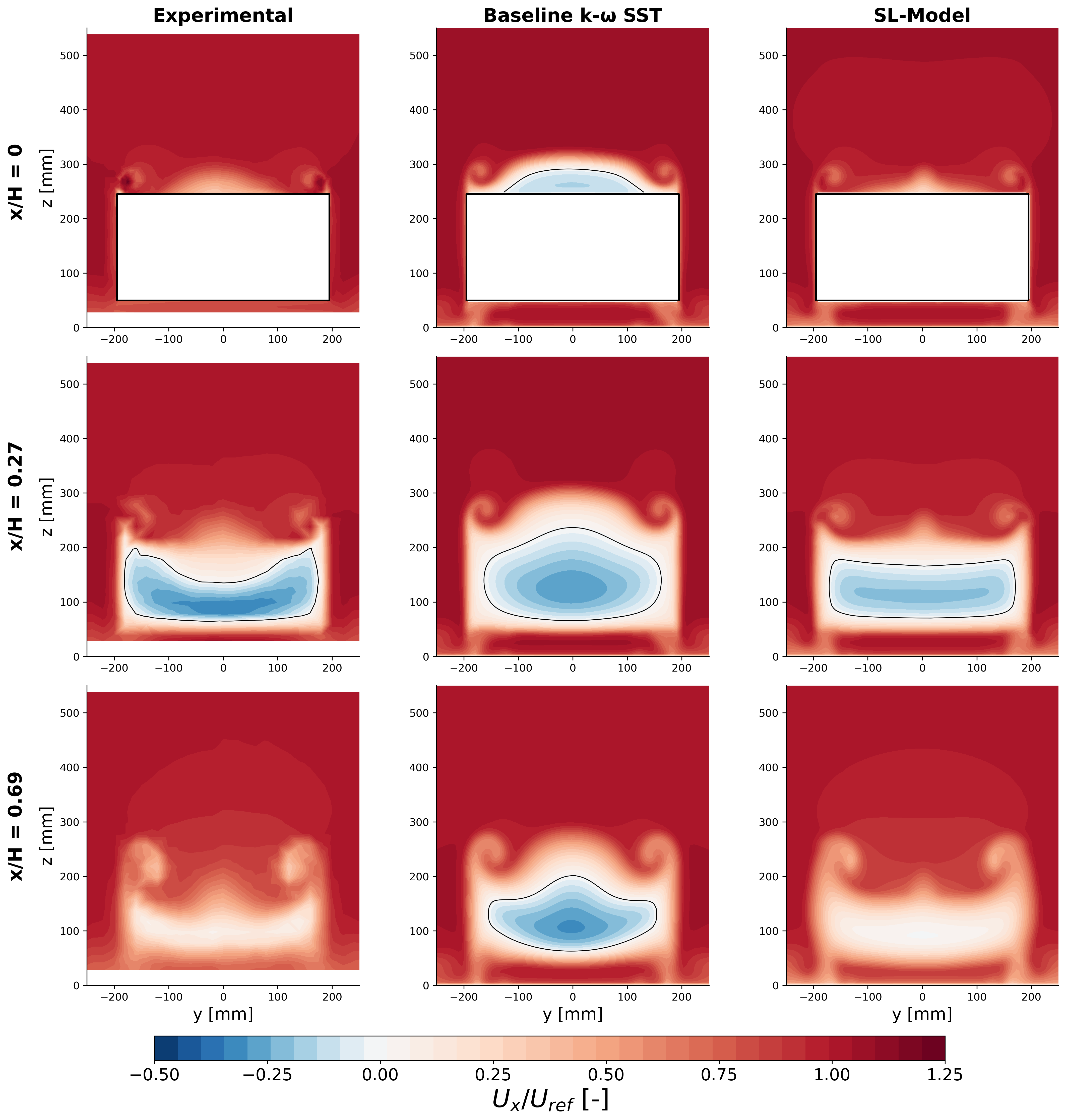}
    \caption{Comparison of wake structure predictions at streamwise planes (x/H = 0, 0.27, 0.69). Columns show experimental data (left), baseline $k-\omega$ SST model (middle), and zonally augmented $k-\omega$ SST model (right). Contours show streamwise velocity normalized by freestream velocity.Black contour lines indicate where streamwise velocity is zero, delineating the separation region.}
    \label{fig:Ahmed-Wake-figures}
\end{figure}

Figure \ref{fig:Ahmed-Wake-figures} illustrates the wake structure development at three streamwise locations behind the Ahmed body. At x/H = 0 (immediately behind the body), all three cases show the initial formation of the wake, with the baseline $k-\omega$ SST model predicting a larger recirculation region (blue contour enclosed by the black zero-velocity line) that is not present in the experimental data. The SL-Model shows improved prediction of the initial wake formation by correctly capturing the absence of this immediate recirculation.

At x/H = 0.27, a pronounced recirculation zone is visible in all three cases, clearly defined by the black zero-velocity contour lines. The experimental data reveals an asymmetric recirculation bubble with varying depths across the span. The baseline model overpredicts both the size and intensity of this recirculation region, showing a more symmetric, deeper, and more concentrated negative velocity region. In contrast, the SL-Model produces a shallower recirculation zone with dimensions and intensity that more closely match the experimental measurements, particularly in capturing the characteristic flattened shape of the upper boundary of the recirculation region.

Further downstream at x/H = 0.69, where wake recovery begins, significant differences persist. The baseline model continues to predict an overly strong and concentrated recirculation region, as indicated by the extent of its zero-velocity contour, while the experimental data shows a weakening and more diffuse negative velocity region. The SL-Model maintains better agreement with experimental data by correctly predicting a more moderate recirculation intensity and better capturing the spatial extent of the wake. This demonstrates the SL-Model's superior ability to predict both the initial formation and subsequent evolution of the wake structure, while appropriately preserving baseline model behavior in areas dominated by vortical structures outside the primary recirculation region.

\begin{figure}[H]
    \centering
    \begin{subfigure}[b]{0.49\textwidth}
        \centering
        \includegraphics[width=\textwidth]{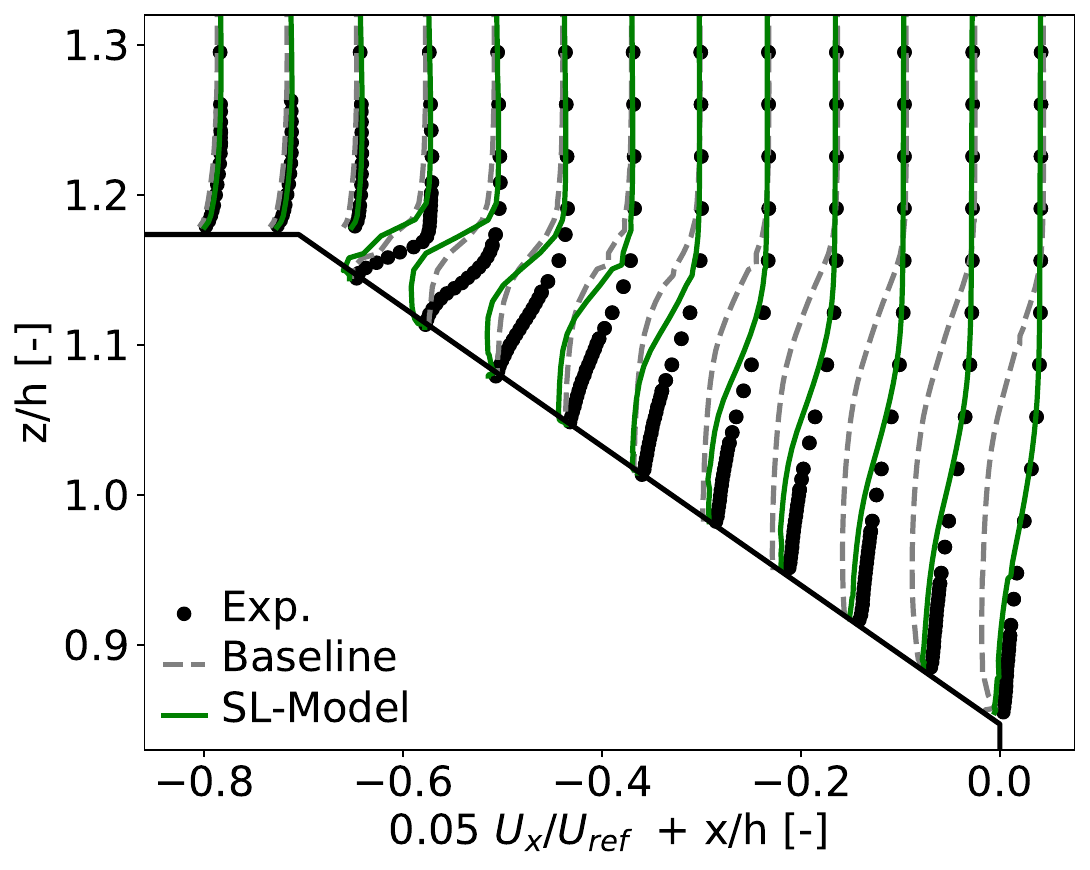}
        \caption{Axial velocity along slant.}
        \label{fig:Ahmed_U_slant}
    \end{subfigure}
    \hfill
    \begin{subfigure}[b]{0.49\textwidth}
        \centering
        \includegraphics[width=\textwidth]{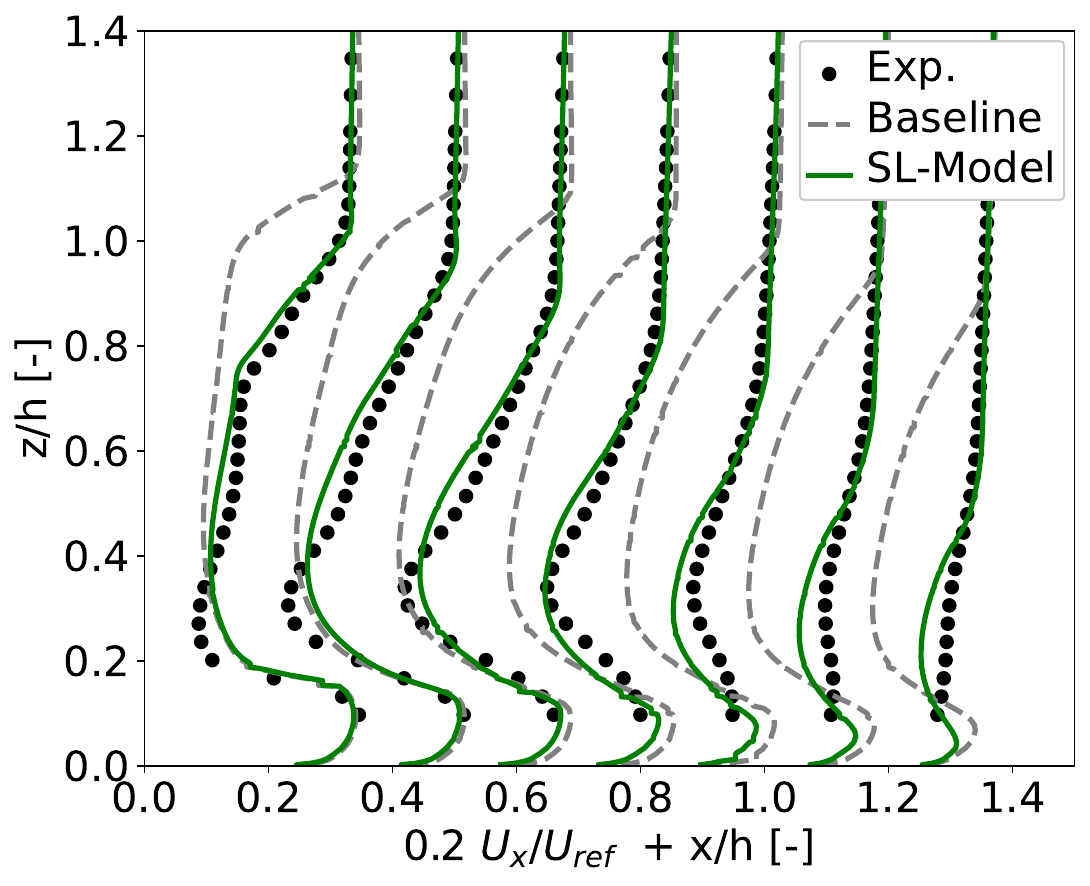} 
        \caption{Axial velocity along wake.}
        \label{fig:Ahmed_U_wake}
    \end{subfigure}
    
    \begin{subfigure}[b]{0.49\textwidth}
        \centering
        \includegraphics[width=\textwidth]{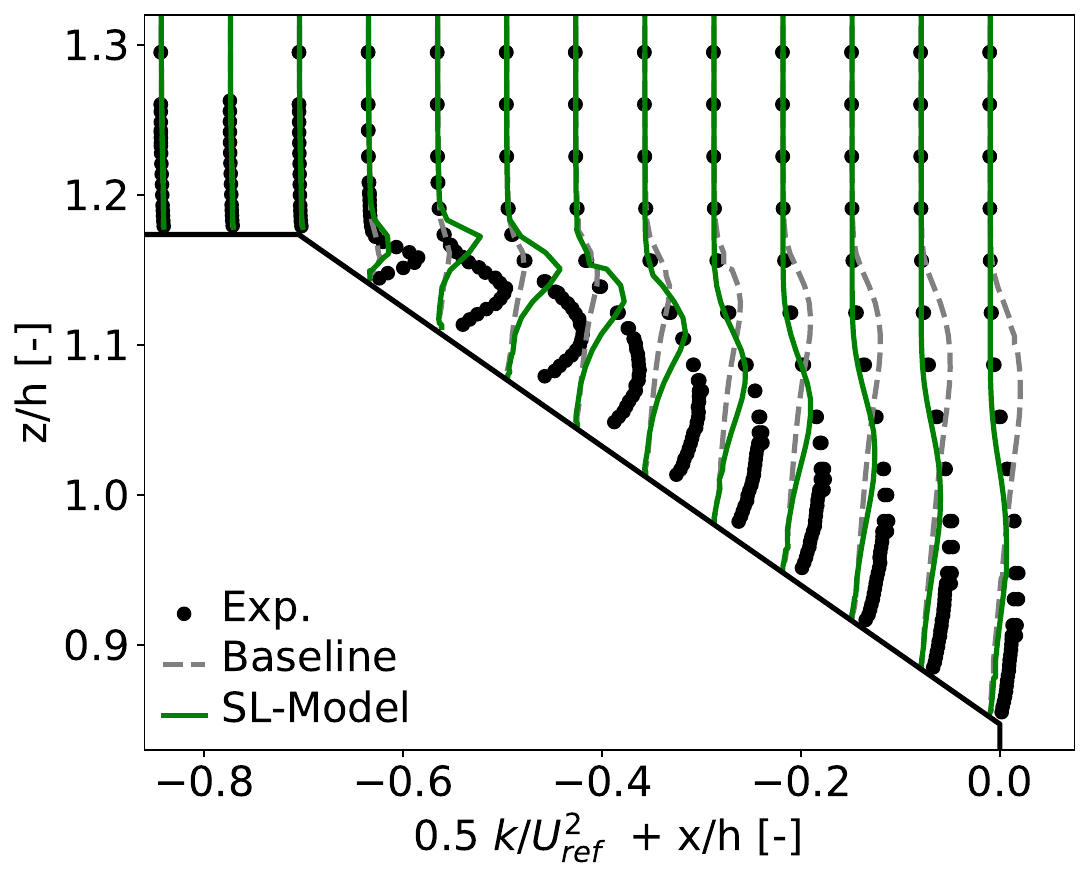}
        \caption{Turbulent kinetic energy along slant.}
        \label{fig:Ahmed_k_slant}
    \end{subfigure}
    \hfill
    \begin{subfigure}[b]{0.49\textwidth}
        \centering
        \includegraphics[width=\textwidth]{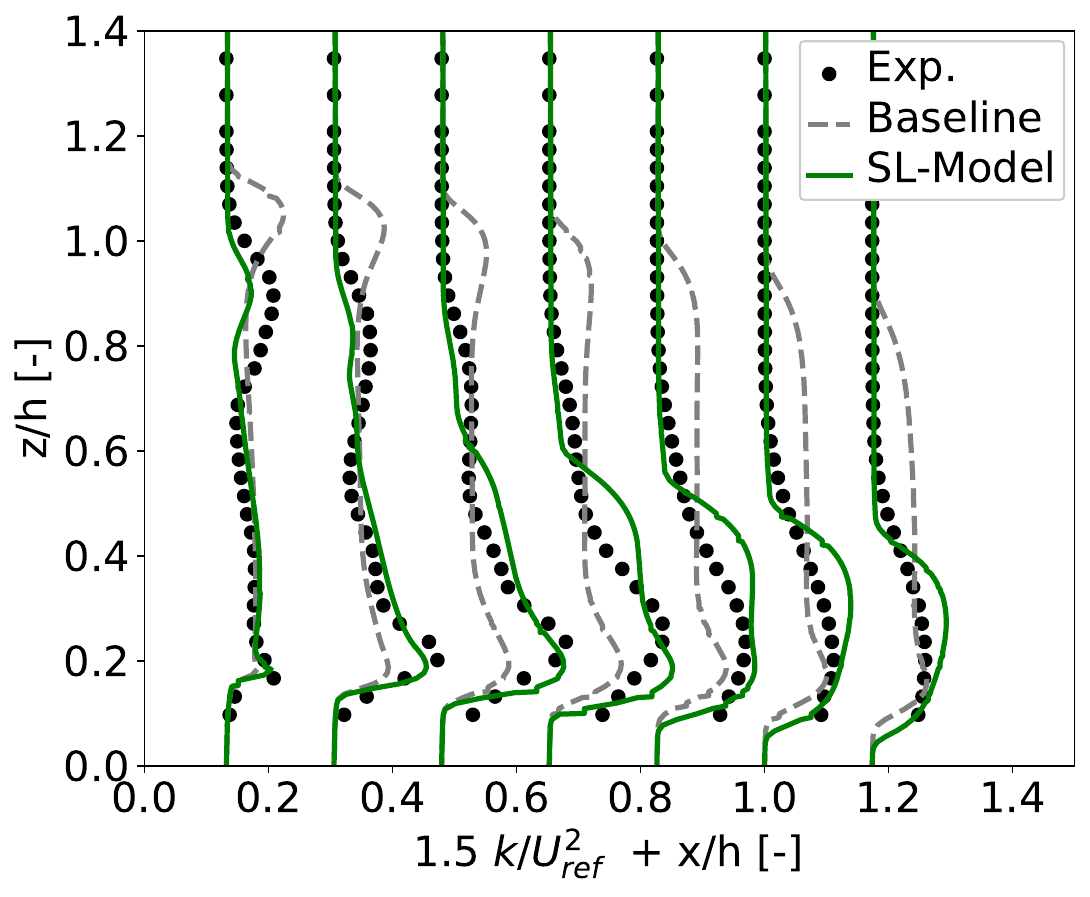} 
        \caption{Turbulent kinetic energy along wake.}
        \label{fig:Ahmed_k_wake}
    \end{subfigure}
    \caption{Axial velocity, a \& b and turbulent kinetic energy profiles, c \& d on the Ahmed body case at the symmetry plane. }
    \label{fig:Ahmed-body-profiles}
\end{figure} 

Quantitative validation through symmetry plane profiles in Figure \ref{fig:Ahmed-body-profiles} demonstrates these improvements. The velocity profiles along the slant angle (-0.7 $<$ x/h $<$ 0) show better prediction of shear layer development, particularly evident near x/h = -0.65 where the baseline model underpredicts the velocity gradient. In the wake region (x/h $>$ 0), the SL-Model achieves closer agreement with experimental data for both velocity deficit magnitude and recovery rate.

The turbulent kinetic energy profiles at the symmetry plane maintain trends observed in previous cases. While showing characteristic elevated levels compared to experimental data, the SL-Model better captures the spatial distribution of turbulence in the shear layer region. This indicates the correction mechanisms appropriately enhance mixing where intended, though with some overprediction of turbulence kinetic energy.

The model's successful prediction of separated shear layer features, combined with its appropriate preservation of baseline behavior in vortical regions, demonstrates the effectiveness of the zonal correction approach. This selective enhancement suggests potential for developing targeted corrections for specific flow features, such as streamwise vortices, in future work.

\section{Conclusion}\label{sec4}

This study introduced the Relative Importance Term Analysis (RITA) methodology, a physics-based approach for systematically identifying and correcting regions where RANS models require enhancement in separated flows. Building upon the SpaRTA framework, we developed a zonal augmentation strategy that combines a frame-invariant classifier with physically consistent correction terms. The RITA methodology demonstrates an effective approach for enhancing RANS predictions in separated flows while preserving the baseline model's established accuracy in fundamental cases such as boundary layers and channel flows. This selective correction strategy ensures that model performance is maintained in well-predicted regions, addressing a key challenge in data-driven turbulence modeling where universal corrections often compromise performance in canonical flows.

The RITA classifier, using dimensionless ratios derived from the turbulence kinetic energy equation, successfully identified shear layer regions requiring correction across diverse flow configurations. Its physics-based criteria of destruction-to-production ratio, Turbulent-to-Total Kinetic Energy Ratio (TTKER), and vorticity Reynolds number demonstrated remarkable adaptability from simple 2D geometries to complex 3D configurations. Importantly, the classifier appropriately remained inactive in regions dominated by streamwise vortices, such as those developing from the slant edges of the Ahmed body, highlighting its ability to distinguish between shear layer separation and vortical regions. This selective activation ensures corrections are applied only where intended, preserving the baseline model's behavior in regions where it performs adequately.

The discovered correction terms, $b_{ij}^{\Delta}$ for anisotropic stress and $R$ for turbulent production, worked in tandem to improve flow predictions through complementary mechanisms. The anisotropic correction enhanced shear layer prediction by better capturing the interaction between mean strain and rotation rates, while the scalar correction augmented turbulent transport through increased production. These mechanisms systematically improved mean flow predictions across all test configurations, from parametric variations of periodic hills to fully three-dimensional cases like the Faith Hill and Ahmed body.

Mean flow predictions showed marked improvement in all test cases. The zonally augmented model successfully handled parametric variations in separation bubble size and reattachment dynamics, maintained accuracy in quasi-2D regions of 3D flows, and captured complex features like wake structures not present in the training data. This consistent performance validates the physical foundations of our correction approach.

Despite these improvements, certain limitations warrant further investigation. The systematic overprediction of turbulent kinetic energy, while enabling improved mean flow prediction through enhanced mixing, suggests potential refinements in the correction mechanisms. Additionally, the current classifier is not designed to identify streamwise vortices, representing an area for future development of complementary classification criteria.

For future work, several promising directions emerge. First, extending the training to include 3D cases could improve the model's performance in predicting complex three-dimensional features. Second, the development of specialized classifiers for vortical structures could complement the current shear layer focus, allowing comprehensive treatment of both separation and vortical regions. Third, further refinement of the correction terms might address the TKE overprediction while maintaining the improved mean flow predictions.

In conclusion, the RITA methodology and resulting zonally augmented RANS model provide a viable approach for enhancing predictions of separated flows. The framework's success in handling various separation mechanisms while appropriately distinguishing between shear layers and vortical regions demonstrates that selective application of physically motivated corrections offers an effective pathway toward improved RANS modeling for complex aerodynamic configurations.

\section*{Acknowledgements}
The authors gratefully acknowledge Williams Grand Prix Engineering Limited for financial support. 

\appendix

\section{Non-Linear Basis Tensors and Invariants}
\label{app:invariants}

The ten non-linear basis tensors $T_{ij}^{(n)}$ and five invariants $I_m$ are defined following the work of \citet{pope_more_1975}:

\begin{align}
    \begin{split}
        T_{ij}^{(1)} &= S \\
        T_{ij}^{(2)} &= S \Omega - \Omega S \\
        T_{ij}^{(3)} &= S^2 - \frac{1}{3} I \{S^2\}  \\
        T_{ij}^{(4)} &= \Omega^2 - \frac{1}{3} I \{\Omega^2\} \\
        T_{ij}^{(5)} &= \Omega S^2 - S^2 \Omega \\
    \end{split}
    \quad\quad
    \begin{split}
        T_{ij}^{(6)} &= \Omega^2 S + S \Omega^2 - \frac{2}{3} I \{ S \Omega^2 \} \\
        T_{ij}^{(7)} &= \Omega S \Omega^2 - \Omega^2 S \Omega \\
        T_{ij}^{(8)} &= S \Omega S^2 - S^2 \Omega S  \\
        T_{ij}^{(9)} &= \Omega^2 S^2 + S^2 \Omega^2 - \frac{2}{3} I \{S^2 \Omega^2\} \\ 
        T_{ij}^{(10)} &= \Omega S^2 \Omega^2 - \Omega^2 S^2 \Omega
    \end{split}
    \label{eq:appendix-pope-bases}
\end{align}

Where the corresponding five invariants are:

\begin{align}  
    I_1 = \{ S^2 \}, \quad I_2 = \{ \Omega^2 \}, \quad  I_3 = \{ S^3 \}, \quad  I_4 = \{ \Omega^2 S \}, \quad  I_5 = \{ \Omega^2 S^2 \}.
    \label{eq:appendix-pope-bases-invars}
\end{align}

Note: $S$ represents the symmetric part of the velocity gradient tensor, $\Omega$ represents the antisymmetric part, and $I\{\cdot\}$ denotes the isotropic part of the tensor.

\section{Unary Functions for Feature Library}\label{app:unary_functions}
The following unary transformations were applied to the invariants in constructing the feature libraries for $b_{ij}^{\Delta}$ and $R$:

\begin{align}
f_1(x) &= x \quad \text{(Identity)} \\
f_2(x) &= x^2 \quad \text{(Square)} \\
f_3(x) &= |x| \quad \text{(Absolute value)} \\
f_4(x) &= \sqrt{|x|} \quad \text{(Square root of absolute value)} \\
f_5(x) &= \tanh(x) \quad \text{(Hyperbolic tangent)} \\
f_6(x) &= \frac{x}{1 + x^2} \quad \text{(Regularized division)} \\
f_7(x) &= \log(|x| + 1) \quad \text{(Regularized logarithm)}
\end{align}

The regularized functions provide numerical stability by avoiding singularities. An example of a candidate function in the library would be $\sqrt{|q_k|}T^{1}_{ij}$, where $q_k$ is an invariant and $T^{1}_{ij}$ is a basis tensor.

\section{Simulation Setup of Flow Cases}\label{secA1}

\subsection{NASA-Hump}

The NASA-Hump case is part of NASA's 2D-separated flow validation cases \cite{rumsey_nasa_2022}. Based on the Glauert-Goldschmied type body, it follows the experimental setup in \citet{greenblatt_experimental_2006}, with the OpenFOAM configuration from \citet{hoefnagel_multi-flow_2023}. At a reference Mach number of 0.1, the flow is treated as incompressible. The case features a turbulent boundary layer that accelerates over the hump under a favorable pressure gradient, separates at the hump's edge due to an adverse pressure gradient, then reattaches downstream.

\subsubsection{Flow Parameters}

An overview of the main flow parameters specified for the NASA-Hump case is given in Table \ref{tab:NASA-Humptransportproperties}. This case has the highest Reynolds number equal to 936,000, computed based on the chord length of the hump $c$ and the free-stream reference velocity $U_{ref}$.

\begin{table}[H]
    \centering
    \small
    \renewcommand{\arraystretch}{1.3} % Adjusting row spacing
      \captionsetup{justification=raggedright,singlelinecheck=false}
    \caption{Overview of the flow parameters specified for the NASA-Hump case.}
    \begin{tabular}{|c|c|c|}
        \hline
        \textbf{Transport Property} & \textbf{Parameter} & \textbf{Value} \\
        \hline
        Reynolds number based on chord Length & $ Re_c$ & 936,000 \\
        \hline
        Kinetic viscosity & $\nu$ & 1.55 x 10$^{-5}$ m$^{2}$ s$^{-1}$ \\
        \hline
        Free-stream reference velocity & $U_{ref}$ & 34.6 m s$^{-1}$ \\
        \hline
        Reference kinetic energy & $k_{ref}$ & 0.00107 m$^2$ s$^{-2}$ \\
        \hline
        Reference specific dissipation rate & $\omega_{ref}$ & 0.118 s$^{-1}$\\
        \hline
        Reference pressure & $p_{ref}$ & 0 m$^2$ s$^{-2}$ \\
        \hline
    \end{tabular}
    \label{tab:NASA-Humptransportproperties}
\end{table}

\subsubsection{Initial and Boundary Conditions}

The boundary conditions and initial conditions for the NASA-Hump case in OpenFOAM are outlined in Table \ref{tab:boundaryconditions-NASA}. The Frozen simulation uses LES data fields from \citet{uzun_wall-resolved_2017}, with variables $U$ and $k$ derived from these fields, while $\omega$ and $\nu_t$ match the Baseline simulation. The Baseline simulation's initial conditions are determined from the flow parameters in Table \ref{tab:NASA-Humptransportproperties}.

\begin{table}[H]
    \centering
    \small
    \renewcommand{\arraystretch}{1.1} % Adjusting row spacing
    \captionsetup{justification=raggedright,singlelinecheck=false}
    \caption{Boundary and initial conditions for the NASA-Hump case.}
    \begin{tabular}{|l|c|c|c|c|c|}
        \hline
        \multicolumn{6}{|c|}{\textbf{Boundary Conditions}} \\
        \hline
        \textbf{Location} & $U$ [m/s] & $p$ [m$^2$/s$^2$] & $k$ [m$^2$/s$^2$] & $\omega$ [s$^{-1}$] & $\nu_t$ [m$^2$/s] \\
        \hline
        Inlet & fixed & zero & fixed & fixed & calcu- \\
              & Value & Gradient & Value & Value & lated \\
        \hline
        Outlet & zero & fixed & zero & zero & calcu- \\
               & Gradient & Value & Gradient & Gradient & lated \\
        \hline
        Top & \multicolumn{5}{c|}{symmetry} \\
        \hline
        Bottom & no & zero & kqR & omega & nutUSpalding \\
               & Slip & Gradient & W.F. & W.F. & W.F. \\
        \hline
        \multicolumn{6}{|c|}{\textbf{Initial Conditions}} \\
        \hline
        Baseline & [$U_{ref}$ 0 0] & 0 & $k_{ref}$ & $\omega_{ref}$ & 0.009 \\
        \hline
        Frozen & $U_{LES,field}$ & 0 & $k_{LES,field}$ & $\omega_{ref}$ & 0.009 \\
        \hline
    \end{tabular}
    \label{tab:boundaryconditions-NASA}
\end{table}

\subsection{Periodic-Hill}

The Periodic-Hill case setup and the high-fidelity LES data have been obtained from the study of \citet{breuer_flow_2009}, who extensively studied this geometry across various Reynolds numbers \citet{breuer_flow_2009}. This configuration consists of a series of repeating hills separated by a flat surface region. The spacing between the hills is calculated to enable the flow to reattach to the flat surface post-separation before encountering the next hill. 

\subsubsection{Flow Parameters}

An overview of the main flow parameters specified for the Periodic-Hill case is given in Table \ref{tab:PHtransportproperties}. This case has a Reynolds number equal to 10,595, computed based on the hill height $H$ and reference bulk velocity $U_{ref}$.

\begin{table}[H]
    \centering
    \small
    \renewcommand{\arraystretch}{1.3} % Adjusting row spacing
      \captionsetup{justification=raggedright,singlelinecheck=false}
    \caption{Overview of the flow parameters specified for the Periodic-Hill case.}
    \begin{tabular}{|c|c|c|}
        \hline
        \textbf{Transport Property} & \textbf{Parameter} & \textbf{Value} \\
        \hline
        Reynolds number based on hill height & $ Re_H$ & 10,595 \\
        \hline
        Kinetic viscosity & $\nu$ & 9.44 x 10$^{-5}$ m$^2$ s$^{-1}$ \\
        \hline
        Free-stream reference velocity & $U_{ref}$ & 1 m s$^{-1}$ \\
        \hline
        Reference kinetic energy & $k_{ref}$ & 0.00375  m$^2$ s$^{-2}$ \\
        \hline
        Reference specific dissipation rate & $\omega_{ref}$ & 0.110 s$^{-1}$\\
        \hline
        Reference pressure & $p_{ref}$ & 0 m$^2$ s$^{-2}$ \\
        \hline
    \end{tabular}
    \label{tab:PHtransportproperties}
    
\end{table}

\subsubsection{Initial and Boundary Conditions}

The OpenFOAM boundary and initial conditions for this case are outlined in Table \ref{tab:boundaryconditions-PH}. To simulate the series of hills, periodic boundary conditions are applied at the inflow and outflow of the domain. The boundary conditions for $k$ and $\nu_t$ differ from those specified for the NASA-Hump case. Here, $k$ is specified to be zero at the walls. The \texttt{nutLowReWallFunction} also sets $\nu_t$ to zero.

\begin{table}[H]
    \centering
    \small
    \renewcommand{\arraystretch}{1.1} % Adjusting row spacing
    \captionsetup{justification=raggedright,singlelinecheck=false}
    \caption{Boundary and initial conditions for the Periodic-Hill case. Propagation conditions match the Baseline case. Wall Function is abbreviated to W.F.}
    \begin{tabular}{|l|c|c|c|c|c|}
        \hline
        \multicolumn{6}{|c|}{\textbf{Boundary Conditions}} \\
        \hline
        \textbf{Location} & $U$ [m/s] & $p$ [m$^2$/s$^2$] & $k$ [m$^2$/s$^2$] & $\omega$ [s$^{-1}$] & $\nu_t$ [m$^2$/s] \\
        \hline
        Inlet/Outlet & cyclic & cyclic & cyclic & cyclic & cyclic \\
        \hline
        Top/Bottom & no & zero & kqR & omega & nutUSpalding \\
                  & Slip & Gradient & W.F. & W.F. & W.F. \\
        \hline
        \multicolumn{6}{|c|}{\textbf{Initial Conditions}} \\
        \hline
        Baseline & [$U_{ref}$ 0 0] & 0 & $1 \times 10^{-15}$ & $\omega_{ref}$ & 0 \\
        \hline
        Frozen & $U_{LES,field}$ & 0 & $k_{LES,field}$ & $\omega_{ref}$ & 0 \\
        \hline
    \end{tabular}
    \label{tab:boundaryconditions-PH}
\end{table}

\subsection{Curved Backward Facing Step}

The Curved Backward Facing Step case, henceforth referred to as CBFS, closely resembles the NASA-Hump and Periodic-Hill cases. It involves a turbulent boundary layer separating from a curved step under an adverse pressure gradient. The key distinction lies in the contour of the step, which has a much gentler curvature compared to the other two cases, thereby promoting the flow to remain attached for longer. The case setup, the high-fidelity LES data and computational grid have been obtained from the study of \citet{bentaleb_large-eddy_2012}.

\subsubsection{Flow Parameters}

An overview of the main flow parameters specified for the CBFS case is given in Table \ref{tab:CBFStransportproperties}. This case has a Reynolds number equal to 13,700, computed based on the step height $H$ and the free-stream inlet reference velocity $U_{ref}$.

\begin{table}[H]
    \centering
    \small
    \renewcommand{\arraystretch}{1.1} % Adjusting row spacing
      \captionsetup{justification=raggedright,singlelinecheck=false}
    \caption{Overview of the flow parameters specified for the CBFS case.}
    \begin{tabular}{|c|c|c|}
        \hline
        \textbf{Transport Property} & \textbf{Parameter} & \textbf{Value} \\
        \hline
        Reynolds Number based on step height & $ Re_H$ & 13,700 \\
        \hline
        Kinetic viscosity & $\nu$ & 7.23 x 10$^{-5}$ m$^2$ s$^{-1}$ \\
        \hline
        Free-stream reference velocity & $U_{ref}$ & 1 m s$^{-1}$ \\
        \hline
        Reference kinetic energy & $k_{ref}$ & 0.00668 m$^2$ s$^{-2}$\\
        \hline
        Reference specific dissipation rate & $\omega_{ref}$ & 0.110 s$^{-1}$\\
        \hline
        Reference pressure & $p_{ref}$ & 0 m$^2$ s$^{-2}$ \\
        \hline
    \end{tabular}
    \label{tab:CBFStransportproperties}
\end{table}

\subsubsection{Initial and Boundary Conditions}

The OpenFOAM boundary and initial conditions for this case are detailed in Table \ref{tab:boundaryconditions-CBFS}. These are similar to the ones prescribed for the other 2D-separated flow cases.

\begin{table}[H]
    \centering
    \small
    \renewcommand{\arraystretch}{1.1} % Adjusting row spacing
    \captionsetup{justification=raggedright,singlelinecheck=false}
    \caption{Boundary and initial conditions for the CBFS case for different simulation types: Baseline, Frozen and propagation conditions. Propagation conditions match Baseline case.}
    \begin{tabular}{|l|c|c|c|c|c|}
        \hline
        \multicolumn{6}{|c|}{\textbf{Boundary Conditions}} \\
        \hline
        \textbf{Location} & $U$ [m/s] & $p$ [m$^2$/s$^2$] & $k$ [m$^2$/s$^2$] & $\omega$ [s$^{-1}$] & $\nu_t$ [m$^2$/s] \\
        \hline
        Inlet & fixed & zero & fixed & fixed & calcu- \\
              & Value & Gradient & Value & Value & lated \\
        \hline
        Outlet & zero & fixed & zero & zero & calcu- \\
               & Gradient & Value & Gradient & Gradient & lated \\
        \hline
        Top Wall & no & zero & 1 x 10$^{-15}$ & omega & nutLowRe \\
                & Slip & Gradient &  & W.F. & W.F. \\
        \hline
        Bottom Wall & no & zero & 1 x 10$^{-15}$ & omega & nutLowRe \\
                   & Slip & Gradient &  & W.F. & W.F. \\
        \hline
        \multicolumn{6}{|c|}{\textbf{Initial Conditions}} \\
        \hline
        Baseline & [$U_{ref}$ 0 0] & 0 & $k_{ref}$ & $\omega_{ref}$ & 0 \\
        \hline
        Frozen & $U_{LES,field}$ & 0 & $k_{LES,field}$ & $\omega_{ref}$ & 0 \\
        \hline
    \end{tabular}
    \label{tab:boundaryconditions-CBFS}
\end{table}

\subsection{Parameterized Periodic-Hill}

The Parameterized Periodic-Hill case setup and the high-fidelity DNS data have been obtained from the study of \citet{xiao_flows_2020}, who systematically varied the geometry using a hill width-scaling factor ($\alpha = 0.5 - 1.5$). Detail on the flow case Flow parameters and OpenFOAM boundary conditions can be found in database from \url{https://github.com/xiaoh/para-database-for-PIML.git}.  

\subsection{FAITH Hill}

\begin{figure}[H]
    \centering
    \begin{subfigure}[b]{0.49\textwidth}
        \centering
        \includegraphics[width=\textwidth]{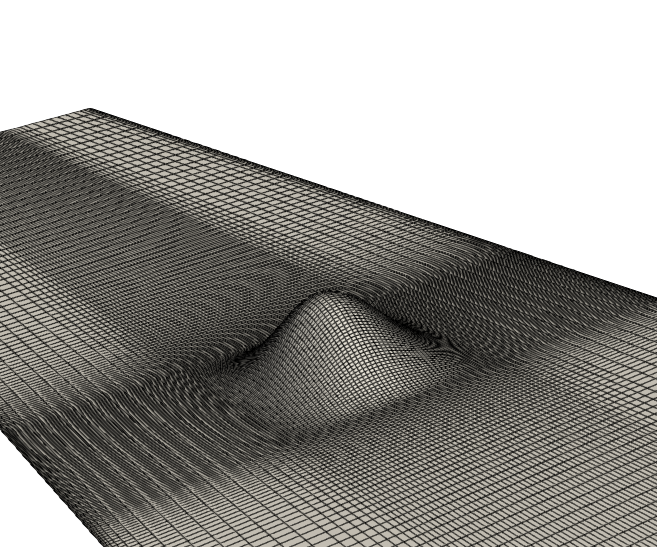} 
        \caption{Surface mesh.}
        \label{fig:faith-surfaceMesh}
    \end{subfigure}
    \hfill
    \begin{subfigure}[b]{0.49\textwidth}
        \centering
        \includegraphics[width=\textwidth]{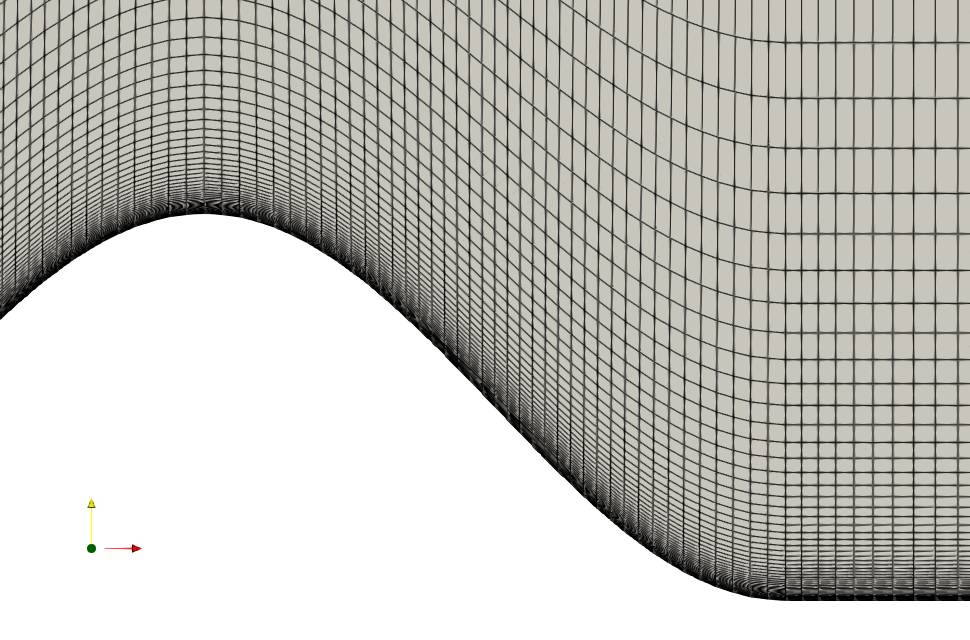} 
        \caption{Volume mesh at symmetry plane.}
        \label{fig:faith-2dMesh}
    \end{subfigure}
    \caption{Surface and volume mesh of Faith Hill case.}
    \label{fig:faith-hill-mesh}
\end{figure}

The 'Fundamental Aero Investigates The Hill' experiment conducted at NASA Ames \cite{bell_surface_2012} is a complex 3-D smooth-body separated aerodynamic flow. The hill has an axisymmetric cosine cross-section, and the oncoming boundary layer thickness was 1/3 the hill height. The computational domain extents were the same as the wind tunnel test section, specifically $L_{x}/H=20$, $L_{y}/H= 5.22$, $L_{z}/H= 8$ where the hill is placed at the center of the section \cite{ho_field_2021}. The velocity at the inlet was prescribed as uniform, and the boundary layer develops naturally resulting in a boundary layer thickness upstream of the hill of $1/3 H$. This matches well with the experimental measurements of \cite{bell_surface_2012}. The turbulence intensity matched the experiment at $Ti$ = 0.13\% and the turbulent length scale used was that of the hill height. A low-$Re$ mesh was used with $1.56\times10^{6}$ cells which was sufficient to obtain grid independence. The surface mesh resolution and volume mesh resolution at the symmetry plane can be seen in Figure \ref{fig:faith-hill-mesh}.

\subsubsection{Flow Parameters}

An overview of the main flow parameters specified for the FAITH hill case is given in Table \ref{tab:FAITHhilltransportproperties}. This case has a Reynolds number equal to 500,000, computed based on the step height $H$ and the free-stream inlet reference velocity $U_{ref}$.

\begin{table}[H]
    \centering
    \small
    \renewcommand{\arraystretch}{1.1} % Adjusting row spacing
      \captionsetup{justification=raggedright,singlelinecheck=false}
    \caption{Overview of the flow parameters specified for the FAITH hill case.}
    \begin{tabular}{|c|c|c|}
        \hline
        \textbf{Transport Property} & \textbf{Parameter} & \textbf{Value} \\
        \hline
        Kinetic viscosity & $\nu$ & 2.0 x 10$^{-6}$ m$^2$ s$^{-1}$ \\
        \hline
        Free-stream reference velocity & $U_{ref}$ & 1 m s$^{-1}$ \\
        \hline
        Reference kinetic energy & $k_{ref}$ & $2.535\times10^{-5}$ m$^2$ s$^{-2}$\\
        \hline
        Reference specific dissipation rate & $\omega_{ref}$ & 0.029 s$^{-1}$\\
        \hline
        Reference pressure & $p_{ref}$ & 0 m$^2$ s$^{-2}$ \\
        \hline
    \end{tabular}
    \label{tab:FAITHhilltransportproperties}
\end{table}

\subsubsection{Initial and Boundary Conditions}

The OpenFOAM boundary and initial conditions for this case are detailed in Table \ref{tab:boundaryconditions-FH}.

\begin{table}[H]
    \centering
    \small
    \renewcommand{\arraystretch}{1.1} % Adjusting row spacing
    \captionsetup{justification=raggedright,singlelinecheck=false}
    \caption{Boundary and initial conditions for the Faith Hill case.}
    \begin{tabular}{|l|c|c|c|c|c|}
        \hline
        \multicolumn{6}{|c|}{\textbf{Boundary Conditions}} \\
        \hline
        \textbf{Location} & $U$ [m/s] & $p$ [m$^2$/s$^2$] & $k$ [m$^2$/s$^2$] & $\omega$ [s$^{-1}$] & $\nu_t$ [m$^2$/s] \\
        \hline
        Inlet & fixed & zero & fixed & fixed & calcu- \\
              & Value & Gradient & Value & Value & lated \\
        \hline
        Outlet & zero & fixed & zero & zero & calcu- \\
               & Gradient & Value & Gradient & Gradient & lated \\
        \hline
        Hill/Walls & no & zero & kqR & omega & nutUSpalding \\
                  & Slip & Gradient & W.F. & W.F. & W.F. \\
        \hline
        Symmetry & symmetry & symmetry & symmetry & symmetry & symmetry \\
                & Plane & Plane & Plane & Plane & Plane \\
        \hline
        \multicolumn{6}{|c|}{\textbf{Initial Conditions}} \\
        \hline
        Baseline & [$U_{ref}$ 0 0] & 0 & $k_{ref}$ & $\omega_{ref}$ & 0 \\
        \hline
    \end{tabular}
    \label{tab:boundaryconditions-FH}
\end{table}

\subsection{Ahmed body}

\begin{figure}[H]
    \centering
    \begin{subfigure}[b]{\textwidth}
        \centering
        \includegraphics[width=0.9\textwidth]{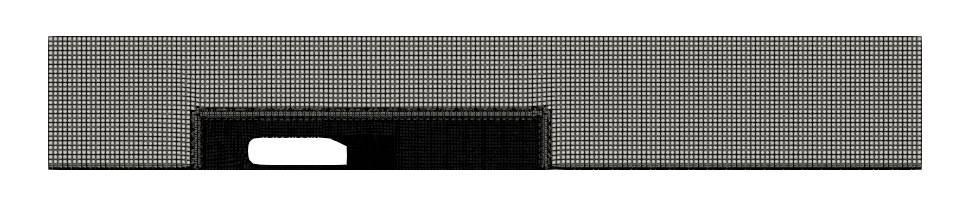}
        \caption{Domain volume refinement at symmetry plane.}
        \label{fig:Ahmed-domain-mesh}
    \end{subfigure}
    \hfill
    \begin{subfigure}[b]{0.49\textwidth}
        \centering
        \includegraphics[width=\textwidth]{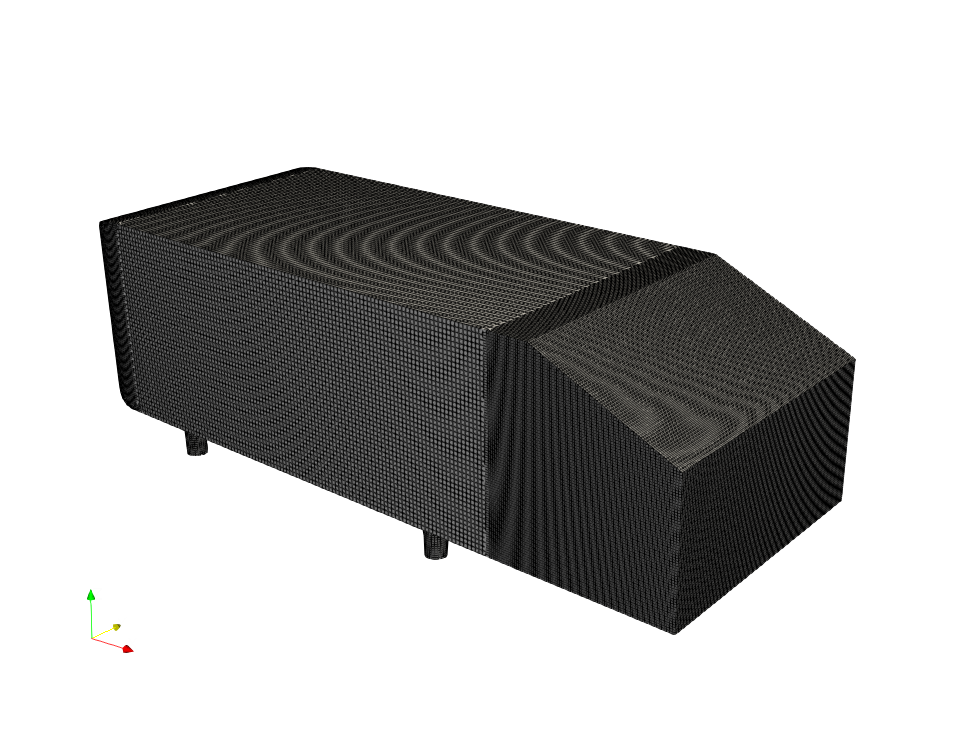} 
        \caption{Body surface mesh.}
        \label{fig:Ahmed-body-surface-mesh}
    \end{subfigure}
    \hfill
    \begin{subfigure}[b]{0.49\textwidth}
        \centering
        \includegraphics[width=\textwidth]{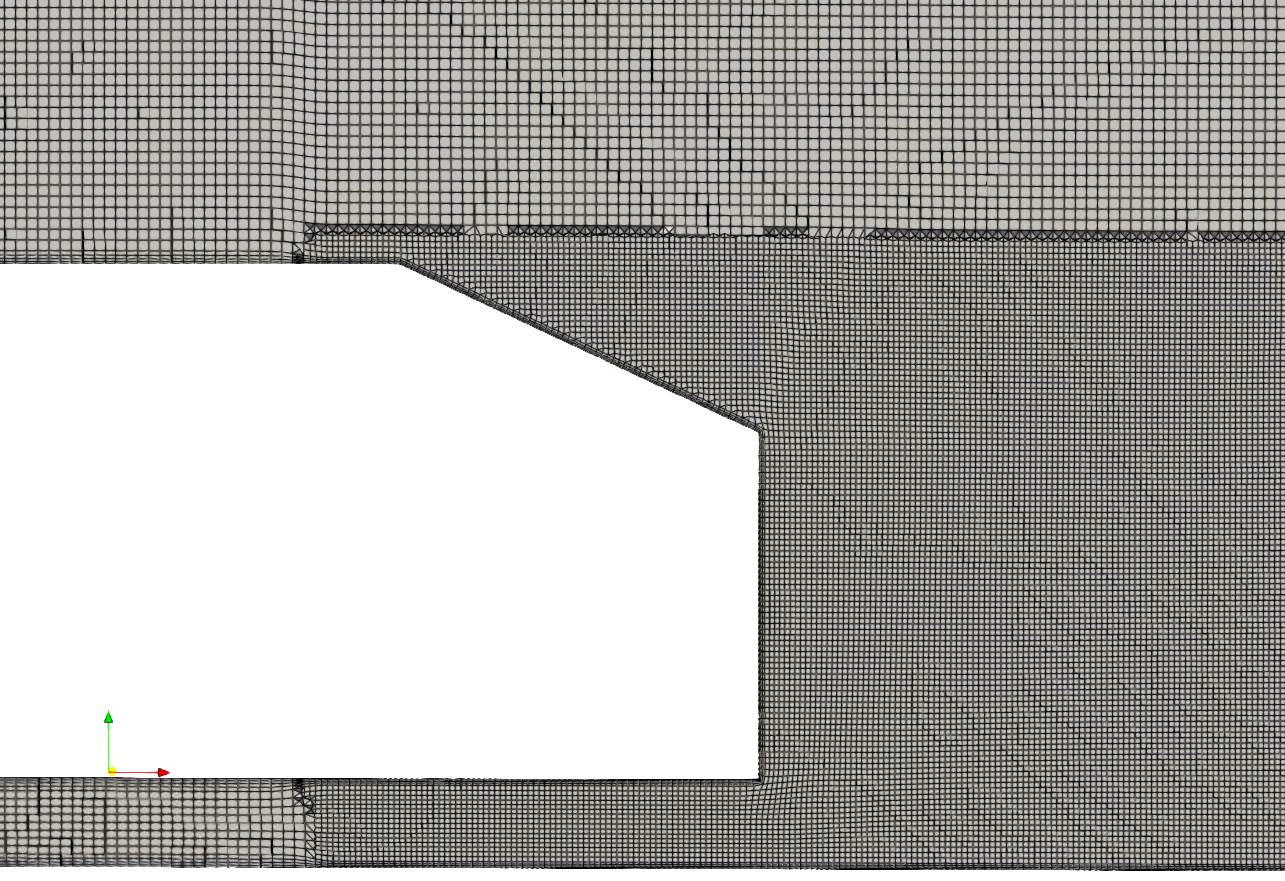} 
        \caption{Close-up of rear refinement.}
        \label{fig:Ahmed-body-surface-rear-mesh}
    \end{subfigure}
    \caption{Surface and volume mesh of Ahmed Body case.}
    \label{fig:ahmed-mesh}
\end{figure}

The Ahmed body (\citet{wagner_flow_2002}) comprises a flat front with rounded corners and a sharp slanted rear upper surface. The 25 degree rear slant angle was simulated including the stilts on which the model was mounted. An inlet condition is imposed 3m upstream of the body and an outlet condition is imposed 6m downstream. A no-slip wall condition is imposed on the ground plane and
car body, with slip walls applied to the wind tunnel walls.
An unstructured mesh of 11M cells was generated using snappyHexMesh (Figure \ref{fig:ahmed-mesh}). A high-y+ approach was taken with 3 prism layers, resulting in a $y^{+}$ range of approximately 50-100 over the top surface of the body.

\subsubsection{Flow Parameters}

An overview of the main flow parameters specified for the Ahmed Body case is given in Table \ref{tab:AhmedBodytransportproperties}. This case has a Reynolds number equal to 760,000 based on the body height $H$ (0.288m) or 278,000 based on body length (1.044m) with the free-stream inlet reference velocity $U_{ref}=40ms^{-1}$.

\begin{table}[H]
    \centering
    \small
    \renewcommand{\arraystretch}{1.1} % Adjusting row spacing
      \captionsetup{justification=raggedright,singlelinecheck=false}
    \caption{Overview of the flow parameters specified for the Ahmed body case.}
    \begin{tabular}{|c|c|c|}
        \hline
        \textbf{Transport Property} & \textbf{Parameter} & \textbf{Value} \\
        \hline
        Kinetic viscosity & $\nu$ & 15 x 10$^{-6}$ m$^2$ s$^{-1}$ \\
        \hline
        Free-stream reference velocity & $U_{ref}$ & 40 m s$^{-1}$ \\
        \hline
        Reference kinetic energy & $k_{ref}$ & 0.00108 m$^2$ s$^{-2}$\\
        \hline
        Reference specific dissipation rate & $\omega_{ref}$ & 0.110 s$^{-1}$\\
        \hline
        Reference pressure & $p_{ref}$ & 0 m$^2$ s$^{-2}$ \\
        \hline
    \end{tabular}
    \label{tab:AhmedBodytransportproperties}
\end{table}

\subsubsection{Initial and Boundary Conditions}

The OpenFOAM boundary and initial conditions for this case are detailed in Table \ref{tab:boundaryconditions-AB}. These are similar to the ones prescribed for the other 2D-separated flow cases.

\begin{table}[H]
    \centering
    \small
    \renewcommand{\arraystretch}{1.1} % Adjusting row spacing
    \captionsetup{justification=raggedright,singlelinecheck=false}
    \caption{Boundary and initial conditions for the Ahmed body case.}
    \begin{tabular}{|l|c|c|c|c|c|}
        \hline
        \multicolumn{6}{|c|}{\textbf{Boundary Conditions}} \\
        \hline
        \textbf{Location} & $U$ [m/s] & $p$ [m$^2$/s$^2$] & $k$ [m$^2$/s$^2$] & $\omega$ [s$^{-1}$] & $\nu_t$ [m$^2$/s] \\
        \hline
        Inlet & fixed & zero & fixed & fixed & calcu- \\
              & Value & Gradient & Value & Value & lated \\
        \hline
        Outlet & zero & fixed & zero & zero & calcu- \\
               & Gradient & Value & Gradient & Gradient & lated \\
        \hline
        Body/Ground & no & zero & kqR & omega & nutUSpalding \\
        Plane & Slip & Gradient & W.F. & W.F. & W.F. \\
        \hline
        External & Slip & zero & zero & omega & zero \\
        Walls &  & Gradient & Gradient & W.F. & Gradient \\
        \hline
        \multicolumn{6}{|c|}{\textbf{Initial Conditions}} \\
        \hline
        Baseline & [$U_{ref}$ 0 0] & 0 & $k_{ref}$ & $\omega_{ref}$ & 0 \\
        \hline
    \end{tabular}
    \label{tab:boundaryconditions-AB}
\end{table}

\section{Model Coefficients and Auxiliary Equations}
\label{app:coef}
\newpage
\begin{tcolorbox}[colback=white,colframe=gray,title=\textbf{Zonally Augmented $k-\omega$ SST Model Coefficients}]
\textbf{Model Coefficients:}
Let $\Phi_1$ represent the coefficients in the $k-\omega$ model and $\Phi_2$ those in the transformed $k-\epsilon$ model. The coefficients $\Phi$ in the $k-\omega$ SST model are found from : $\Phi = F_1\Phi_1 + (1-F_1)\Phi_2$.
The $\Phi_1$ set of coefficients:
\begin{equation}
\sigma_{k1} = 0.85, \quad \sigma_{\omega 1} = 0.5, \quad \beta_1 = 0.0750, \quad \beta^* = 0.09, \quad \gamma_1 = 5/9
\end{equation}
The $\Phi_2$ set of coefficients:
\begin{equation}
\sigma_{k2} = 1.0, \quad \sigma_{\omega 2} = 0.856, \quad \beta_2 = 0.0828, \quad \beta^* = 0.09, \quad \gamma_2 = 0.44
\end{equation}
\textbf{Auxiliary Equations:}
\begin{equation}
\label{eq:arg1}
   F_1 = \tanh{(arg_1^4)} \text{ where } arg_1 = \min\left[\max \left( \frac{\sqrt{k}}{\beta^{*}\omega y}, \frac{500\nu}{y^2\omega} \right), \frac{4\rho \sigma_{\omega_2}k}{CD_{k\omega}y^2}\right]
\end{equation}
\begin{equation}
    CD_{k\omega}=\max\left( 2\rho \sigma_{\omega_2}  \frac{1}{\omega}\frac{\partial k}{\partial x_i}\frac{\partial \omega}{\partial x_i} ,10^{-20}\right)
\end{equation}
\begin{equation}
    F_2 = \tanh{(arg_2^2)}  \quad \text{where} \quad     arg_2 = \max{\left(\frac{2\sqrt{k}}{\beta^*\omega y}, \frac{500\nu}{y^2 \omega} \right)}
\end{equation}
\label{box:New_Model}

\end{tcolorbox}

\section{Training Case Performance for PH and CBFS}
\label{app:Train_Cases}
\subsection{Periodic Hill}

\begin{figure}[H]
    \centering
    \begin{subfigure}[b]{0.75\textwidth}
        \centering
        \includegraphics[width=\textwidth]{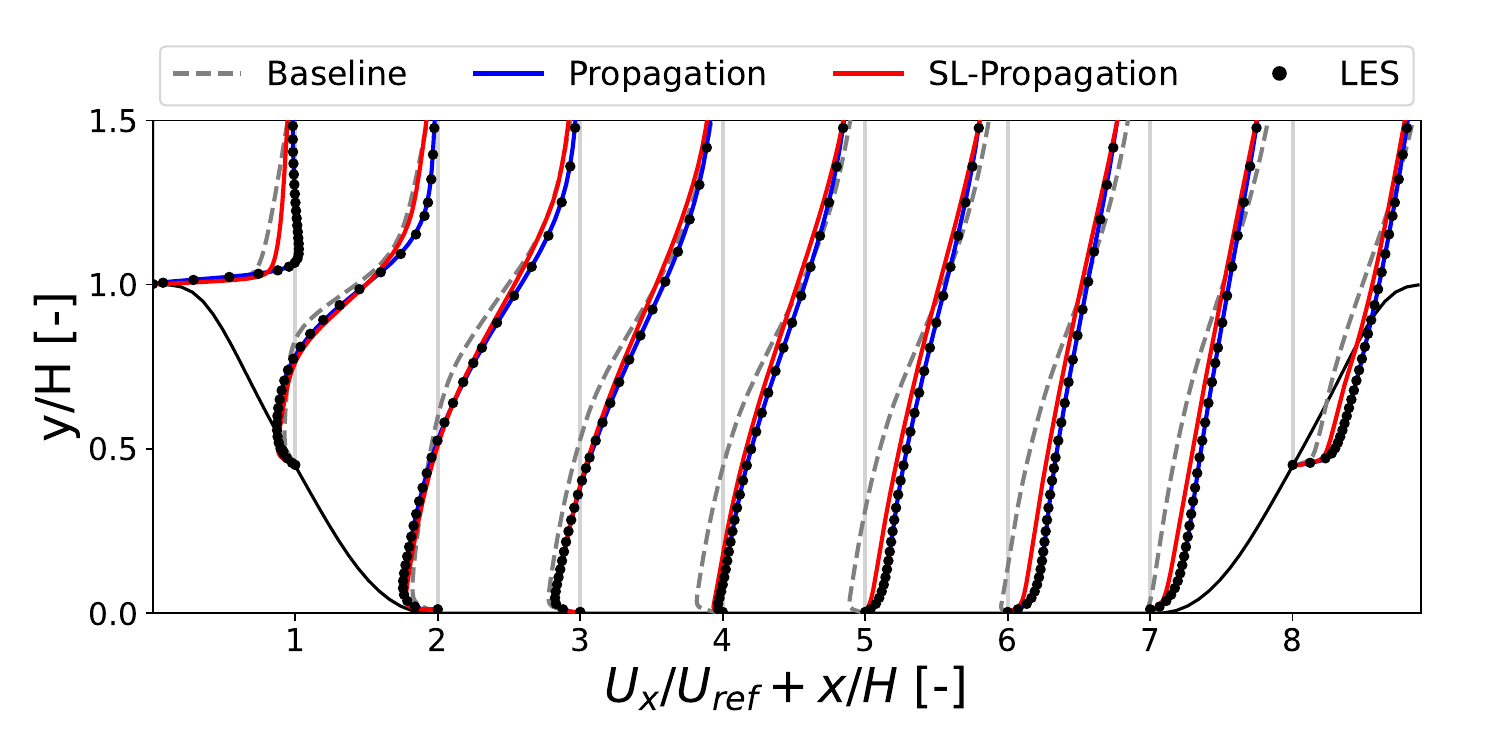} 
        \caption{Axial velocity profiles.}
        \label{fig:ux-prop-PH-final}
    \end{subfigure}
    
    \begin{subfigure}[b]{0.75\textwidth}
        \centering
        \includegraphics[width=\textwidth]{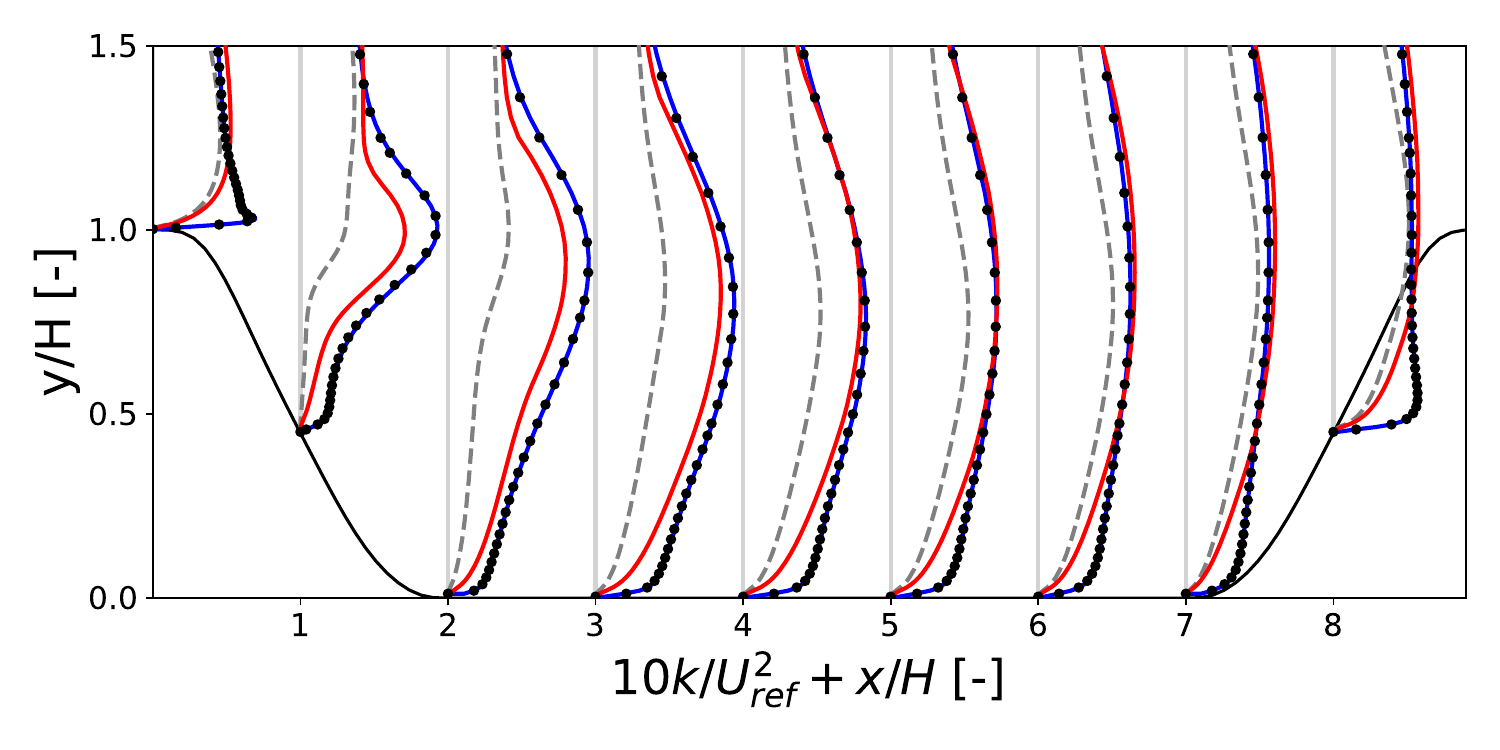}
        \caption{Turbulent kinetic energy profiles.}
        \label{fig:k-prop-PH-final}
    \end{subfigure}
    \hfill
    \begin{subfigure}[b]{0.75\textwidth}
        \centering
        \includegraphics[width=\textwidth]{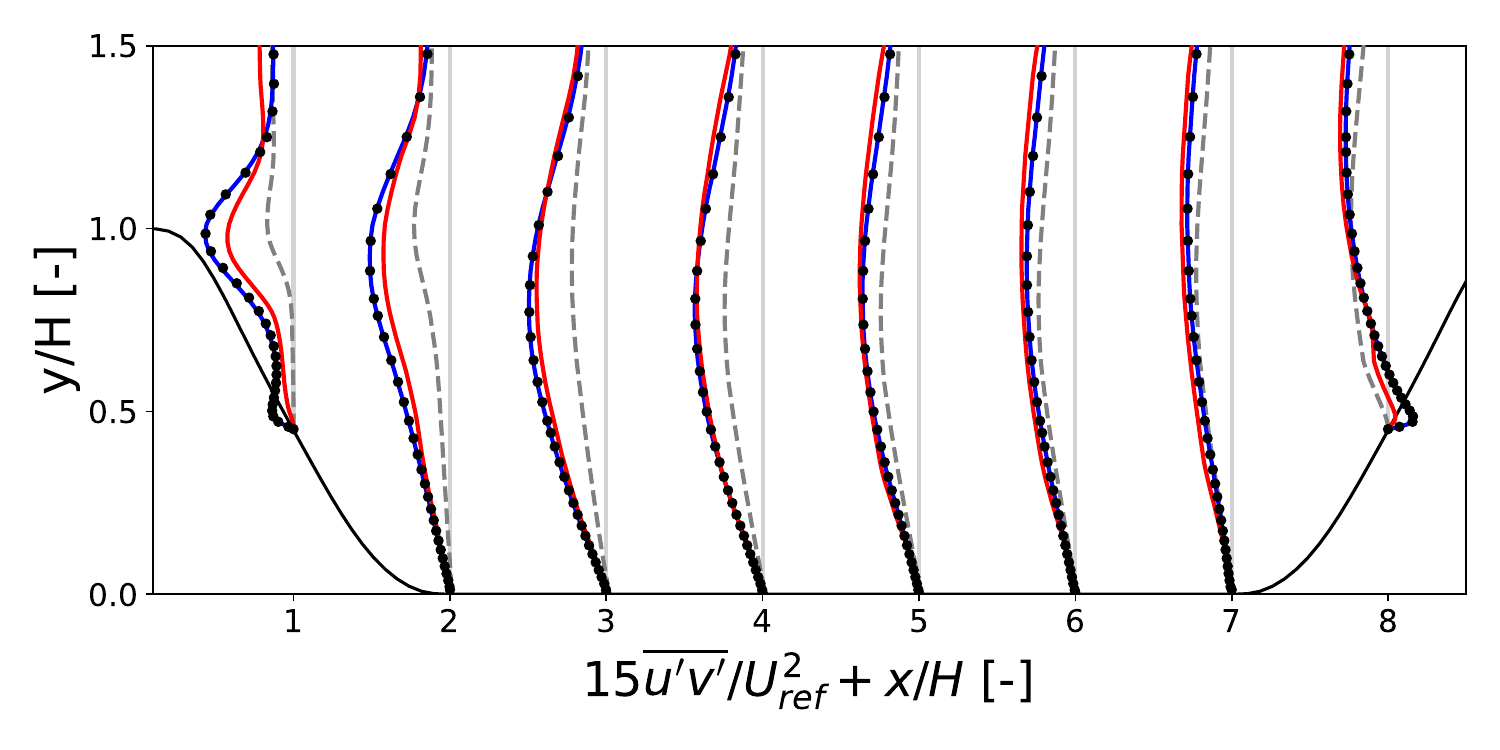} 
        \caption{Reynolds shear stress profiles.}
        \label{fig:upwp-prop-PH-final}
    \end{subfigure}
    
    \caption{Performance comparison between full field propagation and shear layer propagation on the Periodic-Hill training case. }
    \label{fig:prop-PH-final}
\end{figure}

\begin{figure}[H]
    \centering
    \includegraphics[width=\linewidth]{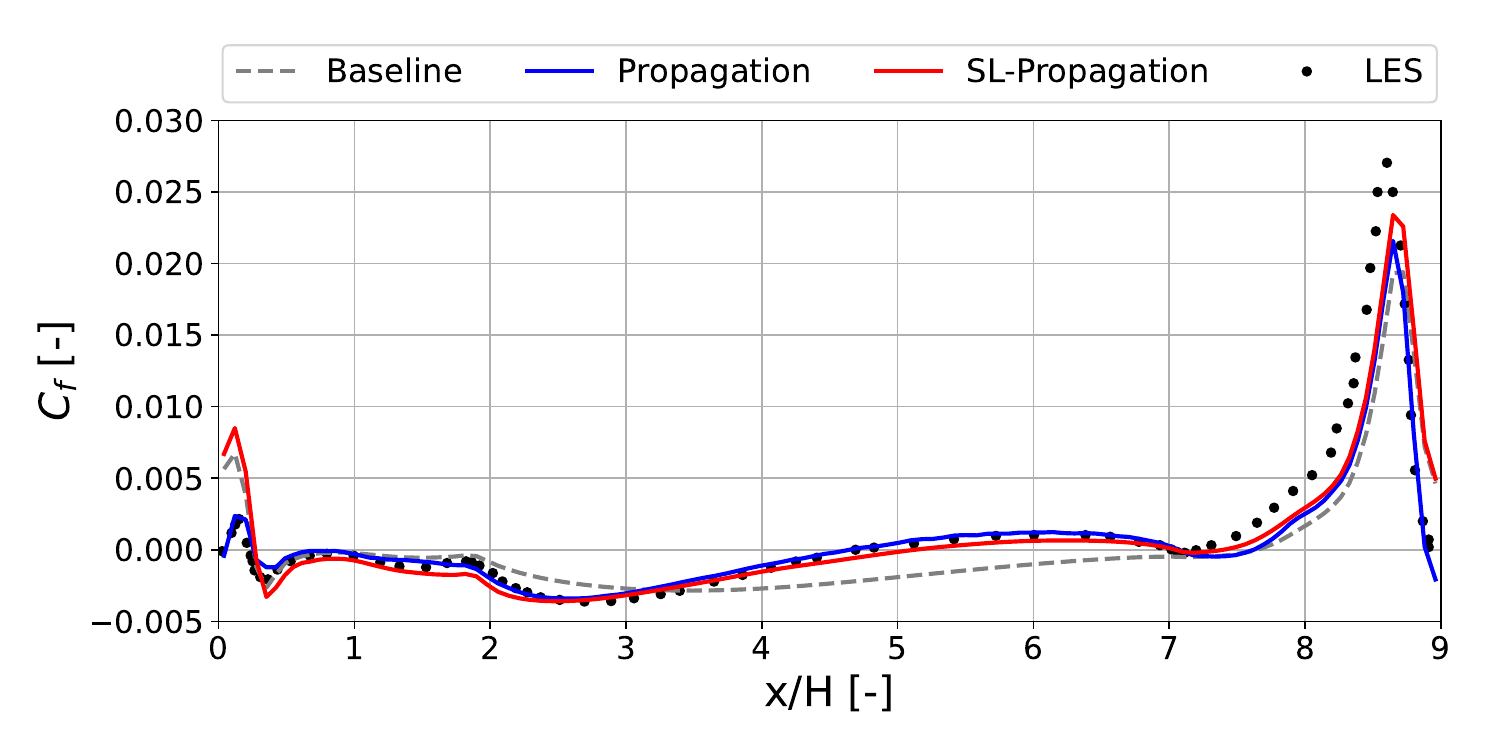}
    \caption{ Skin friction comparison plot between full field propagation and shear layer propagation on the Periodic-Hill training case. }
    \label{fig:Cf-Prop-PH-final}
\end{figure}

In this section, we examine the model's performance on the periodic hill configuration where the flow undergoes sustained cycles of separation and reattachment. Figures \ref{fig:prop-PH-final} and \ref{fig:Cf-Prop-PH-final} compare the performance of full-field and shear layer-targeted correction propagation for this configuration.

\begin{figure}[H]
    \centering
    \begin{subfigure}[b]{0.75\textwidth}
        \centering
        \includegraphics[width=\textwidth]{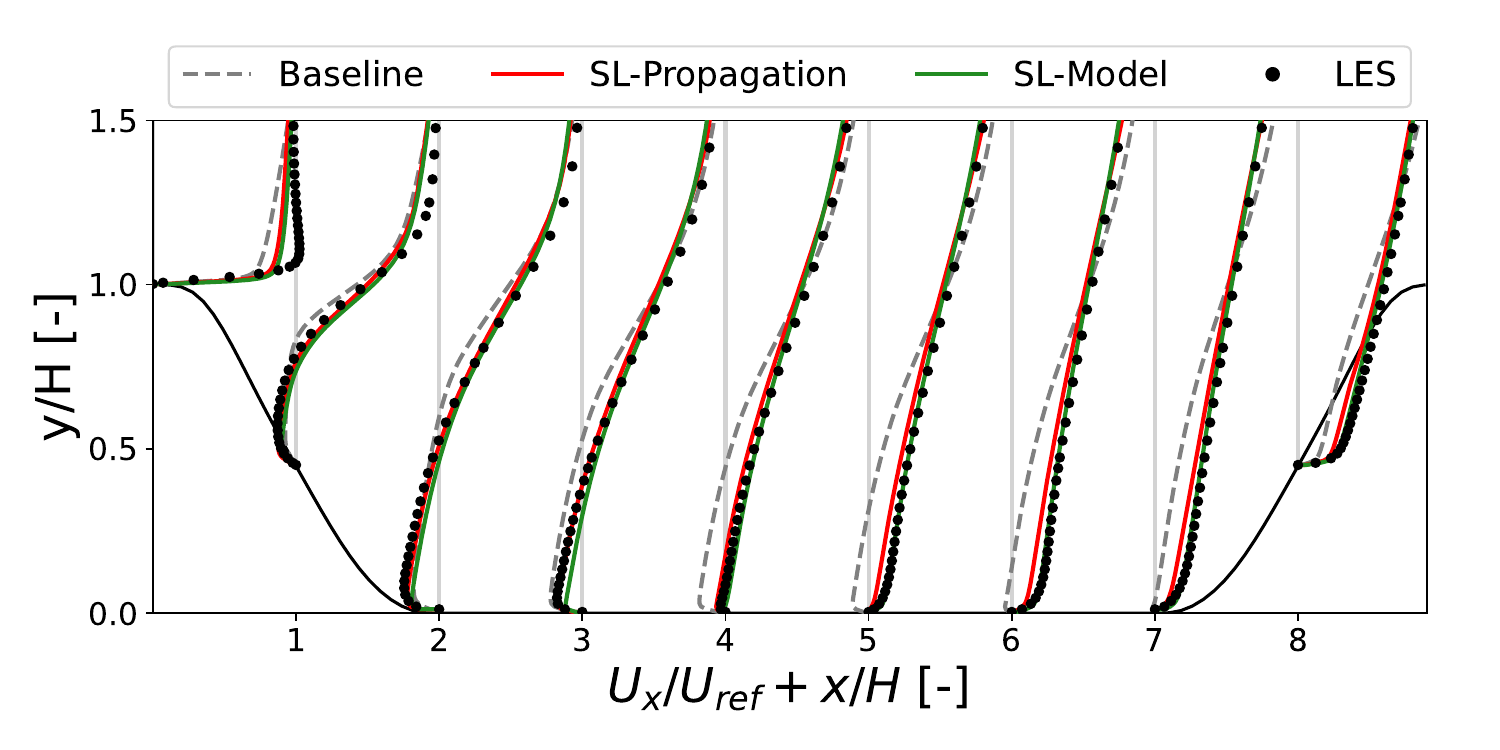} 
        \caption{Axial velocity profiles.}
        \label{fig:ux-prop-PH-model}
    \end{subfigure}
    
    \begin{subfigure}[b]{0.75\textwidth}
        \centering
        \includegraphics[width=\textwidth]{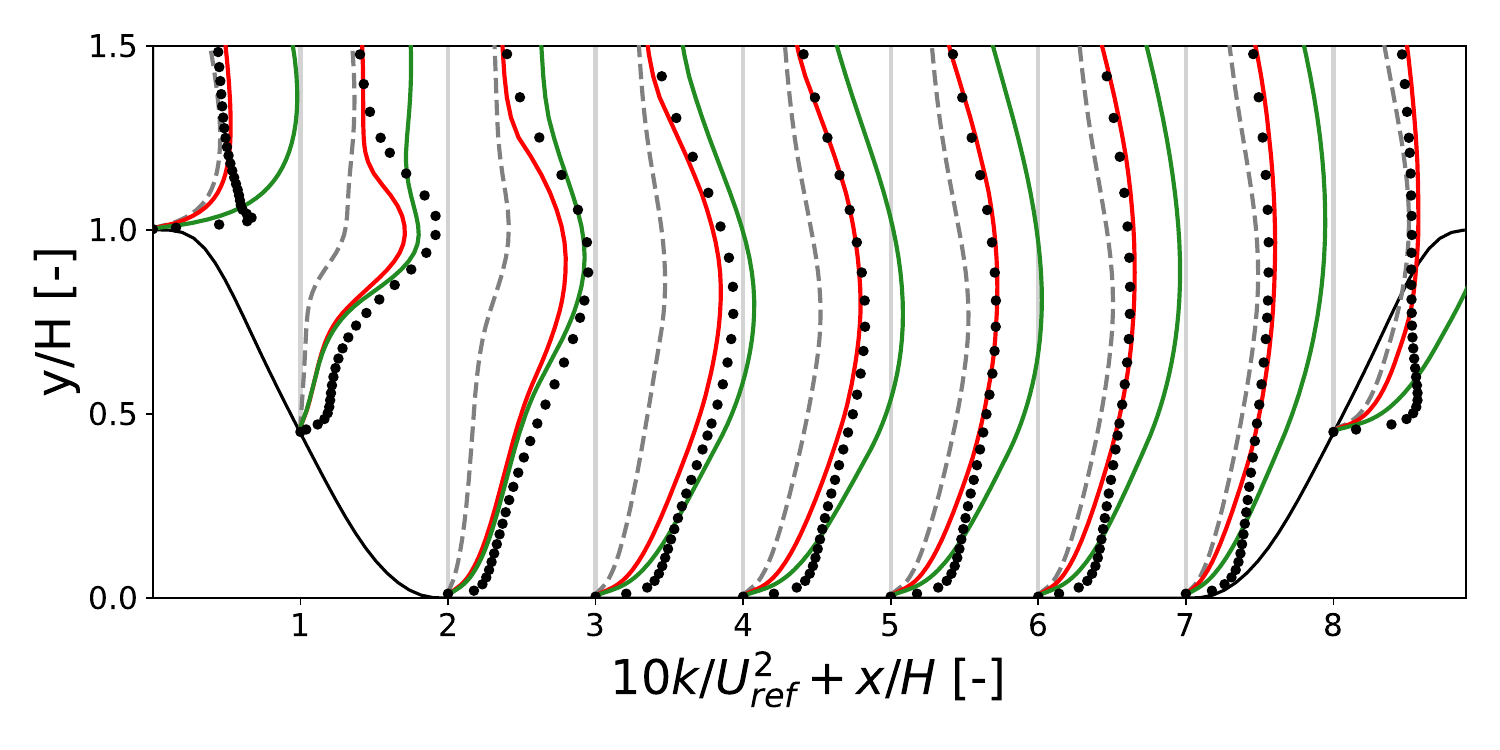}
        \caption{Turbulent kinetic energy profiles.}
        \label{fig:k-prop-PH-model-final}
    \end{subfigure}
    \hfill
    \begin{subfigure}[b]{0.75\textwidth}
        \centering
        \includegraphics[width=\textwidth]{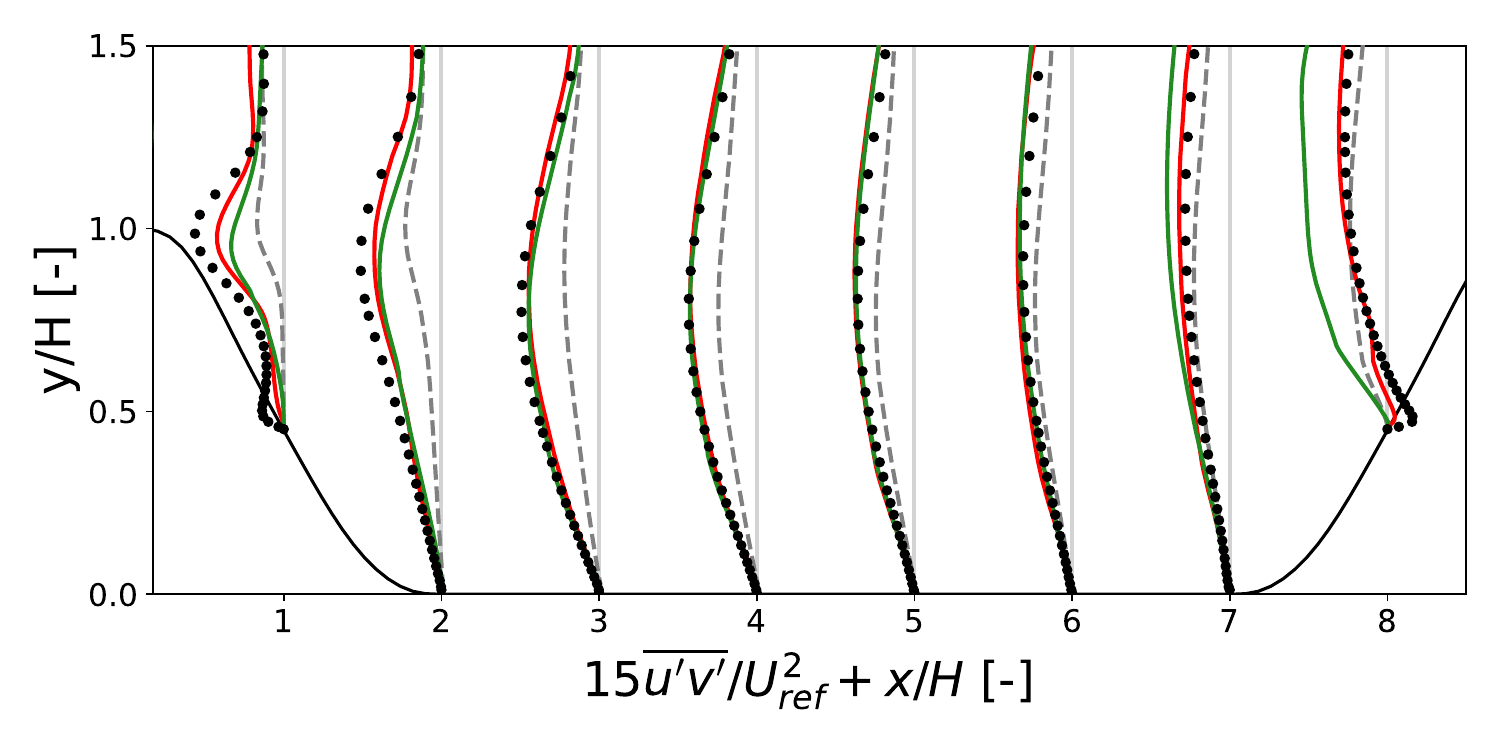} 
        \caption{Reynolds shear stress profiles.}
        \label{fig:upwp-prop-PH-model-final}
    \end{subfigure}
    
    \caption{Performance of the shear layer SpaRTA model on the Periodic-Hill training case.}
    \label{fig:prop-PH-model-final}
\end{figure}

The SL-Propagation case demonstrates comparable performance to full-field propagation, with both approaches improving upon baseline predictions. The velocity profiles show better agreement with LES data throughout the periodic domain, particularly in capturing the recirculation zone extent between hills ($1 < x/H < 3$). The skin friction evolution confirms that the zonal framework maintains accuracy in predicting separation and reattachment behavior.
Building on these results, Figure \ref{fig:prop-PH-model-final} presents the performance of the complete SL-Model implementation.

The SL-Model maintains the improved prediction of separation and reattachment behavior seen in the SL-Propagation case. Analysis of the turbulence quantities reveals characteristics similar to those observed in the NASA Hump case. The turbulent kinetic energy profiles (Figure \ref{fig:k-prop-PH-model-final}) show elevated levels compared to LES data, especially in the shear layer region above the hills, while the Reynolds shear stress distributions (Figure \ref{fig:upwp-prop-PH-model-final}) capture peak stress locations but with magnitudes influenced by the enhanced turbulent kinetic energy prediction. This consistent behavior across both geometries points to an underlying mechanism in the correction approach where mean flow improvements are achieved through intensified turbulent transport.

\begin{figure}[H]
    \centering
    \includegraphics[width=\linewidth]{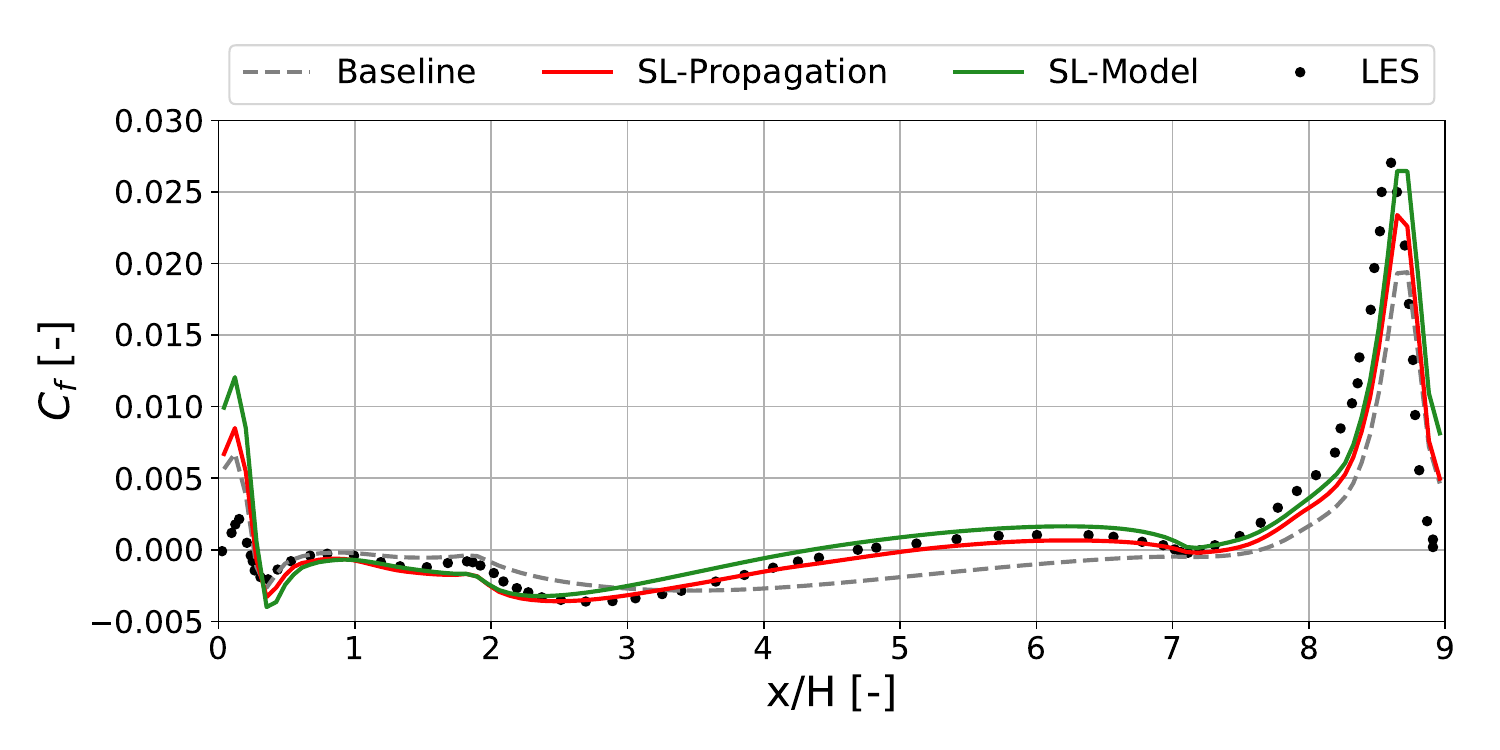}
    \caption{ Skin friction comparison plot for the shear layer SpaRTA model on the Periodic-Hill training case.}
    \label{fig:Cf-Model-PH-final}
\end{figure}

Examination of the skin friction distribution (Figure \ref{fig:Cf-Model-PH-final}) further validates the model's effectiveness across multiple separation-reattachment cycles. The SL-Model accurately reproduces peak skin friction at the hill crest ($x/H \approx 8.5$) while improving predictions in the recovery region between hills.  It successfully captures both negative $C_f$ values in the separation zone and the gradual recovery trend upon reattachment. These results demonstrate that the zonal framework effectively handles the complex physics of periodic separation while maintaining appropriate behavior in attached flow regions.

\subsection{Curved Backwards Facing Step}
\begin{figure}[H]
    \centering
    \begin{subfigure}[b]{0.75\textwidth}
        \centering
        \includegraphics[width=\textwidth]{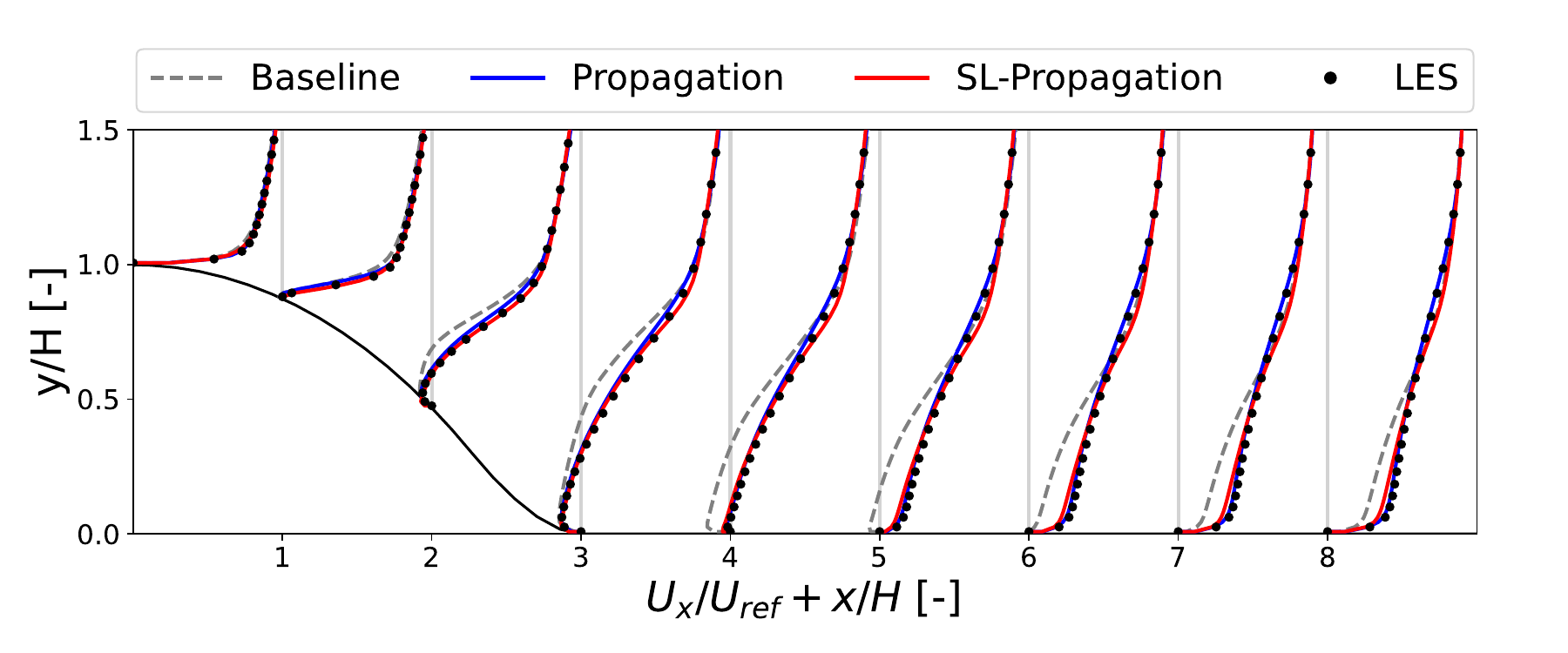} 
        \caption{Axial velocity profiles.}
        \label{fig:ux-prop-CBFS-final}
    \end{subfigure}
    
    \begin{subfigure}[b]{0.72\textwidth}
        \centering
        \includegraphics[width=\textwidth]{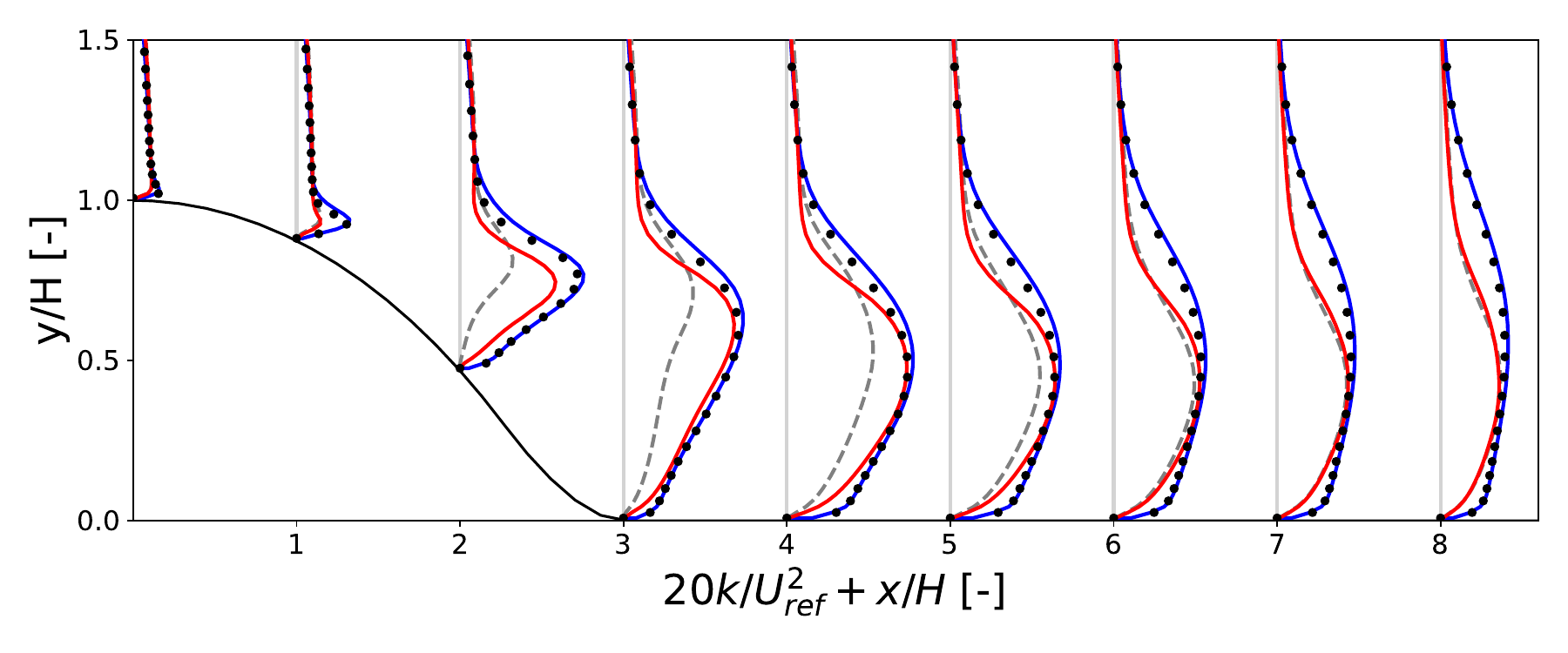}
        \caption{Turbulent kinetic energy profiles.}
        \label{fig:k-prop-CBFS-final}
    \end{subfigure}

    \begin{subfigure}[b]{0.72\textwidth}
        \centering
        \includegraphics[width=\textwidth]{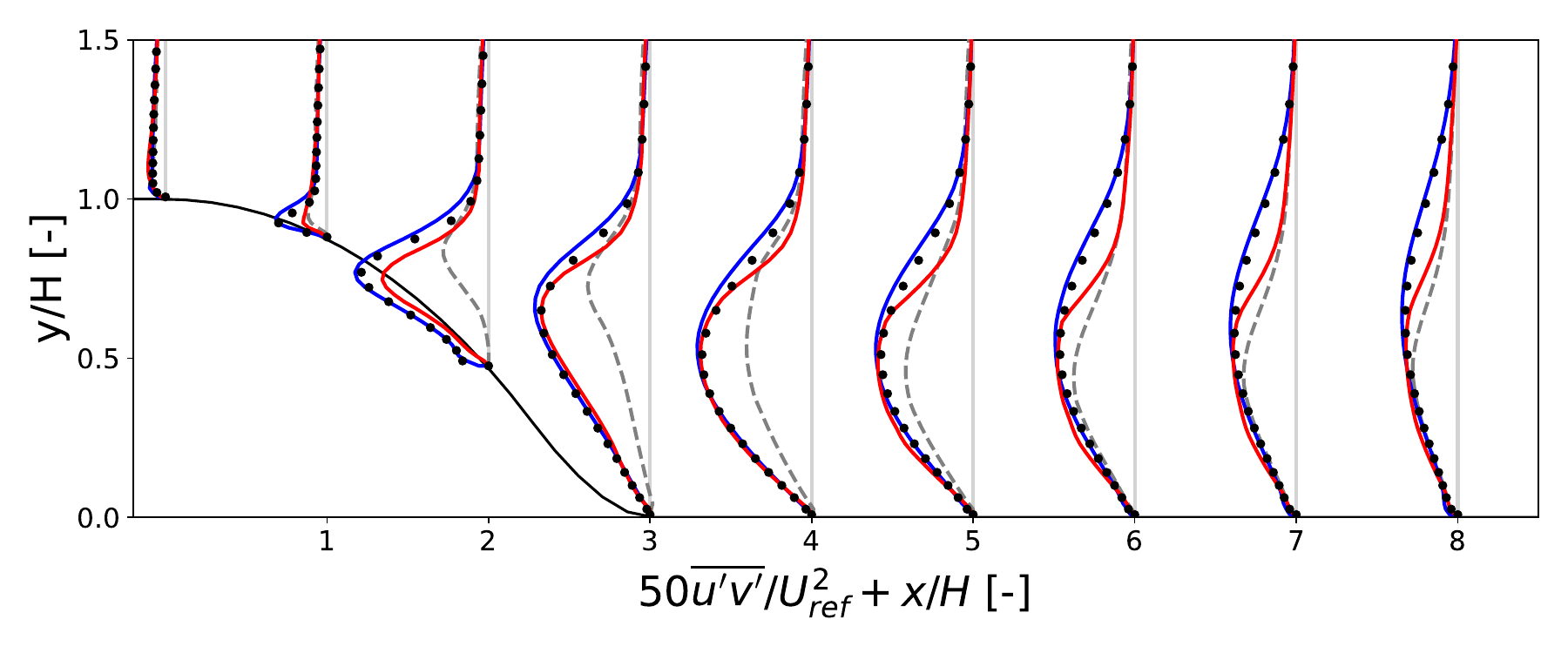} 
        \caption{Reynolds shear stress profiles.}
        \label{fig:upwp-prop-final}
    \end{subfigure}
    
    \caption{Performance comparison between full field propagation and shear layer propagation on the CBFS training case. }
    \label{fig:prop-CBFS-final}
\end{figure}

\begin{figure}[H]
    \centering
    \includegraphics[width=\linewidth]{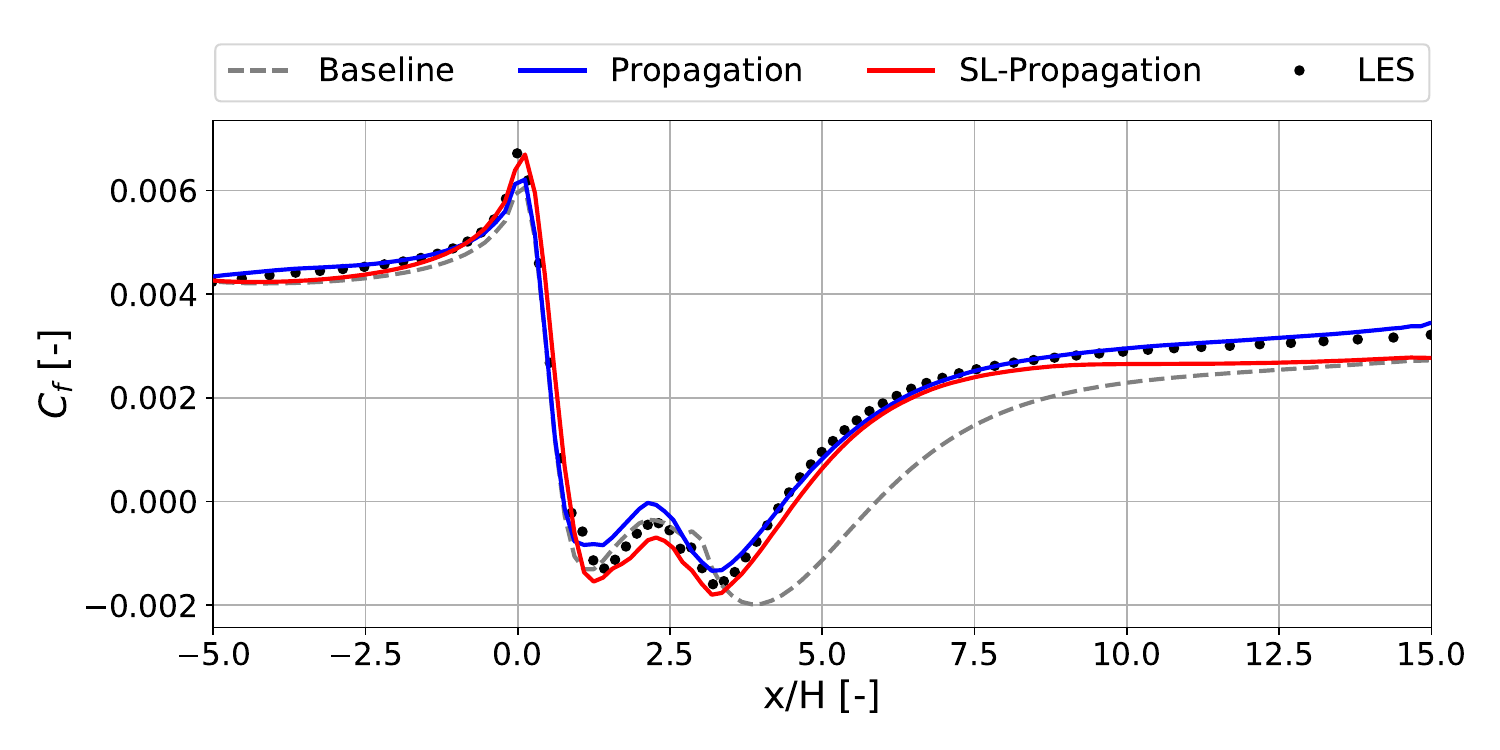}
    \caption{ Skin friction comparison plot between full field propagation and shear layer propagation on the CBFS training case. }
    \label{fig:Cf-Prop-CBFS-final}
\end{figure}

Following validation on smooth and periodic separation, we examine the model's performance on the curved backwards-facing step, where the combination of geometric expansion and surface curvature creates a more complex separation mechanism. Figures \ref{fig:prop-CBFS-final} and \ref{fig:Cf-Prop-CBFS-final} compare the performance of full-field and shear layer-targeted correction propagation for this configuration.

The velocity profiles in Figure \ref{fig:ux-prop-CBFS-final} demonstrate that SL-Propagation achieves comparable flow prediction to full-field propagation, with notable improvements over the baseline model in the post-separation region ($x/H > 2.5$). The turbulent kinetic energy profiles in Figure \ref{fig:k-prop-CBFS-final} show enhanced prediction of mixing layer development downstream of separation, while the Reynolds stress distributions in Figure \ref{fig:upwp-prop-final} better capture the spatial evolution of turbulent transport through the curved section and subsequent expansion. This indicates that the zonal application of corrections effectively maintains the key physical mechanisms captured by full-field propagation.

The skin friction distribution in Figure \ref{fig:Cf-Prop-CBFS-final} validates these improvements quantitatively. Both correction approaches improve the prediction of separation onset and show significantly better agreement with LES data in the recovery region compared to the baseline model. While both correction approaches show similar performance, some differences emerge: SL-Propagation follows LES data more closely near $x/H = 5$, while full-field propagation shows marginally better agreement in the early wake region ($7.5 \leq x/H \leq 10$). The RITA classifier demonstrates effective performance through two key behaviors: maintaining the baseline model's accurate prediction upstream of separation ($x/H < 0$) where corrections aren't needed, and allowing a natural transition back to baseline behavior in the far wake region ($x/H > 10$). The close agreement between SL-Propagation and full-field propagation confirms the RITA classifier's effectiveness in identifying regions requiring correction, particularly through the geometric transition.

\begin{figure}[H]
    \centering
    \begin{subfigure}[b]{0.75\textwidth}
        \centering
        \includegraphics[width=\textwidth]{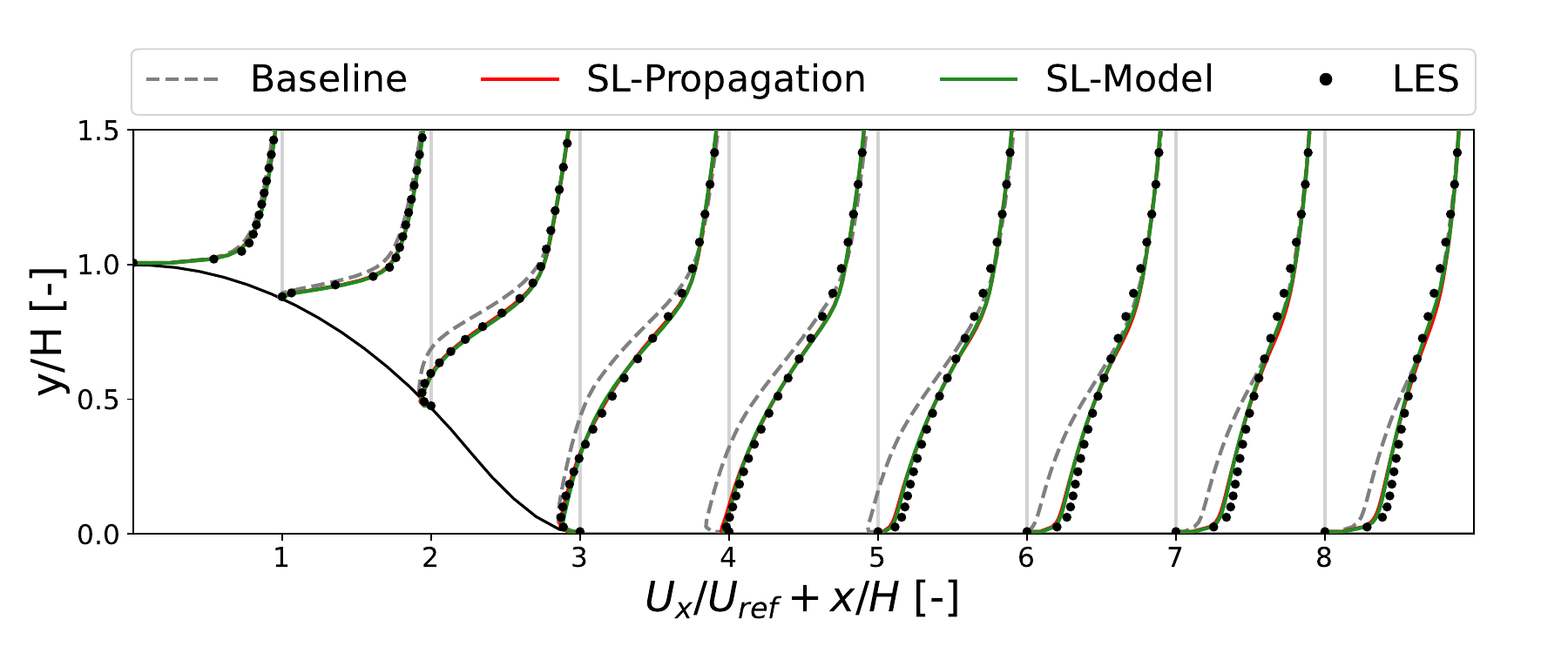} 
        \caption{Axial velocity profiles.}
        \label{fig:ux-prop-CBFS-Model-final}
    \end{subfigure}
    
    \begin{subfigure}[b]{0.72\textwidth}
        \centering
        \includegraphics[width=\textwidth]{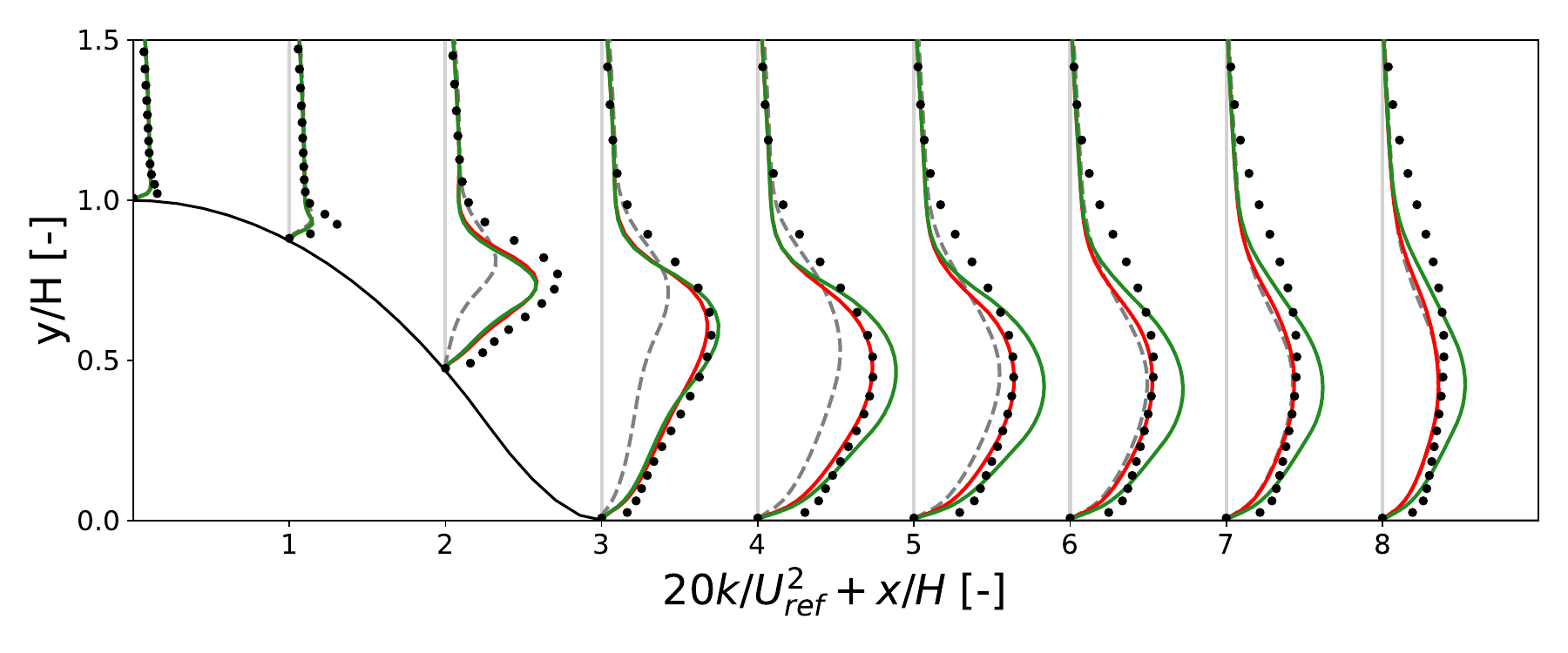}
        \caption{Turbulent kinetic energy profiles.}
        \label{fig:k-prop-CBFS-Model-final}
    \end{subfigure}
    \hfill
    \begin{subfigure}[b]{0.72\textwidth}
        \centering
        \includegraphics[width=\textwidth]{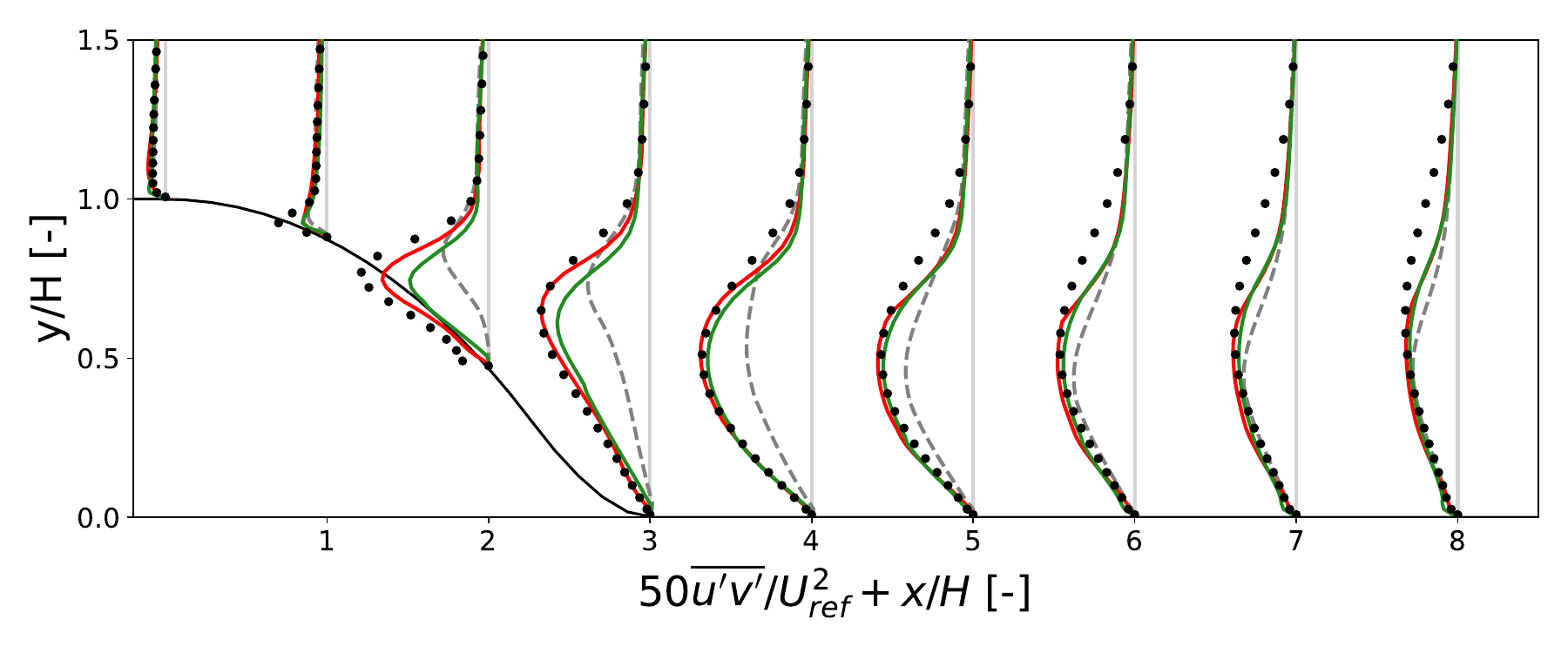} 
        \caption{Reynolds shear stress profiles.}
        \label{fig:upwp-prop-CBFS-Model-final}
    \end{subfigure}
    
    \caption{Performance of the shear layer SpaRTA model on the CBFS training case.}
    \label{fig:prop-CBFS-Model-final}
\end{figure}

Building on these results, the velocity profiles in Figure \ref{fig:ux-prop-CBFS-Model-final} show that the SL-Model successfully maintains the improved prediction of separation bubble size and recovery region. Analysis of the turbulence quantities reveals behavior consistent with previous test cases, but with important differences due to the geometric complexity. The turbulent kinetic energy profiles in Figure \ref{fig:k-prop-CBFS-Model-final} exhibit elevated levels in the shear layer region, particularly pronounced where the flow navigates the curved surface transition. The Reynolds stress distributions in Figure \ref{fig:upwp-prop-CBFS-Model-final} better capture the mixing layer development and its spatial evolution, though their magnitudes reflect the enhanced turbulent kinetic energy predictions seen in all training cases.

\begin{figure}[H]
    \centering
    \includegraphics[width=\linewidth]{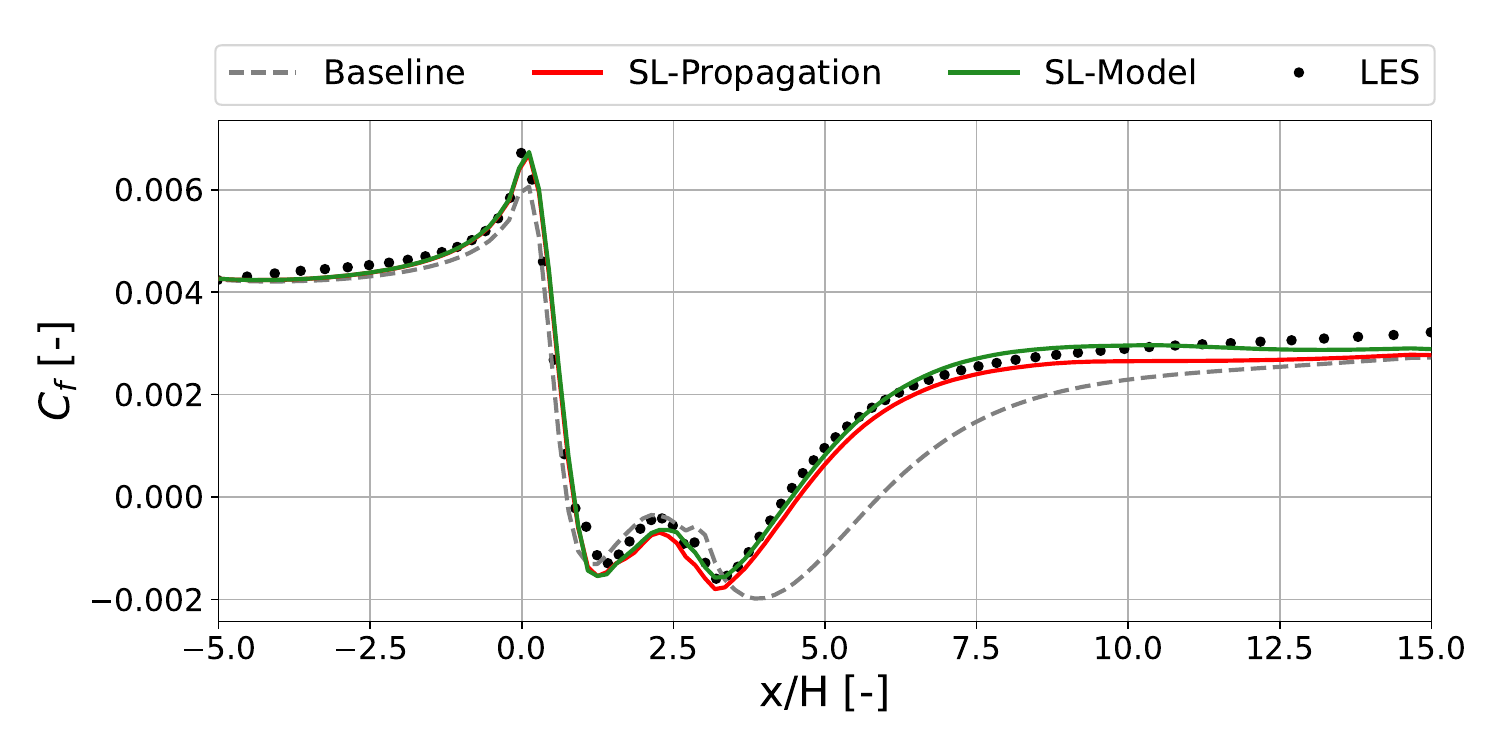}
    \caption{Skin friction comparison plot for the shear layer SpaRTA model on the CBFS training case.}
    \label{fig:Cf-Model-CBFS-final}
\end{figure}

Analysis of the skin friction evolution in Figure \ref{fig:Cf-Model-CBFS-final} confirms the model's robust performance through the geometrically-induced separation. While the baseline model better captures the double-peak structure near $x/H = 2.5$, the SL-Model achieves similar improvements to SL-Propagation in overall separation prediction, particularly in capturing the minimum Cf value ($x/H \approx 3.5$) and recovery rate ($5 < x/H < 7.5$). The model effectively preserves upstream prediction ($x/H < 0$) where the baseline $k-\omega$ SST performs well, and shows marginally better agreement with LES data than SL-Propagation in the far wake region ($x/H > 10$). This demonstrates that the discovered correction terms successfully reproduce the benefits of propagated corrections while maintaining appropriate baseline behavior through geometric transitions.

%%       \citet{<label>} ==> Lamport [21]
%%       \citep{<label>} ==> [21]
%%

\bibliographystyle{elsarticle-num-names} 
\bibliography{zotero_ref_final_v2}



%\begin{thebibliography}{00}


%\bibitem[Lamport(1994)]{lamport94}
%  \textit{\LaTeX: a document preparation system},
%  Addison Wesley, Massachusetts,

%\end{thebibliography}
\end{document}